\documentclass[11pt]{article}

\usepackage{mathtools}
\usepackage{multirow}
\usepackage{graphicx,epsfig}
\usepackage{fancyhdr,fancybox,float}
\usepackage[title]{appendix}
\usepackage{indentfirst}
\usepackage{verbatim}
\usepackage[sort&compress, numbers]{natbib}
\usepackage{geometry}
\usepackage{extarrows,chemarrow}
\usepackage{xcolor}
\usepackage[labelfont=bf]{caption}
\usepackage{bbding}
\usepackage{microtype}
\usepackage{amssymb}
\DisableLigatures[f]{encoding = *, family = *}
\usepackage{times}
\usepackage{amssymb}
\usepackage{amsthm}
\usepackage{bm}
\usepackage[hidelinks]{hyperref} 
\hypersetup{
    colorlinks=true,
    citecolor=blue,
    linkcolor=blue,
    urlcolor=blue,
}
\usepackage[T1]{fontenc}
\usepackage{float}

\newcommand{\paperfont}{\fontsize{10pt}{1.1\baselineskip}\selectfont}
\begin{document}
\theoremstyle{definition}
\makeatletter
\thm@headfont{\bf}
\makeatother
\newtheorem{definition}{Definition}
\newtheorem{example}{Example}
\newtheorem{theorem}{Theorem}
\newtheorem{lemma}{Lemma}
\newtheorem{corollary}{Corollary}
\newtheorem{remark}{Remark}
\newtheorem{proposition}{Proposition}
\lhead{}
\rhead{}
\lfoot{}
\rfoot{}
\renewcommand{\refname}{References}
\renewcommand{\figurename}{Figure}
\renewcommand{\tablename}{Table}
\renewcommand{\proofname}{Proof}
\def\code#1{\texttt{#1}}
\newcommand{\diag}{\mathrm{diag}}
\newcommand{\tr}{\mathrm{tr}}
\newcommand{\re}{\mathrm{Re}}
\newcommand{\one}{\mathbbm{1}}
\newcommand{\Pnum}{\mathbb{P}}
\newcommand{\Enum}{\mathbb{E}}
\newcommand{\Rnum}{\mathbb{R}}
\newcommand{\dnum}{\mathrm{d}}
\newcommand{\hyper}{{}_2F_1}
\newcommand{\confl}{{}_1F_1}

\renewcommand{\figurename}{\textbf{Fig.}} 
\captionsetup[figure]{labelsep=space} 

\newmuskip\pFqskip
\pFqskip=1mu
\mathchardef\pFcomma=\mathcode`, 

\newcommand*\pFq[5]{%
  \begingroup
  \begingroup\lccode`~=`,
    \lowercase{\endgroup\def~}{\pFcomma\mkern\pFqskip}%
  \mathcode`,=\string"8000
  {}_{#1}F_{#2}({#3})%
  \endgroup
}
	
\title{\textbf{Analysis of a detailed multi-stage model of stochastic gene expression using queueing theory and model reduction}}

\author{Muhan Ma$^{1,\dag}$,\;\;\;Juraj Szavits-Nossan$^{1,\dag}$,\;\;\;Abhyudai Singh$^{2}$,\;\;\;Ramon Grima$^{1,*}$\\
\footnotesize $^1$ School of Biological Sciences, University of Edinburgh, Edinburgh EH9 3BF, UK. \\
\footnotesize $^2$ Department of Electrical and Computer Engineering, University of Delaware, Newark DE 19716, USA.\\
\footnotesize $^*$ Correspondence: \href{mailto:ramon.grima@ed.ac.uk}{ramon.grima@ed.ac.uk};\\
\footnotesize $^\dag$ These authors contributed equally to this work.}

\date{}
\maketitle
\thispagestyle{plain} 
\paperfont
\captionsetup{font=footnotesize}
{\abstract
We introduce a biologically detailed, stochastic model of gene expression describing the multiple rate-limiting steps of transcription, nuclear pre-mRNA processing, nuclear mRNA export, cytoplasmic mRNA degradation and translation of mRNA into protein. The processes in sub-cellular compartments are described by an arbitrary number of processing stages, thus accounting for a significantly finer molecular description of gene expression than conventional models such as the telegraph, two-stage and three-stage models of gene expression. We use two distinct tools, queueing theory and model reduction using the slow-scale linear-noise approximation, to derive exact or approximate analytic expressions for the moments or distributions of nuclear mRNA, cytoplasmic mRNA and protein fluctuations, as well as lower bounds for their Fano factors in steady-state conditions. We use these to study the phase diagram of the stochastic model; in particular we derive parametric conditions determining three types of transitions in the properties of mRNA fluctuations: from sub-Poissonian to super-Poissonian noise, from high noise in the nucleus to high noise in the cytoplasm, and from a monotonic increase to a monotonic decrease of the Fano factor with the number of processing stages. In contrast, protein fluctuations are always super-Poissonian and show weak dependence on the number of mRNA processing stages. Our results delineate the region of parameter space where conventional models give qualitatively incorrect results and provide insight into how the number of processing stages, e.g. the number of rate-limiting steps in initiation, splicing and mRNA degradation, shape stochastic gene expression by modulation of molecular memory.}

\section{Introduction}
\label{sec1}

Chemical dynamics is stochastic~\cite{van2007stochastic}. The Stochastic Simulation  Algorithm (SSA)~\cite{gillespie1977exact,gillespie2007stochastic}, a useful tool to simulate stochastic chemical reaction systems, also provides a simple means to understand how the stochasticity in molecule numbers emerges from the stochasticity in timing events. Given the molecule numbers of all chemical species at time $t$ and the rate constants of all chemical reactions, two random numbers are generated, one determining which of the possible reactions will occur and the other determining the time $t + \Delta t$ at which the reaction will fire, causing the molecule numbers to change. A major source of this uncertainty in timing events is diffusion: many reactions occur once a molecule has bound with another one and the diffusive process bringing two molecules together, Brownian motion, is a stochastic process. The size of the resulting discrete fluctuations in molecule numbers, i.e. the standard deviation divided by the mean, is roughly inversely proportional to the square root of the mean number of molecules~\cite{van2007stochastic}, and therefore intrinsic noise is particularly important for subcellular processes in living cells because the number of gene copies and messenger mRNA (mRNA) molecules per cell can be very low~\cite{milo2015cell}. For example, most mRNAs in {\it{E. coli}} have copy numbers per cell less than one~\cite{taniguchi2010quantifying} and in mouse fibroblasts less than 100~\cite{schwanhausser2011global}. Intrinsic noise is, to some extent, responsible for the observed heterogeneity in mRNA and protein numbers between cells, typically deduced using fluorescent reporter measurements~\cite{swain2002intrinsic,golding2005real}. Many studies have shown that the steady-state distribution predicted by the simple telegraph model of gene expression~\cite{peccoud1995markovian,raj2006stochastic} provides a good fit to the experimentally measured distributions of mRNA molecules per cell (see for example~\cite{zenklusen2008single,halpern2015bursty}). In this Markovian model, it is assumed that the gene switches between two states, an inactive and active one from which mRNA is produced; the mRNA subsequently degrades. This implies that while the gene is active, a geometrically distributed number of molecules are transcribed and the time between successive bursts of transcription is exponentially distributed, properties that are in agreement with experiments~\cite{golding2005real}. The telegraph model also predicts three distinct types of mRNA count distributions~\cite{jiao2015distribution}; the same categories have been found from experiments using embryonic stem cells~\cite{singer2014dynamic}. 

Importantly, the telegraph model predicts that the Fano factor of mRNA molecule numbers, defined as the variance divided by the mean, is greater than or equal to 1 for all values of the rate parameters. Note that in this model, a Fano factor of 1, i.e. a Poisson distribution of mRNA counts, is only obtained when the gene is always active. In the limit that the gene spends most of its time in the off state (as commonly inferred for eukaryotic genes; see Table I of Ref.~\cite{cao2020analytical} for a summary of estimates from various papers), expression occurs in isolated bursts, a process that is often referred to as bursty transcription. In this case, the model predicts that the Fano factor is equal to 1 plus the mean burst size (mean number of mRNA molecules transcribed when the gene is active), hence a Fano factor greater than 1 typically is taken to imply bursty transcription. Of course, the simplicity of the telegraph model necessarily means that it excludes the description of various biologically important processes. Hence, it has been argued that the larger than one value of the measured Fano factor of mRNA fluctuations is not simply due to transcriptional bursting but also due to other noise sources such as the doubling of the gene copy number during DNA replication, the partitioning of molecules between daughter cells during cell division, the variability in the cell cycle duration time, the coupling of gene expression to cell size or cell-cycle phase and cell-to-cell variation in transcriptional parameters~\cite{foreman2020mammalian,jia2023coupling,ham2020extrinsic,peterson2015effects}. These noise sources can be collectively described as extrinsic noise, since they arise independently of a gene of interest~\cite{swain2002intrinsic,hilfinger2011separating}.  

Despite the clear trend in the literature of considering transcription as an inherently bursty process~\cite{tunnacliffe2020transcriptional}, there is also evidence to the contrary. Poissonian distributions (Fano factor equal to 1) have been measured for a number of genes~\cite{zenklusen2008single,gandhi2011transcription} and there have even been isolated reports from bacteria and hints for single genes in eukaryotes of Fano factors below 1~\cite{sun2020size,muthukrishnan2012dynamics,lionnet2010nuclear} though without well-controlled confirmation. These low noise genes have not received much attention until recently, when it was 
demonstrated beyond any reasonable doubt that several constitutive (non-regulated) cell division genes in fission yeast exhibit mRNA variances significantly below the mean (Fano factors as low as approximately 0.5)~\cite{weidemann2023minimal}. The strength of this study is the relatively large sample size (which leads to small confidence intervals for the Fano factor) and the proper accounting for extrinsic noise, which artificially amplifies the Fano factor of mRNA fluctuations. Clearly, these observations cannot be predicted by the telegraph model or its myriad modifications (described in the previous paragraph) because these models exclusively predict super-Poissonian noise. We note that while models have shown that this type of noise can be obtained by negative autoregulatory feedback~\cite{ramos2015gene} or by steric hindrances between RNA polymerases~\cite{szavitsnossangrima2023,voliotis2008fluctuations}, these cannot explain the observations in non-regulated genes that are infrequently transcribed (such as cell division genes in fission yeast). Hence, a different stochastic model of transcription was proposed in Ref.~\cite{weidemann2023minimal} which reproduces the observed low-noise, sub-Poissonian expression. While it is clear that the model's predictions for the moments of the mRNA can fit those measured from single-cell data, its detailed mathematical analysis and extension to also predict the commonly observed super-Poissonian fluctuations remains missing, principally because of its complexity. As well, it is unclear how the sub-Poissonian character of mRNA fluctuations influences protein fluctuations --- the standard two-stage and three-stage models of gene expression~\cite{bokes2012exact,paulsson2005models,shahrezaei2008analytical,popovic2016geometric} that predict protein fluctuations are extensions of the telegraph model and therefore cannot be used to study this question.  

In this paper, we undertake a rigorous analytical study of a generalized version of the model proposed in Ref.~\cite{weidemann2023minimal}. The model under consideration is a multi-stage and multi-compartment model, meaning that we model the mRNA fluctuations in both the nucleus and the cytoplasm, and in each of these compartments there are several processing stages each described by a different labelled species. In principle, because the propensities of the stochastic model are linear in the molecule numbers, all the moments of molecule numbers can be derived in closed-form. Unfortunately, in practice, it becomes impossible to write compact expressions, from which any meaning can be deduced, when the number of species exceeds two or three. Hence, in this paper to make progress, we resort to two powerful but different analytical techniques: queueing theory~\cite{jia2011intrinsic,kumar2015transcriptional,szavits2023charting} and stochastic model reduction using the slow-scale linear noise approximation~\cite{thomas2012slow,thomas2012rigorous,eilertsen2022stochastic}. The paper is divided as follows. The model and its detailed biological interpretation are introduced in Section~\ref{section_model}. This is followed by the model's analysis using queueing theory in Section~\ref{qtheory} and using model reduction in Section~\ref{modred}. The theoretical results are then confirmed by stochastic simulations in Section~\ref{sims}, and finally we conclude by a summary and discussion of the results in Section~\ref{Conc}. 

\section{The model and its biological interpretation }
\label{section_model}

\begin{figure}[hp]
    \centering       
    \includegraphics[width=0.9\textwidth]{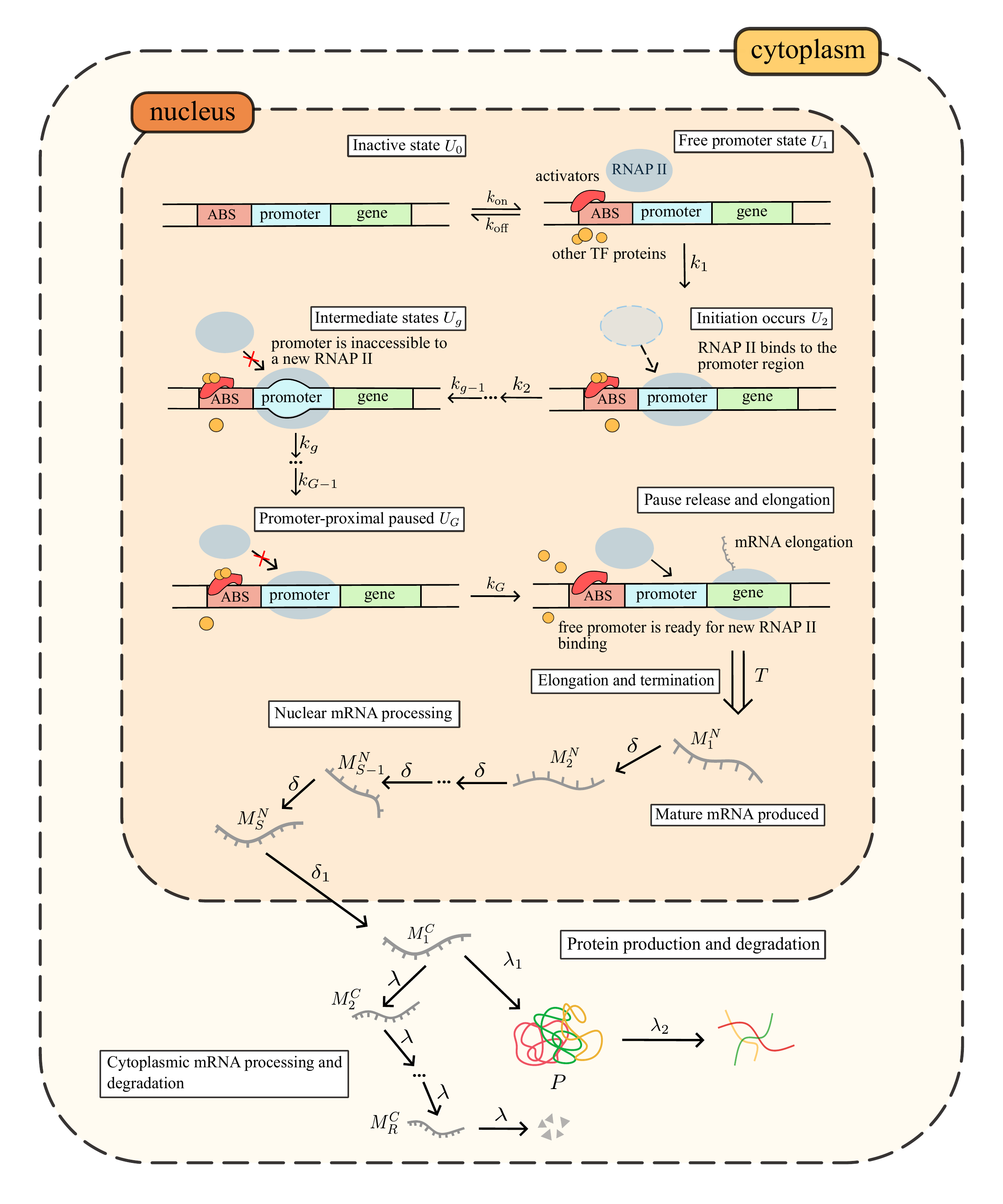}
    \caption{\label{fig:cartoon} \textbf{Illustration of the model.} The state $U_0$ denotes the gene-off state, where activators and other transcription factors are not in the vicinity of the promoter. In the state $U_1$, the activator binds to the promoter and RNAP II can access the promoter region. Initiation begins ($U_2$) once RNAP II binds to the promoter region. It temporarily pauses in the state $U_G$ due to promoter-proximal pausing. The states $U_g, g=3,4,...,G-1$ denote multiple long-lived intermediate states between binding and pausing. After the pause is released, nascent mRNA $M_0^N$ is produced (not shown), followed by elongation and termination with a deterministic time delay $T$, after which the pre-mRNA $M_1^N$ is produced. This undergoes several processing steps (each associated with a stage, $M_i^N, i=2,...,S$) before being exported to the cytoplasm as $M_1^C$. Once in the cytoplasm, $M_1^C$ serves as a template for protein ($P$) translation. The several stages of the mRNA degradation process are modelled by $M_i^C, i=2,...,R$.}
\end{figure}

The model consists of the following set of (effective) reactions:
\begin{equation}\label{model_reactions}
\begin{aligned}
    &U_{0} \xrightleftharpoons[k_{\rm{off}}]{k_{\rm{on}}} U_{1},\\
    &U_{1}\xrightarrow{k_1} U_2 \xrightarrow{k_2} ...\xrightarrow{k_{G-1}} U_G \xrightarrow{k_{G}} U_{1}+M^{N}_0,\\
    &M^{N}_0 \xRightarrow{T} M^N_1,\\
    &M^N_1\xrightarrow{\delta}M^N_2\xrightarrow{\delta}...\xrightarrow{\delta}M^{N}_{S-1}\xrightarrow{\delta}M^N_{S},\\
    &M^N_S\xrightarrow{\delta_1}M^C_1,\\
    &M^C_1\xrightarrow{\lambda}M^C_2\xrightarrow{\lambda}...\xrightarrow{\lambda}M^C_{R-1}\xrightarrow{\lambda}M^C_R\xrightarrow{\lambda}\varnothing,\\
    &M^C_1\xrightarrow{\lambda_1}M^C_1+P,\quad P\xrightarrow{\lambda_2}\varnothing,
\end{aligned}
\end{equation}
where $U_i$ denote promoter states, $M_i^N$ denote nuclear mRNA species, $M_i^C$ denote cytoplasmic mRNA species and $P$ denotes protein species. Note that this model is a generalization of the one described in Ref.~\cite{weidemann2023minimal} (which assumed $S=1$, $k_{\rm{off}}=0$ and there was no protein description). The model is illustrated by a cartoon in Fig.~\ref{fig:cartoon}.

The states and reactions have the following biological interpretation. $U_0$ denotes a closed chromatin state that impairs activator binding and therefore prevents RNA polymerase II (RNAP) from accessing the promoter region~\cite{fuda2009defining,voss2014dynamic}. $U_1$ signifies a state in which activator binding has reshaped the nearby nucleosome structure~\cite{voss2014dynamic}. This restructuring enables RNAP to reach the promoter region alongside all the necessary elements for transcription initiation, such as transcription factors, co-activators and initiation factors. Once RNAP binds to the promoter, initiation occurs and the state changes to $U_2$ (closed RNAP-promoter complex). The next step is for the RNAP to open the DNA double helix, a process that includes several long-lived intermediate states, which we denote by $U_g$ ($g=3,4,..., G-1$)~\cite{friedman2012mechanism,lloyd2016dissecting}. Finally, an open complex results and the RNAP begins mRNA elongation but pauses shortly after (promoter proximal pausing~\cite{core2008transcription,bartman2019transcriptional}); this state is described by $U_G$. Once the pause is released, RNAP begins moving away from the promoter region, thus starting productive elongation that leads to an RNAP molecule with a nascent mRNA tail ($M_0^N$). Simultaneously, since the promoter region is now cleared of the RNAP, a new RNAP can bind, hence the gene state changes back to $U_1$~\cite{shao2017paused,roussel2006validation} (volume exclusion can prevent a new RNAP binding event if another already bound RNAP is very close to the promoter).

After a fixed time delay $T$ (during which elongation followed by termination occur), RNAP detaches from the DNA and the mRNA strand is complete ($M_1^N$). Note that the modelling of elongation plus termination as a step occurring after a fixed delay is justified by microscopic arguments: the time for a particle (RNAP) hopping in one direction on a lattice with $N$ sites (nucleotides) to exit the last site is a random variable distributed according to an Erlang distribution with a coefficient of variation (ratio of standard deviation and mean) equal to $1/\sqrt{N}$.  Hence, if $N$ is large, the time for elongation plus termination to finish is approximately deterministic. Experimental support for the deterministic nature of the elongation and termination processes is provided in Ref.~\cite{larson2011real}.  

In eukaryotes, the new mRNA ($M_1^N$) needs processing before it is ready for translation. Hence, at this stage it is called a pre-mRNA. The processing steps that it must go through while in the nucleus include the addition of a 5' cap, splicing, editing, and 3' polyadenylation (poly-A) tail~\cite{alberts2022molecular}. Note that for some of these processes, such as splicing, there are opposing models suggesting they can occur either before or after the RNA is detached from DNA~\cite{khodor2011nascent,coulon2014kinetic}. The pre-mRNA in these various stages is described by $M_s^N$ ($s=2,...,S$). From the last stage, nuclear export occurs, resulting in the first stage of cytoplasmic mRNA, $M_1^C$. We assume that cytoplasmic mRNA goes through several other stages, $M_r^C$ ($r=2,...,R$), before it finally gets completely degraded. Each of the latter stages is associated with a different process in the complex mRNA degradation pathway  ~\cite{cao2001computational}. This also means that the cytoplasmic mRNA lifetime distribution is generally not exponential ~\cite{deneke2013complex}. 

Finally, we assume that translation can only easily proceed before the mRNA becomes targeted for degradation and therefore only the cytoplasmic species from the first stage, $M_1^C$, can lead to protein production. The evidence for this is that mRNA decapping (a critical step in the mRNA decay pathway) is significantly enhanced when translation initiation is inhibited~\cite{coller2005general} although the interaction between translation and degradation is in reality much more complex than this~\cite{huch2014interrelations}. Protein decay is modelled by  a simple first-order reaction; this effectively models dilution due to the partitioning of protein molecules between two daughter cells when cell-division occurs.

\section{Results using queueing theory}
\label{qtheory}

In this section, instead of using the standard chemical master equation (CME)~\cite{gillespie2007stochastic} approach to study the stochastic properties of the reaction system~\eqref{model_reactions}, we will make use of an alternative powerful approach based on queueing theory (for applications of this theory to solve problems in gene expression, see for example~\cite{jia2011intrinsic,fralix2023markovian,dean2022noise,elgart2010applications,choubey2018nascent,thattai2016universal,mather2013translational}). Our aim is to derive expressions for the Fano factor of mRNA fluctuations in the nucleus and the cytoplasm in steady-state conditions, and to use these to derive insight into the relationship between the two. We also obtain steady-state distributions of total nuclear and cytoplasmic mRNA for the special case in which their processing and degradation times are deterministic. 

\subsection{Mapping the model to a queueing system}

We begin by mapping the model~(\ref{model_reactions}) to a queueing system in which mRNA molecules are the customers, transcription is the arrival process and the processing of mRNA is the service process. The arrival of nascent mRNA is described by a Markov jump process that starts from state $U_1$, the state the gene switches to immediately after producing a nascent mRNA molecule, and ends in state $U_G$ from which a new nascent mRNA molecule is produced. After the production of nascent mRNA, the gene switches back to state $U_1$, and the process is repeated. This process therefore constitutes a renewal process, meaning that the interarrival times are independent and identically distributed random variables. 

Once a nascent mRNA molecule is produced, it is processed into a nuclear mRNA molecule after a fixed time $T$. It is important to emphasize that in the queueing theory approach we do not consider neither individual nuclear mRNA species $M_{1}^{N},\dots,M_{S}^{N}$ nor individual cytoplasmic mRNA species $M_{1}^{C},\dots,M_{R}^{C}$. Instead, the ``customers" in our queueing system are individual nuclear and cytoplasmic mRNA molecules, which are distinguished only by the (random) time it takes to process them. This is different from the CME formalism, which also keeps track of the particular stage of each mRNA molecule. Hence, the system of chemical reactions that we consider using the queueing theory approach can be better described as
\begin{equation}\label{model_reactions-queueing}
\begin{aligned}
    &U_{0} \xrightleftharpoons[k_{\rm{off}}]{k_{\rm{on}}} U_{1},\quad U_{1}\xrightarrow{k_1} U_2 \xrightarrow{k_2} ...\xrightarrow{k_{G-1}} U_G \xrightarrow{k_{G}} U_{1}+M^{N}_0,\\
    & M^{N}_0\xRightarrow{T} M^N,\quad M^N\rightsquigarrow M^C,\quad M^C\rightsquigarrow \varnothing,\\
\end{aligned}
\end{equation}
where $M^N=M_{1}^N+\dots+M_{S}^N$ and $M^C=M_{1}^C+\dots+M_{R}^C$ denote the total nuclear and cytoplasmic mRNA, respectively, and the symbol $\rightsquigarrow$ means that the reaction occurs after some random service time. The usual Markovian case is obtained when the service time distribution is exponential (in our model, that happens only when $S=R=1$). The reaction system  (\ref{model_reactions-queueing}) can be considered as a queueing system consisting of two queues in a tandem. The first queue produces a nuclear mRNA molecule $M^N$, which is then turned into a cytoplasmic mRNA molecule after some random time $t^N$, which in turn is degraded after some other random time $t^C$ (the distributions of $t^N$ and $t^C$ will be discussed later). The departure process of the first queue is therefore an arrival process of the second queue. 

Tandem queues are in general difficult to study analytically, because the departures of all but the simplest queues are generally intractable. We will therefore focus only on the total nuclear mRNA $M^N$, which has the same arrival process as the nascent mRNA, since the time $T$ is fixed. We will consider total cytoplasmic mRNA $M^C$ only in the special case in which the processing time of nuclear mRNA is deterministic, since in that case both queues share the same arrival process. We also note that each mRNA molecule (nuclear or cytoplasmic) is processed independently of other mRNA molecules. In queueing theory, this is equivalent to saying that both queues have infinitely many servers. The number of customers (the number of mRNA molecules) in each queue (the queue length) is therefore equal to the number of busy servers.

Queueing systems are usually described using Kendall's notation $A/S/c$, where $A$ denotes the arrival process, $S$ denotes the service process, and $c$ denotes the number of servers. The queueing system describing the production and processing of nuclear mRNA is a $G/G/\infty$ queue, where the first $G$ refers to renewal arrivals with general interarrival distribution, the second $G$ refers to general service time distribution, and there are infinitely many servers. This classic queueing system was analysed in detail in Ref.~\cite{Takacs_1958}, and analytical results were obtained for the moments of the queue length distribution in terms of the interarrival and service time distributions. For convenience, we rephrase these results here.

Let $t_i$ and $\chi_i$ for $i=1,2,\dots$ denote the interarrival and service times of customers arriving to a $G/G/\infty$ queue, respectively, and set $P(t_i\leq x)=F(x)$ and $P(\chi_i\leq x)=H(x)$ to be the interarrival and service time cumulative distribution functions, respectively. Let $\eta(t)$ denote the queue length at time $t$ such that $\eta(0)=0$, and let $P^{*}_{k}=\lim_{t\rightarrow\infty}P(\eta(t)=k)$ denote the steady-state distribution of $\eta(t)$. Define the $r$th binomial moment of $P^{*}_{k}$ as
\begin{equation}
    B^{*}_{r}=\sum_{k=r}^{\infty}\binom{k}{r}P^{*}_{k},\quad r=0,1,2\dots, 
\end{equation}
Then it follows from Ref.~\cite{Takacs_1958} that
\begin{equation}
    \label{B-r-steady-state}
    B_{r}^{*}=\frac{1}{\alpha}\int_{0}^{\infty}dt B_{r-1}(t)[1-H(t)], \quad r=1,2,3,\dots,
\end{equation}
where $\alpha$ is the mean interarrival time, and $B_r(t)$ is defined as
\begin{equation}
    \label{B-r-time}
    B_0(t)\equiv 1,\quad B_{r}(t)=\int_{0}^{t}dR(x) B_{r-1}(t-x)[1-H(t-x)],\quad r=1,2,3\dots.
\end{equation}
The function $R(x)$, which is called the renewal function, is given by
\begin{equation}
    \label{renewal-function-def}
    R(x)=\sum_{n=1}^{\infty}F_n(x),
\end{equation}
where $F_n(x)=\int_{0}^{x}dx'f^{*n}(x')$, $f(x)=dF/dx$ and $f^{*n}(x)$ is the $n$th iterated convolution of $f(x)$. From this result, knowing $F(x)$ and $H(x)$, one can compute all steady-state moments recursively. A generalization of this result to batch arrivals (multiple customers arriving at the queue at the same time) can be found in Ref.~\cite{Liu_1990}.

We now apply this result to total nuclear mRNA $M^N$. We have previously established that the interarrival time of total nuclear mRNA $M^N$ is the same as of nascent mRNA $M_{0}^{N}$, except for the first arrival. Since the steady-state queue length distribution is independent of the distribution of the first arrival time~\cite{Takacs_1958}, the above results are applicable to total nuclear mRNA. We are interested in the Fano factor, which is a measure of the spread of fluctuations, and is defined as the ratio of the mean and the variance of mRNA fluctuations,
\begin{equation}
    \label{FF-N-G-S}
    FF_{G,S}^{N}=\frac{\sum_{k=0}^{\infty}k^2 P_{k}^{*}-B_{1}^{*2}}{B_{1}^{*}}=1+\frac{2 B_{2}^{*}}{B_{1}^{*}}-B_{1}^{*},
\end{equation}
where the subscripts $G$ and $S$ denote the number of gene states and the number of post-transcriptional processing steps in the nucleus, and $B_{1}^{*}$ and $B_{2}^{*}$ are the first two binomial moments of the total nuclear mRNA number distribution in the steady state, 
\begin{equation}
    \label{B1-B2}
    B_{1}^{*}=\sum_{k=1}^{\infty}k P_{k}^{*}, \quad B_{2}^{*}=\frac{1}{2}\sum_{k=2}^{\infty}k(k-1)
    P_{k}^{*}.
\end{equation}
According to queueing theory, binomial moments of the total nuclear mRNA number distribution depend only on the interarrival time distribution (through the renewal function $R(x)$ in Eq. (\ref{renewal-function-def})), and the service time distribution. The following result establishes the distribution of the interarrival times of nascent mRNA.

\begin{lemma}
\label{lemma_LT-pdf-interarrival-time}

Let $f(t)$ denote the probability density function of the interarrival times between successive arrivals of nascent mRNA. The Laplace transform of $f(t)$ is given by
\begin{equation}
    \label{LT-pdf-interarrival-time}
    \phi(s)=\int_{0}^{\infty}dt f(t)e^{-st}=\frac{k_1(s+k_{\rm{on}})}{s^2+s(k_1+k_{\rm{on}}+k_{\rm{off}})+k_1 k_{\rm{on}}}\prod_{i=2}^{G}\frac{k_i}{s+k_i}.
    \end{equation}
\end{lemma}

\noindent\textit{\textbf{Proof of Lemma~\ref{lemma_LT-pdf-interarrival-time}}.} The pdf $f(t)$ can be found by solving the first passage time problem for the system of reactions
\begin{equation}
    U_{0} \xrightleftharpoons[k_{\rm{off}}]{k_{\rm{on}}} U_{1}\xrightarrow{k_1} U_2 \xrightarrow{k_2} ...\xrightarrow{k_{G-1}} U_G \xrightarrow{k_{G}}U_1+M_{0}^{N},  
\end{equation}
where the starting point is state $U_1$ (the state the gene goes to after initiation), and the ending point is the creation of nascent mRNA, i.e. reaction $U_G\rightarrow U_1+M_{0}^{N}$. Let $P_i(t)$ denote the probability that the system is in state $i$ at time $t$, and set $P_i(0)=\delta_{i,1}$, where $\delta_{i,j}$ is the Kronecker delta. The master equation for $P_i(t)$ for $i=0,\dots,G$ is given by
\begin{align}
 \label{Peqs}
    & \frac{dP_0}{dt}=k_{\rm{off}}P_1-k_{\rm{on}} P_0,\\ \nonumber
    & \frac{dP_1}{dt}=k_{\rm{on}}P_0+k_G P_G-(k_{\rm{off}}+k_1) P_1,\\ \nonumber
    & \frac{dP_i}{dt}=k_{i-1}P_{i-1}-k_i P_i,\quad i=2,\dots,G.
\end{align}
The probability density function $f(t)$ can be computed from 
\begin{equation}
\label{fpt}
    f(t)=k_G P_G(t).
\end{equation}
The result in Eq.~(\ref{LT-pdf-interarrival-time}) follows from solving Eq.~\eqref{Peqs} for $P_G(t)$ and substituting this result in the Laplace transform of Eq.~\eqref{fpt}.~$\square$

From Eq.~(\ref{LT-pdf-interarrival-time}) we get the following expression for the mean interarrival time $\alpha$,
\begin{equation}
    \label{mean-interarrival-time}
    \alpha=-\left.\frac{d\phi}{ds}\right\vert_{s=0}=\frac{k_{\rm{on}}+k_{\rm{off}}}{k_{\rm{on}} k_1}+\sum_{i=2}^{G}\frac{1}{k_i}.
\end{equation}
To compute $B_{r}(t)$, we need to compute the renewal function $R(x)$, and the service time distribution. These results are established in Lemma~\ref{lemma_renewal-function} and Lemma~\ref{lemma_H-function}, respectively.  

\begin{lemma}
\label{lemma_renewal-function}

Let $f(t)$ denote the probability density function whose Laplace transform $\phi(s)$ is given by Eq.~(\ref{LT-pdf-interarrival-time}), and let $p(s)$ and $q(s)$ denote two coprime polynomials such that
\begin{equation}
    \frac{\phi(s)}{s[1-\phi(s)]}=\frac{p(s)}{q(s)}=\frac{p(s)}{s^2(s+s_1)\dots(s+s_M)},
\end{equation}
where $M=\text{deg}(q)-2$, and $-s_1,\dots,-s_M$ are non-zero zeros of $q(s)$. Then the renewal function defined in Eq.~(\ref{renewal-function-def}) is given by
\begin{equation}
    \label{renewal-function}
    R(x)=\frac{x}{\alpha}-\sum_{r=1}^{M}A_r\left(1-e^{-s_r x}\right),
\end{equation}
where $A_r=p(-s_r)/q'(-s_r)$.
\end{lemma}

\noindent\textit{\textbf{Proof of Lemma~\ref{lemma_renewal-function}}.} It is a well-known result from renewal theory (see for example Chapter 4 in Ref.~\cite{Cox_1967}) that the Laplace transform of the renewal function $R(x)$ is given by
\begin{equation}
    \label{LT-renewal-function}
    \mathcal{L}[R(x)](s)=\frac{\phi(s)}{s[1-\phi(s)]},
\end{equation}
where $\phi(s)$ is the Laplace transform of $f(t)=dF(t)/dt$. Since $\phi(s)$ is a rational function, we can always write
\begin{equation}
    \label{LT-renewal-function-2}
    \mathcal{L}[R(x)](s)=\frac{p(s)}{q(s)},\quad q(s)=s^2\prod_{r=1}^{M}(s+s_r),
\end{equation}
where in the last step we factorized the denominator and kept only factors that are not present in $p(s)$, hence $p(s)$ are $q(s)$ are coprime. The Laplace transform in Eq.~(\ref{LT-renewal-function-2}) can be inverted using partial fraction decomposition,
\begin{equation}
    \label{LT-renewal-function-3}
    \mathcal{L}[R(x)](s)=\frac{c_1}{s}+\frac{c_2}{s^2}+\sum_{r=1}^{M}\frac{A_r}{s+s_r}.
\end{equation}
From here it follows that
\begin{equation}
    p(s)=c_1 s\prod_{r=1}^{M}(s+s_r)+c_2\prod_{r=1}^{M}(s+s_r)+s^2\sum_{r=1}^{M}A_r\prod_{\genfrac{}{}{0pt}{}{j=1}{j\neq r}}^{M}(s+s_j).
\end{equation}
The polynomial on the right-hand side has a degree of $M+2$. However, it can be shown that $\text{deg}(p)<M+1$, from where it follows that
\begin{equation}
    c_1=-\sum_{r=1}^{M}A_r.
\end{equation}
On the other hand,
\begin{equation}
    c_1=\lim_{s\rightarrow 0}\frac{d}{ds}\left[\frac{s\phi(s)}{1-\phi(s)}\right]=\frac{\alpha_2-2\alpha^2}{2\alpha^2}=\frac{\text{CV}_{f}^{2}-1}{2},
\end{equation}
where $\alpha_2$ is the second moment of the interarrival time distribution, and $\text{CV}_{f}^{2}=(\alpha_2-\alpha^2)/\alpha^2$ is the corresponding coefficient of variation squared. The coefficient $c_2$ follows from
\begin{equation}
    c_2=\lim_{s\rightarrow 0}\frac{s\phi(s)}{1-\phi(s)}=\frac{1}{\alpha}.
\end{equation}
Finally, the coefficients $A_r$ for $r=1,\dots,M$ are given by
\begin{equation}
    A_r=\frac{p(-s_r)}{q'(-s_r)}.
\end{equation}
From here the main statement of the lemma follows by inverting the Laplace transform $\mathcal{L}\{R(x)\}(s)$ in Eq.~(\ref{LT-renewal-function-3}).\;$\square$

\begin{lemma}
\label{lemma_H-function}

Let $t^N$ denote the random time it takes to process nuclear mRNA and export it to cytoplasm, and let $H^N(t)=P(t^N\leq t)$ denotes its cumulative distribution function. The expression for $H^N(t)$ is given by
\begin{align}
\label{H-function}
    H^N(t)=\frac{\gamma(S-1,\delta t)}{\Gamma(S-1)}-\left(\frac{\delta}{\delta-\delta_1}\right)^{S-1}\frac{\gamma(S-1,(\delta-\delta_1)t)}{\Gamma(S-1)}e^{-\delta_1 t},
\end{align}
where $\gamma(n,t)$ is the lower incomplete Gamma function, 
\begin{equation}
    \gamma(n,t)=\int_{0}^{t}dx\; x^{n-1}e^{-x}=\Gamma(n)\left(1-e^{-t}\sum_{i=0}^{n-1}\frac{t^i}{i!}\right), \quad n=1,2\dots,
\end{equation}
and $\Gamma(n)$ is the Gamma function.
\end{lemma}

\noindent\textit{\textbf{Proof of Lemma~\ref{lemma_H-function}}.} According to the reaction system~(\ref{model_reactions}), the nuclear mRNA processing time is a sum of two random variables: one that is Erlang distributed with shape $S-1$ and rate parameter $\delta$, and the other that is exponentially distributed with rate parameter $\delta_1$. The cumulative distribution function of the service time is therefore a convolution of these two probability distributions,
\begin{align}
    H^N(t)=&\int_{0}^{t}dx\;e^{-\delta_1(t-x)}\frac{\delta^{S-1} x^{S-2}}{\Gamma(S-1)}e^{-\delta x},\nonumber\\
    &=\frac{\gamma(S-1,\delta t)}{\Gamma(S-1)}-\left(\frac{\delta}{\delta-\delta_1}\right)^{S-1}\frac{\gamma(S-1,(\delta-\delta_1)t)}{\Gamma(S-1)}e^{-\delta_1 t}.\;\square
\end{align}

We now have all the ingredients to compute $B_r(t)$ and $B_{r}^{*}$ for any $r$, and therefore the Fano factor $FF_{G,S}^{N}$. According to Eq.~(\ref{B-r-steady-state}), $B_{1}^{*}$ reads
\begin{equation}
    B_{1}^{*}=\frac{1}{\alpha}\int_{0}dt [1-H^N(t)]=\frac{\rho}{\alpha},
\end{equation}
where $\alpha$ and $\rho$ are the mean interarrival and service times, respectively. The mean interarrival time $\alpha$ has been computed in Eq.~(\ref{mean-interarrival-time}). The mean service time $\rho$ follows from Lemma~\ref{lemma_H-function},
\begin{equation}
    \rho=\frac{S-1}{\delta}+\frac{1}{\delta_1}.
\end{equation}
The following Proposition gives the expression for $B_{1}(t)$.

\begin{proposition}
\label{prop_B-1-time}
    
Given $R(x)$ and $H^N(t)$ obtained in Eqs. (\ref{renewal-function}) and (\ref{H-function}), the expression for $B_{1}(t)$ is given by
\begin{align}
    \label{B-1-time}
    & B_{1}(t)=R(t)-\frac{1}{\alpha}\left[\frac{\varphi(S-1,\delta,0,t)}{\Gamma(S-1)}-\left(\frac{\delta}{\delta-\delta_1}\right)^{S-1}\frac{\varphi(S-1,\delta-\delta_1,-\delta_1,t)}{\Gamma(S-1)}e^{-\delta_1 t}\right]\nonumber\\
    &+\sum_{r=1}^{M}A_r s_r\left[\frac{\varphi(S-1,\delta,s_r,t)}{\Gamma(S-1)}-\left(\frac{\delta}{\delta-\delta_1}\right)^{S-1}\frac{\varphi(S-1,\delta-\delta_1,s_r-\delta_1,t)}{\Gamma(S-1)}e^{-\delta_1 t}\right]
\end{align}
where the function $\varphi(n,a,b,t)$ is defined as
\begin{align}
    \label{phi-function}
    & \varphi(n,a,b,t)=\begin{cases}
    \frac{\gamma(n,at)}{b}-\left(\frac{a}{a-b}\right)^n\frac{\gamma(n,(a-b)t)}{b}e^{-bt}, & b\neq 0\\
    t\gamma(n,at)-\frac{\gamma(n+1,at)}{a}, & b=0.
    \end{cases}
\end{align}
\end{proposition}

\noindent\textit{\textbf{Proof of Proposition~\ref{prop_B-1-time}.}} For $r=1$ in Eq.~(\ref{B-r-time}) we get
\begin{equation}
    \label{B-1-time-def}
    B_1(t)=\int_{0}^{\infty}dx \frac{dR(x)}{dx}[1-H^N(t-x)]
\end{equation}
The result in Eq.~(\ref{B-1-time}) can be obtained by inserting $dR(x)/dx$ and $H^N(t-x)$ into Eq.~(\ref{B-1-time-def}), and using the following identity,
\begin{equation}
    \varphi(n,a,b,t)\equiv \int_{0}^{t}dx \gamma(n,a(t-x))e^{-bx},
\end{equation}
where the expression for $\varphi(n,a,b,t)$ is given in Eq.~(\ref{phi-function}). This integral is easily solved by integration by parts.~$\square$

To compute $B_{2}^{*}$ and finally the Fano factor $FF_{G,S}^{N}$, we need to perform the integral in Eq.~(\ref{B-r-steady-state}) for $r=2$ using $B_{1}(t)$ and $H^N(t)$ computed in Lemma~\ref{lemma_H-function} and Proposition~\ref{prop_B-1-time}, respectively. Although the integral can be carried out analytically, the calculation is quite tedious for general $S$. Instead, we demonstrate the calculation for the special cases $S=1$ and $S=2$.

\begin{corollary}
\label{cor_FF-N-S-1}

For $S=1$, the Fano factor of the total nuclear mRNA reads
\begin{equation}
    FF_{G,1}^{N}=1-\sum_{r=1}^{M}\frac{A_r s_r}{s_r+\delta_1}=\frac{1}{1-\phi(\delta_1)}-\frac{1}{\alpha\delta_1},
\end{equation}
where $\phi(s)$ is the Laplace transform of the interarrival time distribution given by Eq.~(\ref{LT-pdf-interarrival-time}).
\end{corollary}

\noindent\textit{\textbf{Proof of Corollary~\ref{cor_FF-N-S-1}.}} The $S=1$ case is a special limit of the $S=2$ case when $\delta\rightarrow\infty$, hence the proof will be given in Corollary~\ref{cor_FF-N-S-2}.~$\square$

\begin{corollary}
\label{cor_FF-N-S-2}

For $S=2$, the Fano factor of the total nuclear mRNA reads
\begin{equation}
    \label{FF-N-G-2}
    FF_{G,2}^{N}=1-\sum_{r=1}^{M}A_r \frac{s_r(\delta+\delta_1)^3+s_{r}^{2}(\delta^2+3\delta\delta_1+\delta_{1}^{2})}{(\delta+\delta_1)^2(\delta+s_r)(\delta_1+s_r)}.
\end{equation}
\end{corollary}

\noindent\textit{\textbf{Proof of Corollary~\ref{cor_FF-N-S-2}.}} For $S=2$ the cumulative distribution function $H^N(t)$ is given by
\begin{equation}
    H^N(t)=1-\frac{\delta}{\delta-\delta_1}e^{-\delta_1 t}+\frac{\delta_1}{\delta-\delta_1}e^{-\delta t},
\end{equation}
The mean service time $\rho$ reads
\begin{equation}
    \rho=\frac{1}{\delta}+\frac{1}{\delta},
\end{equation}
which gives 
\begin{equation}
    \label{B-1-steady-state-S-1}
    B_{1}^{*}=\frac{1}{\alpha}\left(\frac{1}{\delta}+\frac{1}{\delta_1}\right).   
\end{equation}
Next, we insert $H^N(t)$ into Eq.~(\ref{B-1-time}), and then insert the resulting expression into Eq.~(\ref{B-r-steady-state}) for $r=2$, which yields
\begin{equation}
    \label{B-2-steady-state-S-2}
    B_{2}^{*}=\frac{1}{2}B_{1}^{*2}-\frac{1}{\alpha}\sum_{r=1}^{M}A_r \frac{s_r(\delta+\delta_1)^3+s_{r}^{2}(\delta^2+3\delta\delta_1+\delta_{1}^{2})}{(\delta+\delta_1)^2(\delta+s_r)(\delta_1+s_r)}.
\end{equation}
Inserting Eqs. (\ref{B-1-steady-state-S-1}) and (\ref{B-2-steady-state-S-2}) into Eq.~(\ref{FF-N-G-S}) yields the stated expression for the Fano factor. To prove the result in Corollary~\ref{cor_FF-N-S-1}, we set $\delta\rightarrow\infty$ in Eq.~(\ref{FF-N-G-2}). The second equality follows from Eqs. (\ref{LT-renewal-function}) and (\ref{LT-renewal-function-3}).~$\square$ 

\textit{Application of queueing theory to the total mRNA.} We note that the results for the $G/G/\infty$ queue can be applied to compute the Fano factor of the total mRNA number, which includes both nuclear and cytoplasmic mRNA. In that case, we need to compute the service time distribution of the total time $t^{N+C}$ it takes the cell to process nuclear mRNA, export it to cytoplasm, and degrade it. The distribution of $t^{N+C}$ is given by
\begin{equation}
    \label{H-C-function}
    H^{N+C}(t)=P(t^{N+C}\leq t)=\int_{0}^{t}dx \frac{\lambda^{R}x^{R-1}}{\Gamma(R)}e^{-\lambda x}H^{N}(t-x).
\end{equation}
From here it follows that the mean service time $\rho$ is
\begin{equation}
    \rho=\frac{S-1}{\delta}+\frac{1}{\delta_1}+\frac{R}{\lambda},
\end{equation}
and therefore
\begin{equation}
    B_{1}^{*}=\frac{1}{\alpha}\left(\frac{S-1}{\delta}+\frac{1}{\delta_1}+\frac{R}{\lambda}\right),
\end{equation}
where $\alpha$ is given by Eq.~(\ref{mean-interarrival-time}). The next step is to insert the expression for $H^{N+C}(t)$ in Eq.~(\ref{H-C-function}) into Eq.~(\ref{B-r-time}) to compute $B_{1}(t)$, using $R(x)$ given by Eq.~(\ref{renewal-function}). Inserting the resulting expression for $B_{1}(t)$ into Eq.~(\ref{B-r-steady-state}) for $r=2$ yields $B_{2}^{*}$, which in turn can be used to compute the Fano factor of the nuclear and cytoplasmic mRNA combined. This calculation is omitted here as it is quite tedious.

In the rest of this section, we consider a case in which the service process consists of an infinitely many steps, such that the service time becomes deterministic. In that case, it is possible to compute the steady-state distribution of the total nuclear or cytoplasmic mRNA.

\begin{proposition}
\label{prop_nuclear-total-delay-prob}
    
Let $k_{\rm{off}} = 0$, $\delta_1=\delta$ and $k_i = k$ for $i=1,..,G$. Assume that $S\rightarrow\infty$ and $\delta\rightarrow\infty$ such that the mean time to process nuclear mRNA, $T^N=S/\delta$, is finite. In that case the distribution of $t^N$ becomes deterministic, i.e. the probability density function of $t^N$ is given by a Dirac delta function, $h^N(t)=dH^N/dt=\delta(t-T^N)$. Under these conditions, the steady-state distribution of the total nuclear mRNA $M^N$, $P_{G,\infty}^N(m)$, is given by
\begin{subequations}
\label{prob-nuclear-total-delay}
\begin{align}
    & P_{G,\infty}^N(0)=\frac{e^{-k T^N}}{G}\sum_{j=0}^{G-1}(G-j)\frac{(k T^N)^{j}}{j!},\\
    & P_{G,\infty}^N(m)=\frac{e^{-k T^N}}{G}\left\{\sum_{j=0}^{G(m+1)-1}[G(m+1)-j]\frac{(k T^N)^{j}}{j!}\right.\nonumber\\
    & \left.-2\sum_{j=0}^{Gm-1}(Gm-j)\frac{(k T^N)^{j}}{j!}+\sum_{j=0}^{G(m-1)-1}[G(m-1)-j]\frac{(k T^N)^{j}}{j!}\right\},\quad m\geq 1.
\end{align}    
\end{subequations}
\end{proposition}

\noindent\textit{\textbf{Proof of Proposition~\ref{prop_nuclear-total-delay-prob}}.} This problem can be solved using renewal theory as described in Ref.~\cite{Szavits_2023}. Since the service time is fixed, the number of nuclear mRNA at some time $t$ in the steady state is equal to the number of nuclear mRNA that arrived between $t-T^N$ and $t$. The Laplace transform of this probability distribution reads~\cite{Cox_1967}
\begin{equation}
    \label{LT-prob-nuclear-total-delay}
    \mathcal{L}[P_{G,\infty}^N(m)](s)=\int_{0}^{\infty}dT^N P_{G,\infty}^N(m)e^{-sT^N}=\begin{dcases}
    \frac{\alpha s-1+\phi(s)}{\alpha s^2}, & m=0\\
    \frac{[1-\phi(s)]^2[\phi(s)]^{m-1}}{\alpha s^2},& m\geq 1,\end{dcases}
\end{equation}
where $\phi(s)$ is the Laplace transform of the probability density function of the interarrival time of nuclear mRNA, and $\alpha$ is the mean interarrival time. For the Erlang distribution,
\begin{equation}
    \label{LT-pdf-interarrival-time-Erlang}
    \phi(s)=\left(\frac{k}{s+k}\right)^G, \quad \alpha=G/k.
\end{equation}
The inverse of Eq.~(\ref{LT-prob-nuclear-total-delay}) using Eq.~(\ref{LT-pdf-interarrival-time-Erlang}) has been computed in Appendix B of Ref.~\cite{Szavits_2023}, yielding the result in Eq.~(\ref{prob-nuclear-total-delay}).~$\square$

\begin{corollary}
\label{cor_delay-nuclear-cytoplasmic-prob}

Let $t^C$ denote the time it takes to process and degrade cytoplasmic mRNA. Let $k_{\rm{off}} = 0$, $k_i = k$ for $i=1,..,G$ and $\delta_1=\delta$. Set $S\rightarrow \infty$ and $\delta\rightarrow\infty$ such that the mean nuclear mRNA processing time, $T^N=S/\delta$, is finite. Similarly, set $R\rightarrow \infty$ and $\lambda\rightarrow\infty$ such that the mean time to process and degrade cytoplasmic mRNA, $T^C = R/\lambda$, is finite. In that case the distribution of $t^C$ becomes deterministic, i.e. the probability density function of $t^C$ becomes a Dirac delta function, $h^C(t)=dH^C/dt=\delta(t-T^C)$. Under these conditions, the steady-state probability distribution of cytoplasmic mRNA $M^C$, $P_{G,\infty}^{C}(m)$, is given by Eq.~(\ref{prob-nuclear-total-delay}) in which $T^N$ is replaced by $T^C$.
\end{corollary}

\noindent\textit{\textbf{Proof of Corollary~\ref{cor_delay-nuclear-cytoplasmic-prob}}.} 
Since the processing of nuclear mRNA is deterministic, the interarrival time of cytoplasmic mRNA is the same as the interarrival time of nuclear mRNA. The degradation of cytoplasmic mRNA is also deterministic, hence the steady-state distribution of the total number of cytoplasmic mRNA is the same as the steady-state distribution of the total number of nuclear mRNA, except that $T^N$ is replaced by $T^C$.~$\square$

\begin{proposition}
\label{prop_delay-nuclear-cytoplasmic-FF}

Under the conditions of Corollary~\ref{cor_delay-nuclear-cytoplasmic-prob}, $FF_{G,\infty}^{N}$ decays monotonically with $T^{N}$. Since $FF_{G,\infty}^{C}$ has the same dependence on $T^C$ as $FF_{G,\infty}^{N}$ has on $T^N$, $FF_{G,\infty}^{C}$ decays monotonically with $T^{C}$.

\end{proposition}

\noindent\textit{\textbf{Proof of Proposition~\ref{prop_delay-nuclear-cytoplasmic-FF}}.} Let $\Psi_{G,\infty}^N(z)$ denote the probability generating function of total nuclear mRNA number in the stationary limit,
\begin{equation}
    \Psi_{G,\infty}^{N}(z)=\sum_{m=0}^{\infty}z^m P_{G,\infty}^N(m).
\end{equation}
Using Eq.~(\ref{LT-prob-nuclear-total-delay}), the Laplace transform of $\Psi_{G,\infty}^N(z)$ with respect to $T^N$ is given by
\begin{equation}
    \mathcal{L}[\Psi_{G,\infty}^N(z)](s)=\int_{0}^{\infty}dT^N \Psi_{G,\infty}^N(z)e^{-sT^N}=\frac{1}{s}+\frac{(z-1)[1-\phi(s)]}{\alpha s^2[1-z\phi(s)]}.
\end{equation}
From here it follows that the mean and the variance of the total nuclear mRNA number are given by
\begin{equation}
    \mu^N=\frac{T^N }{\alpha}, \quad (\sigma^{N})^2=\mathcal{L}^{-1}\left\{\frac{1+\phi(s)}{\alpha s^2[1-\phi(s)]}\right\}(T^N)-\left(\frac{T^N}{\alpha}\right)^2,
\end{equation}
and therefore the Fano factor $FF_{G,\infty}^N$ reads
\begin{equation}
    \label{delay-nuclear-cytoplasmic-FF}
    FF_{G,\infty}^N=\frac{1}{T^N}\mathcal{L}^{-1}\left\{\frac{1+\phi(s)}{s^2[1-\phi(s)]}\right\}(T^N)-\frac{T^N}{\alpha}.
\end{equation}
The first term in Eq.~(\ref{delay-nuclear-cytoplasmic-FF}) can be computed using partial fraction decomposition,
\begin{equation}
    \label{LT-delay-nuclear-partial-fractions}
    \frac{1+\phi(s)}{s^2[1-\phi(s)]}=\frac{(k+s)^G+k^G}{s^2[(k+s)^G-k^G]}=\frac{c_1}{s}+\frac{c_2}{s^2}+\frac{c_3}{s^3}+\sum_{r=1}^{G-1}\frac{A_r}{s+s_r},
\end{equation}
where $s_r$ are defined as
\begin{equation}
    s_r=k(1-\epsilon_r),\quad \epsilon_r=e^{\frac{2\pi i r}{G}},\quad r=1,\dots,G-1.
\end{equation}
From here we get that
\begin{align}
    \label{LT-delay-nuclear-partial-fractions-2}
    & A_r=\frac{2}{kG}\frac{\epsilon_r}{(1-\epsilon_r)^2},\quad c_1=-\sum_{r=1}^{G-1}A_r,\quad c_2=1-\frac{2}{G}\sum_{r=1}^{G-1}\frac{1}{1-\epsilon_r},\quad c_3=\frac{2k}{G}.
\end{align}
Inserting Eq.~(\ref{LT-delay-nuclear-partial-fractions-2}) into Eq.~(\ref{LT-delay-nuclear-partial-fractions}) and inverting the Laplace transform yields
\begin{equation}
    \label{delay-nuclear-cytoplasmic-FF-2}
    FF_{G,\infty}^N=1+\frac{2}{G}\sum_{r=1}^{G-1}\frac{\epsilon_r}{1-\epsilon_r}\left\{1-\frac{1}{kT^N(1-\epsilon_r)}\left[1-e^{-kT(1-\epsilon_r)}\right]\right\},
\end{equation}
where we also used the following result, 
\begin{equation}
    \sum_{r=1}^{G-1}\frac{1}{1-\epsilon_r}=-\sum_{r=1}^{G-1}\frac{\epsilon_r}{1-\epsilon_r}. 
\end{equation}
which is easy to verify. Namely, if $G$ is odd, then $\epsilon_1,\dots,\epsilon_{G-1}$ are positioned symmetrically in the complex plane with respect to the real axis, meaning that $\epsilon_r=1/\epsilon_{G-r}$. In that case,
\begin{equation}
    \sum_{r=1}^{G-1}\frac{1+\epsilon_r}{1-\epsilon_r}=\sum_{r=1}^{\frac{G-1}{2}}\left[\frac{1+\epsilon_r}{1-\epsilon_r}+\frac{1+1/\epsilon_r}{1-1/\epsilon_r}\right]=0.
\end{equation}
On the other hand, if $G$ is even, then there is an extra point $\epsilon_{r}=-1$ for $r=G/2$, which satisfies $1/(1-\epsilon_r)=-\epsilon_r/(1-\epsilon_r)$. 

Let $\theta_r=2\pi r/G$. The expression in Eq.~(\ref{delay-nuclear-cytoplasmic-FF-2}) simplifies to
\begin{equation}
    \label{FF-delay}
    FF_{G,\infty}^N=\begin{dcases}
    \frac{1}{G}\left[1+2\sum_{r=1}^{\frac{G-1}{2}}g(kT^N,\theta_r)\right], & \text{$G$ is odd}\\
    \frac{1}{G}\left[1+2\sum_{r=1}^{\frac{G-2}{2}}g(kT^N,\theta_r)+g(kT^N,\pi)\right], & \text{$G$ is even},
    \end{dcases}
\end{equation}
where $g(x,y)$ is defined as
\begin{equation}
    g(x,y)=\frac{1-cos(x sin(y))e^{-x(1-cos(y))}}{x(1-cos(y))},\quad y\neq 0.
\end{equation}
Next, we show that $\partial g/\partial x<0$ for $x>0$ and $y\neq 0$, which means that $g(x,y)$ is monotonically decreasing for positive $x$. The partial derivative of $g(x,y)$ with respect to $x$ is given by
\begin{equation}
    \frac{\partial g}{\partial x}=-\frac{1+\left[xcos(y+xsin(y))-(1+x)cos(xsin(y))\right]e^{-x(1-cos(y))}}{x^2(1-cos(y))}.
\end{equation}
Since $1-cos(y)>0$ for $y\neq 0$, the factor in front of the curly brackets is always negative. On the other hand, the expression in the curly brackets is positive for any $x>0$ and $y\neq 0$, since in that case $e^{-x(1-cos(y))}<1$ and
\begin{equation}
    xcos(y+xsin(y))-(1+x)cos(xsin(y))>-x-(1-x)=-1,
\end{equation}
where we used that $cos(y+xsin(y))>-1$ and $cos(xsin(y))<1$ for $y\neq 0$. The proof is the same for total cytoplasmic mRNA, since it has the same arrival process as the total nuclear mRNA. ~$\square$

\begin{corollary}
    \label{cor_FF-N-C}
    Under the conditions of Corollary~\ref{cor_delay-nuclear-cytoplasmic-prob}, if $T^N < T^C$ then $FF_{G,\infty}^N>FF_{G,\infty}^C$ and \textit{vice versa}.
\end{corollary}

\begin{corollary}
    \label{cor_CV-N-C}
    Let $CV^{N}_{G,\infty}=\sqrt{\langle m^2\rangle^{N}_{G,\infty}-(\langle m\rangle^{N}_{G,\infty})^2}/\langle m\rangle^{N}_{G,\infty}$ and $CV^{C}_{G,\infty}=\sqrt{\langle m^2\rangle^{C}_{G,\infty}-(\langle m\rangle^{C}_{G,\infty})^2}/\langle m\rangle^{C}_{G,\infty}$ denote the coefficients of variation of the total nuclear and cytoplasmic mRNA, respectively. Then $CV^{N}_{G,\infty}$ decays monotonically with $T^N$, and $CV^{C}_{G,\infty}$ decays monotonically with $T^C$. Furthermore, if $T^N < T^C$ then $CV_{G,\infty}^N>CV_{G,\infty}^C$ and \textit{vice versa}.
\end{corollary}

\noindent\textit{\textbf{Proof of Corollary~\ref{cor_CV-N-C}}.} Note that $CV^{N}_{G,\infty}=\sqrt{FF^{N}_{G,\infty}/\langle m\rangle^{N}_{G,\infty}}$, where $\langle m\rangle^{N}_{G,\infty}$ is the mean number of total nuclear mRNA, $\langle m\rangle^{N}_{G,\infty}=T^N/\alpha=T^N k/G$. Since $\langle m\rangle^{N}_{G,\infty}\propto T^N$, and $FF^{N}_{G,\infty}$ decays monotonically with $T^N$, it follows immediately that $CV^{N}_{G,\infty}$ decays monotonically with $T^N$. The same argument applies to total cytoplasmic mRNA with $T^N$ replaced by $T^C$. Since $CV^{N}_{G,\infty}$ and $CV^{C}_{G,\infty}$ have the same functional dependence on $T^N$ and $T^C$, respectively, then $T^N < T^C$ implies $CV_{G,\infty}^N>CV_{G,\infty}^C$ and \textit{vice versa}. $\square$

In summary, we mapped the model in Eq.~(\ref{model_reactions}) to the model in Eq. (\ref{model_reactions-queueing}) describing total nuclear and cytoplasmic mRNA, and reframed it as a $G/G/\infty$ queue. As our main result, we computed the Fano factor of the total nuclear mRNA $M^N$ using results of Ref.~\cite{Takacs_1958}. In the special limit in which the processing of both nuclear and cytoplasmic mRNA becomes deterministic, we computed the distributions of the total nuclear and cytoplasmic mRNA numbers, and proved that their Fano factors decay monotonically with their respective processing times. In Section~\ref{sims}, we will test these results using stochastic simulations.

\section{Results using stochastic model reduction}
\label{modred}

In this section, we use a completely different method of mathematical analysis than the previous section. We utilize the slow-scale linear-noise approximation (ssLNA)~\cite{thomas2012slow,thomas2012rigorous}, which provides a rigorous method of model reduction for systems with linear propensities (such as ours) when there exists timescale separation between species. Specifically, the ssLNA provides an analytical recipe to compute the first and second moments of the number of molecules of the slow species. It provides accurate results whenever the timescales of the transients in the mean molecules of different species are well separated. Other methods that provide a reduced stochastic description of stochastic reaction kinetics have also been developed; see for example~\cite{kim2014validity,kim2017reduction,mastny2007two,herath2018reduced,kang2019quasi}. 

\subsection{Determining the timescales for each species} \label{section_timescales}

Before we can determine the timescales, we need the time-evolution equations for the mean molecule numbers of the reaction system~\eqref{model_reactions}. These can be derived directly from the CME, though in this case it is simpler to state them directly using the law of mass action because since each reaction is first-order then the time-evolution equations for the means are precisely the same as the deterministic rate equations:
\begin{equation}
\label{ode}
\begin{aligned}
    &\frac{d[U_0]}{dt}=k_{\rm{off}}[U_1]-k_{\rm{on}}[U_0],\\
    &\frac{d[U_1]}{dt}=k_{\rm{on}}[U_0]+k_G(1-[U_0]-...-[U_{G-1}])-k_{\rm{off}}[U_1]-k_1[U_1],\\
    &\frac{d[U_g]}{dt}=k_{g-1}[U_{g-1}]-k_g[U_g],\  g=2,3,...,G-1,\\
    &\frac{d[M^N_1]}{dt}=k_G(1-[U_0]-...-[U_{G-1}])-\delta [M^N_1],\\
    &\frac{d[M^N_s]}{dt}=\delta [M^N_{s-1}]-\delta [M^N_s],\  s=2,3,...,S-1,\\ 
    &\frac{d[M^N_S]}{dt}=\delta [M^N_{S-1}]-\delta_1 [M^N_S],\\
    &\frac{d[M^C_1]}{dt}=\delta_1 [M^N_S]-\lambda [M^C_1],\\
    &\frac{d[M^C_r]}{dt}=\lambda [M^C_{r-1}]-\lambda [M^C_r],\ r=2,3,...,R,\\
    &\frac{d[P]}{dt}=\lambda_1 [M^C_1]-\lambda_2[P],
\end{aligned}
\end{equation}
where $[x]$ denotes the mean molecule numbers of species $x$, and $[U_G]=1-\sum_{g}^{G-1}[U_g]$ follows from the conservation law of gene states (assuming one gene copy). 

Note that we have not included a rate equation for the nascent mRNA species $M_0^N$. This is not important in the determination of the mean molecule numbers of other species because due to the deterministic nature of elongation (and termination), the time between two subsequent $M_0^N$ production events is precisely the same as the time between two subsequent $M_1^N$ production events. Hence, from a steady-state perspective, one could as well replace the set of reactions $U_G \rightarrow U_{1}+M^{N}_0, M^{N}_0 \Rightarrow M^N_1$ by the simpler reaction $U_G \rightarrow U_{1}+M^N_1$. This argument holds not only for the rate equations but also for the CME description of the system, hence in all that follows we do not track the nascent mRNA species. 

The steady state mean molecule numbers of each species are obtained by setting the time derivative to zero and solving the equations simultaneously. Before presenting this solution, we define the elementary symmetric polynomial, since it enables the results to be presented compactly.

\begin{definition}[\textbf{Elementary symmetric polynomial}]
\label{sym_poly}

The elementary symmetric polynomial $e_g(k_1,...,k_G)$ in $G$ variables for $g=1,2,...,G$ is defined as~\cite{macdonald1998symmetric}
\begin{equation}
\begin{aligned}
    e_g(k_1,...,k_G)=\sum_{1\leq j_1<j_2...<j_g\leq G}k_{j_1}\cdot\cdot\cdot k_{j_g}.
\end{aligned}
\end{equation}
Note that $e_g(k_1,...,k_G)=1$ for $g=0$, whereas $e_g(k_1,...,k_G)=0$ for $g<0$ or when $g$ is larger than the number of variables.
\end{definition}

The mean concentrations of each species are then given by
\begin{equation}
\label{mean}
\begin{aligned}
    &[U_0]=\frac{k_{\rm{off}}e_{G-1}(k_2,...,k_G)}{k_{\rm{on}} e_{G-1}(k_1,...,k_G)+k_{\rm{off}}e_{G-1}(k_2,...,k_G)},\\
    &[U_g]=\frac{e_{G+1}(k_{\rm{on}},...,k_G)/k_g}{k_{\rm{on}} e_{G-1}(k_1,...,k_G)+k_{\rm{off}}e_{G-1}(k_2,...,k_G)},\ g=1,2,...G, \\
    &[M^N_s]=\frac{e_{G+1}(k_{\rm{on}},...,k_G)}{\delta(k_{\rm{on}} e_{G-1}(k_1,...,k_G)+k_{\rm{off}}e_{G-1}(k_2,...,k_G))},\ s=1,2,...,S-1,\\
    &[M^N_S]=\frac{e_{G+1}(k_{\rm{on}},...,k_G)}{\delta_1(k_{\rm{on}} e_{G-1}(k_1,...,k_G)+k_{\rm{off}}e_{G-1}(k_2,...,k_G))},\\
    &[M^C_r]=\frac{e_{G+1}(k_{\rm{on}},...,k_G)}{\lambda(k_{\rm{on}} e_{G-1}(k_1,...,k_G)+k_{\rm{off}}e_{G-1}(k_2,...,k_G))},\ r=1,2,...,R,\\
    &[P]=\frac{\lambda_1e_{G+1}(k_{\rm{on}},...,k_G)}{\lambda_2\lambda(k_{\rm{on}} e_{G-1}(k_1,...,k_G)+k_{\rm{off}}e_{G-1}(k_2,...,k_G))}.
\end{aligned}
\end{equation} 

Next, we construct the Jacobian matrix associated with the deterministic rate equations. If each species is assigned a number, then the $(i,j)$ element of the Jacobian matrix is obtained by differentiating the right-hand side of the time-evolution equation for the mean of the $i$th species with respect to the mean of the $j$th species. We do not write explicit equations for this matrix since it is cumbersome, but it can be done easily using Eq.~\eqref{ode}.  Finally, the timescale of each species can be determined from the Jacobian matrix and the steady-state solution Eq.~\eqref{mean}, as follows.   

\begin{definition}[\textbf{Timescales}]
\label{def_timescales}
Let $\tau^G_{g}, \tau^N_{s}, \tau^C_{r}$, and $\tau^P$ denote the timescales of gene states $g=0,...,G$, nuclear mRNA species $s=1,...,S$, cytoplasmic mRNA species $r=1,...,R$ and proteins, respectively. The timescale of each species is an inverse of an eigenvalues of the Jacobian matrix evaluated at the steady-state mean molecule number solution of the deterministic rate equations. Furthermore, we define a timescale separation parameter between species $i$ and $j$,
\begin{equation}
    \label{lambda-i-j}
    \Lambda^{i,j}=\frac{\max[\tau^{i}]}{\min[\tau^{j}]},
\end{equation}
where $i,j=G,N,C,P$. A species $i$ is fast compared to species $j$ if $\Lambda^{i,j} \ll 1$.
\end{definition}

Given a set of parameter values, using the above definition, it is always possible to numerically find the timescales for each species. Provided a slow species can be identified, then the ssLNA is applicable and the equations for the steady-state means and variances of the slow species can be determined in a closed form. A brief summary of the ssLNA method can be found in Appendix~\ref{app_ssLNA}. 

These timescales are roughly known for mammalian cells. The median lifetimes of cytoplasmic mRNA and protein are about 9 hours and 46 hours~\cite{schwanhausser2011global}, respectively, hence protein species are the slowest of the two. The nuclear mRNA lifetime (retention time) has a median of 20 minutes~\cite{battich2015control}, hence cytoplasmic mRNA species are slower than nuclear mRNA species. Finally, gene timescales are very short, of the order of seconds to few minutes~\cite{lammers2020matter}, and therefore gene species can be considered faster than both mRNA and protein species. Given the natural timescale separation between various species, next we apply the ssLNA to derive expressions for the statistics of mRNA and protein fluctuations.

\subsection{Applying the ssLNA to reaction system~(\ref{model_reactions})}

\begin{proposition}
\label{prop_cmrna_cov}

Let $FF^C_{s}$ denote the Fano factor of total cytoplasmic mRNA $M^C=M_{1}^{C}+\dots+M_{R}^{C}$ where the subscript $s$ refers to the result being obtained using the ssLNA. Under the assumption that the timescale of each cytoplasmic mRNA species is significantly larger than the timescales of all gene and nuclear mRNA species, the Fano factor of total cytoplasmic mRNA in steady-state conditions is given by
\begin{equation}
\label{FFcmrna}
\begin{aligned}
    FF^C_{s}
    =&1+2e_G(k_1,...,k_G)\left(1-\frac{(2R-1)!!}{(2R)!!}\right)\\
    &\times\frac{\left(k_{\rm{off}}e_{G-1}(k_2,...k_G)-k_{\rm{on}}k_{\rm{off}}e_{G-2}(k_2,...,k_G)- k_{\rm{on}}^2e_{G-2}(k_1,...k_G)\right)}{(k_{\rm{off}}e_{G-1}(k_2,...k_G)+k_{\rm{on}}e_{G-1}(k_1,...k_G))^2}.
\end{aligned}
\end{equation}
\end{proposition}

Note that an analogous analysis can be conducted for nuclear mRNA, assuming that nuclear mRNA exhibits sufficiently slower dynamics compared to gene states, and that the export rate ($\delta_1$) is equal to the nuclear mRNA processing rate ($\delta$). In that case, the resulting Fano factor for nuclear mRNA is the same as that for cytoplasmic mRNA, i.e. Eq.~\eqref{FFcmrna}, with $R$ replaced by $S$.\\

\noindent\textit{\textbf{Sketch of the derivation of Proposition~\ref{prop_cmrna_cov}.}} The ssLNA states that the covariance matrix of the slow variables obeys a Lyapunov equation of the form
\begin{equation}
    \label{cmrna_Lyapunov}
    \mathbf{J}_s^C \mathbf{C}_s^C+ \mathbf{C}_s^C [\mathbf{J}_s^{C}]^T+\mathbf{D}_{s}^C=\mathbf{0},
\end{equation}
where $\mathbf{C}_s^C$ is the covariance matrix of cytoplasmic mRNA species, $\mathbf{J}_s^C$ is the reduced Jacobian matrix and $\mathbf{D}_s^C$ is the reduced diffusion matrix, which are defined below. 

In Appendix~\ref{app_cmrna_FF}, we show that the non-zero elements of $\mathbf{J}_s^C$ are given by
\begin{equation}
\label{cmrna_Jacob}
\begin{aligned}
    &\mathbf{J}^C_{s(1,1)}=-\lambda,\\
    &\mathbf{J}^C_{s(i,i-1)}=\lambda,\; \mathbf{J}^C_{s(i,i)}=-\lambda,\;\; \text{for}\; i=2,3,...,R,
\end{aligned}
\end{equation}
and the non-zero elements of $\mathbf{D}_s^C$ are given by
\begin{equation}
\label{cmrna_diff}
\begin{aligned}
    &\mathbf{D}^C_{s(1,1)}=\lambda[M_1^C]+e_{G+1}(k_{\rm{on}},k_1,...,k_G)\\
    &\qquad\qquad\times\frac{\left(k_{\rm{on}}^2 e_{G-1}(k_1^2,...,k_G^2)+e_{G-1}(k_2^2,...,k_G^2)(k_{\rm{off}}^2+2k_{\rm{off}}k_{\rm{on}}+2k_{\rm{off}}k_1)\right)}{(k_{\rm{off}}e_{G-1}(k_2,...,k_G)+k_{\rm{on}}e_{G-1}(k_1,...,k_G))^3},\\
    &\mathbf{D}^C_{s(i-1,i)}=-\lambda[M_{i-1}^C],\; \mathbf{D}^C_{s(i,i)}=\lambda([M_{i-1}^C]+[M_{i}^C]),\; \mathbf{D}^C_{s(i,i+1)}=-\lambda[M_i^C],\\
    &\qquad\qquad\text{for}\; i=2,3,...,R-1,\\
    &\mathbf{D}^C_{s(R,R-1)}=-\lambda[M_{R-1}^C],\; \mathbf{D}^C_{s(R,R)}=\lambda([M_{R-1}^C]+[M_{R}^C]).
\end{aligned}
\end{equation}

Directly solving the Lyapunov Eq.~\eqref{cmrna_Lyapunov}  for arbitrary values of the model parameters is too difficult. Instead, we solved the equation explicitly for several small-species systems, e.g. by setting $(G = 2, R = 1)$, $(G = 2, R = 2)$ and so on, from which we deduced a general form for $\mathbf{C}_s^C$: 
\begin{align}
\label{covareq}
    \mathbf{C}^C_{s(i,j)}=&\frac{(k_{\rm{off}}e_{G-1}(k_2,...,k_G)-k_{\rm{on}}k_{\rm{off}}e_{G-2}(k_2,...,k_G)-k^2_{\rm{on}}e_{G-2}(k_1,...,k_G))}{\lambda(k_{\rm{on}} e_{G-1}(k_1,...,k_{G})+k_{\rm{off}}e_{G-1}(k_2,...,k_G))^3}\nonumber\\
    &\times\frac{2^{2-i-j}\Gamma(i+j-1)}{\Gamma(i)\Gamma(j)}k_{\rm{on}}e_G(k_1^2,...,k_G^2),\quad\text{for}\; i\neq j,\nonumber\\
    \mathbf{C}^C_{s(i,j)}=&e_{G+1}(k_{\rm{on}},k_1,...,k_G)\left(\frac{k_{\rm{on}}^2e_{G-1}(k_1^2,...,k_G^2)+\frac{(2R-3)!!}{(2R-2)!!}k_{\rm{off}}k_1e_{G-1}(k_2^2,...,k_G^2)}{\lambda(k_{\rm{on}} e_{G-1}(k_1,...,k_{G})+k_{\rm{off}}e_{G-1}(k_2,...,k_G))^3}\right.\\
    &+\frac{\left(2-\frac{(2R-3)!!}{(2R-2)!!}\right)e_{G+1}(k_{\rm{on}},k_1,...,k_G)(k_{\rm{off}}e_{G-2}(k_2,...,k_G)+k_{\rm{on}}e_{G-2}(k_1,...,k_G))}{\lambda(k_{\rm{on}} e_{G-1}(k_1,...,k_{G})+k_{\rm{off}}e_{G-1}(k_2,...,k_G))^3}\nonumber\\
    &+\left.\frac{e_{G-1}(k_2^2,...,k_G^2)(2k_{\rm{off}}k_{\rm{on}}+k_{\rm{off}}^2)}{\lambda(k_{\rm{on}} e_{G-1}(k_1,...,k_{G})+k_{\rm{off}}e_{G-1}(k_2,...,k_G))^3}\right)\nonumber,\quad\text{for}\; i=j.
\end{align}
We then verified the solution by substituting it in the left-hand side of Eq.~\eqref{cmrna_Lyapunov}, and showing that it leads to zero for an arbitrary set of model parameters.

The variance of total cytoplasmic mRNA is equal to the sum of the covariances of each pair of cytoplasmic mRNA species (the sum over all $i$ and $j$). On the other hand, the mean of total cytoplasmic mRNA is equal to the sum of the mean of the cytoplasmic mRNA species, as given by Eq.~\eqref{mean}. Hence, by dividing the variance by the mean, we obtain the Fano factor of total cytoplasmic mRNA. We note that the Fano factor of total cytoplasmic mRNA in Eq.~(\ref{FFcmrna}) is independent of the parameters of reactions involving nuclear mRNA, i.e. $\delta$ (the nuclear processing rate) and $\delta_1$ (the nuclear export rate). 

\begin{corollary}
\label{cor_cmrna_thre}
If the activation rate $k_{\rm{on}}$ is large enough such that the condition $e_{G-1}(k_2,...,k_G)-k_{\rm{on}}e_{G-2}(k_2,...,k_G)\le 0$ holds, then the fluctuations in the total cytoplasmic mRNA are sub-Poissonian for all values of the deactivation rate $k_{\rm{off}}$, i.e. $FF^C_{s} < 1$. On the other hand, if $k_{\rm{on}}$ is small enough such that the condition $e_{G-1}(k_2,...,k_G)-k_{\rm{on}}e_{G-2}(k_2,...,k_G)>0$ holds, then the fluctuations change from sub-Poissonian ($FF^C_s < 1$) to super-Poissonian ($FF^C_s > 1$) as the deactivation rate $k_{\rm{off}}$ crosses a threshold $k_{\rm{off}}^\star$ given by
\begin{equation}
    \label{thre1}
    k_{\rm{off}}^\star=\frac{k_{\rm{on}}^2e_{G-2}(k_1,...,k_G)}{e_{G-1}(k_2,...,k_G)-k_{\rm{on}}e_{G-2}(k_2,...,k_G)}. 
\end{equation}
Furthermore, when $k_1=k_2=..=k_{G}=k$, the condition $e_{G-1}(k_2,...,k_G)-k_{\rm{on}}e_{G-2}(k_2,...,k_G)>0$ reduces to $k > k_{\rm{on}}\cdot(G-1)$, and the threshold simplifies to
\begin{equation}
\label{crit}
    k_{\rm{off}}^\star=\frac{1}{2} \frac{G(G-1)k_{\rm{on}}^2}{k - k_{\rm{on}}(G-1)}.
\end{equation}
\end{corollary}

\begin{corollary}
\label{cor_cmrna_cov_positive}
Covariances of cytoplasmic mRNA states, denoted by $C^C_{s(i,j)}$, where $i\neq j\in\{1,\dots,R\}$, are positive when $FF^{C}_{s}$ is super-Poissonian ($FF^{C}_{s}>1$) and negative when $FF^{C}_{s}$ is sub-Poissonian ($FF^{C}_{s}<1$).
\end{corollary}

\noindent\textit{\textbf{Proof of Corollary~\ref{cor_cmrna_cov_positive}}.} Given Eq.~\eqref{covareq} and Corollary~\ref{cor_cmrna_thre}, it follows that if $k_{\rm{on}}$ is large enough such that $e_{G-1}(k_2,...,k_G)-k_{\rm{on}}e_{G-2}(k_2,...,k_G)\le 0$ holds, i.e. $FF^C_s<1$, then
\begin{equation}
    k_{\rm{off}}(e_{G-1}(k_2,...,k_G)-k_{\rm{on}}e_{G-2}(k_2,...,k_G))-k^2_{\rm{on}}e_{G-2}(k_1,...,k_G)<0,
\end{equation}
and therefore $C^C_{s(i,j)}<0$. If $k_{\rm{on}}$ is small enough such that $e_{G-1}(k_2,...,k_G)-k_{\rm{on}}e_{G-2}(k_2,...,k_G)>0$ holds, and if $k_{\rm{off}}\geq k_{\rm{off}}^\star$, i.e. if $FF^C_s\geq 1$, then
\begin{equation}
    k_{\rm{off}}(e_{G-1}(k_2,...,k_G)-k_{\rm{on}}e_{G-2}(k_2,...,k_G))-k^2_{\rm{on}}e_{G-2}(k_1,...,k_G)\geq 0,
\end{equation}
from where it follows that $C^C_{s(i,j)}\geq0$. On the hand, if $k_{\rm{off}}<k_{\rm{off}}^\star$, i.e. if $FF^C_s<1$, then
\begin{equation}
    k_{\rm{off}}(e_{G-1}(k_2,...,k_G)-k_{\rm{on}}e_{G-2}(k_2,...,k_G))-k^2_{\rm{on}}e_{G-2}(k_1,...,k_G)< 0,
\end{equation}
from where it follows that $C^C_{s(i,j)}<0$. Therefore, $C^C_{s(i,j)}>0$ when $FF^C_s>1$ and $C^C_{s(i,j)}<0$ when $FF^C_s<1$.~$\square$

\begin{corollary}
\label{cor_cmrna_FFmono}
The Fano factor of the total cytoplasmic mRNA, $FF^C_{s}$, increases with the number of processing steps $R$ in the cytoplasm provided $FF^C_{s}$ is super-Poissonian ($FF^C_{s}>1$), and decreases with $R$ provided $FF^C_{s}$ is sub-Poissonian ($FF^C_{s}<1$).  
\end{corollary}

\noindent\textit{\textbf{Proof of Corollary~\ref{cor_cmrna_FFmono}.}}
For any $R\in \mathbb Z+$, we will show that the sign of $FF^C_s(R+1)-FF^C_s(R)$ depends on whether $FF^C_s$ is super-Poissonian or sub-Poissonian. The difference $FF^C_s(R+1)-FF^C_s(R)$ is given by
\begin{equation}
\label{FFcmrna_diff_R}
\begin{aligned}
&2e_{G}(k_1,...,k_G)\left(\frac{(2R-1)!!}{(2R)!!}-\frac{(2R+1)!!}{(2R+2)!!}\right)\\
    &\times\frac{\left(k_{\rm{off}}e_{G-1}(k_2,...k_G)-k_{\rm{on}}k_{\rm{off}}e_{G-2}(k_2,...,k_G)- k_{\rm{on}}^2e_{G-2}(k_1,...k_G)\right)}{(k_{\rm{off}}e_{G-1}(k_2,...k_G)+k_{\rm{on}}e_{G-1}(k_1,...k_G))^2}\\
    =&2e_{G}(k_1,...,k_G)\left(\frac{(2R)!}{(2R+2)!!(2R)!!}\right)\\
    &\times\frac{\left(k_{\rm{off}}e_{G-1}(k_2,...k_G)-k_{\rm{on}}k_{\rm{off}}e_{G-2}(k_2,...,k_G)- k_{\rm{on}}^2e_{G-2}(k_1,...k_G)\right)}{(k_{\rm{off}}e_{G-1}(k_2,...k_G)+k_{\rm{on}}e_{G-1}(k_1,...k_G))^2}.
\end{aligned}
\end{equation}
From here we conclude that the sign of $FF^C_s(R+1)-FF^C_s(R)$ depends on the sign of the expression in the last row. If $e_{G-1}(k_2,...,k_G)-k_{\rm{on}}e_{G-2}(k_2,...,k_G)\le 0$ holds such that $FF^C_{s}<1$, then
\begin{equation}
    k_{\rm{off}}e_{G-1}(k_2,...k_G)-k_{\rm{on}}k_{\rm{off}}e_{G-2}(k_2,...,k_G)-k_{\rm{on}}^2e_{G-2}(k_1,...k_G)<0,   
\end{equation}
i.e. $FF^C_s(R+1)-FF^C_s(R)<0$.

From Eq.~(\ref{thre1}) it follows that if $e_{G-1}(k_2,...,k_G)-k_{\rm{on}}e_{G-2}(k_2,...,k_G)>0$ and $k_{\rm{off}}\geq k_{\rm{off}}^\star$ such that $FF^C_{s}\geq 1$, then
\begin{equation}
    k_{\rm{off}}e_{G-1}(k_2,...,k_G)-k_{\rm{on}}k_{\rm{off}}e_{G-2}(k_2,...,k_G))- k_{\rm{on}}^2e_{G-2}(k_1,...,k_G)\geq 0,
\end{equation}
i.e. $FF^C_s(R+1)-FF^C_s(R)\geq 0$, and when $e_{G-1}(k_2,...,k_G)-k_{\rm{on}}e_{G-2}(k_2,...,k_G)>0$, and $k_{\rm{off}}< k_{\rm{off}}^\star$ such that $FF^C_{s}<1$, we have
\begin{equation}
    k_{\rm{off}}e_{G-1}(k_2,...,k_G)-k_{\rm{on}}k_{\rm{off}}e_{G-2}(k_2,...,k_G))-k_{\rm{on}}^2e_{G-2}(k_1,...,k_G)<0,  
\end{equation}
i.e. $FF^C_s(R+1)-FF^C_s(R)<0$.
Therefore, the sign of Eq.~(\ref{FFcmrna_diff_R}) is positive when $FF^C_{s}>1$ and negative when $FF^C_{s}<1$. ~$\square$  

\begin{corollary}\label{cor_cmrna_CVmono}
The squared of the coefficient of variation for the total cytoplasmic mRNA is given by
\begin{equation}
\label{CVcmrna}
\begin{aligned}
    \left[CV^C_s\right]^2&=\frac{\lambda(k_{\rm{off}}e_{G-1}(k_2,...k_G)+k_{\rm{on}}e_{G-1}(k_1,...k_G))}{R e_{G+1}(k_{\rm{on}},...,k_G)}\\
    &+2\left(1-\frac{(2R-1)!!}{(2R)!!}\right)\frac{\lambda\left(k_{\rm{off}}e_{G-1}(k_2,...k_G)-k_{\rm{on}}k_{\rm{off}}e_{G-2}(k_2,...,k_G)- k_{\rm{on}}^2e_{G-2}(k_1,...k_G)\right)}{Rk_{\rm{on}}(k_{\rm{off}}e_{G-1}(k_2,...k_G)+k_{\rm{on}}e_{G-1}(k_1,...k_G))}.
\end{aligned}
\end{equation}
This increases with the number of processing steps $R$ in the cytoplasm provided fluctuations are super-Poissonian and decreases with $R$ if they are sub-Poissonian.
\end{corollary}

\noindent\textit{\textbf{Proof of Corollary~\ref{cor_cmrna_CVmono}.}} The formula Eq.~(\ref{CVcmrna}) follows immediately by the fact that $\left[CV^C_s\right]^2$ equals the ratio of $FF^C_s$ and the mean total cytoplasmic mRNA counts $(R\times[M^C_s])$.
The dependence on the number of processing steps $R$ can be proved in a similar way to the Fano factor case (Proof of Corollary~\ref{cor_cmrna_FFmono}). ~$\square$  

\begin{corollary}
\label{cor_cmrna_cases}
Let
\begin{equation}
\label{def_conditions}
\begin{aligned}
    &k_1^\star:=\frac{e_{G-1}(k_2,...,k_G)e_{G-2}(k_2,...,k_G)}{e_{G-2}(k_2^2,...,k_G^2)},\\ 
    &k_{\rm{on},1}^\star:=\frac{e_{G-1}(k_2,...,k_G)}{e_{G-2}(k_2,...,k_G)},\\
    &k_{\rm{on},2}^\star:=\frac{e_{G-1}(k_2,...,k_G)e_{G-1}(k_1,...,k_G)}{k_1e_{G-2}(k_2^2,...,k_G^2)-e_{G-1}(k_2,...,k_G)e_{G-2}(k_2,...,k_G)}.\\
\end{aligned}
\end{equation}
The Fano factor of cytoplasmic mRNA $FF^C_{s}$ varies with respect to the deactivation rate $k_{\rm{off}}$ according to the following three cases.
\begin{itemize}
    \item Case 1: When $k_1>k_1^\star$ and $k_{\rm{on}}>k_{\rm{on},2}^\star$, the Fano factor $FF^C_{s}$ first decreases until it reaches the critical point $k_{\rm{off},1}$, and then increases, eventually approaching $1$ from below.
    
    \item Case 2: When $k_1\leq k_1^\star$ and $k_{\rm{on}}\geq k_{\rm{on},1}^\star$ or $k_1>k_1^\star$ and $k_{\rm{on},1}^\star\leq k_{\rm{on}}\leq k_{\rm{on},2}^\star$, the Fano factor $FF^C_{s}$ monotonically increases, eventually approaching $1$ from below.
    
    \item Case 3: When $k_{\rm{on}}<k_{\rm{on},1}^\star$, the Fano factor $FF^C_{s}$ first increases until it reaches the critical point $k_{\rm{off},1}$, and then decreases, eventually approaching $1$ from above.
\end{itemize}
The proof can be found in Appendix~\ref{app_cmrna_cases}.
\end{corollary}

\begin{corollary}
\label{cor_cmrna_limit}
The minimum of $FF^C_{s}$ is reached when $k_{\rm{off}}=0$ and $k_1=k_2=...=k_G=k$, and is given by
\begin{equation}
    \label{cmrna_limit}
    \lim_{\substack{k_1=...=k_G=k \\ k_{\rm{off}}= 0}}FF^C_{s}=\frac{1}{G}+\left(1-\frac{1}{G}\right)\frac{(2R-1)!!}{(2R)!!}.
\end{equation} 
\end{corollary}

\noindent\textit{\textbf{Proof of Corollary~\ref{cor_cmrna_limit}.}}
It follows from Corollary~\ref{cor_cmrna_cases} that when varying $k_{\rm{off}}$, the minimum is achieved at $k_{\rm{off}}=0$ in cases 2 and 3. We then fix $k_{\rm{off}}=0$ and show that the Fano factor achieves the minimum at $k_1=k_2=...=k_G=k$. When $k_{\rm{off}}=0$, we have
\begin{equation}
\begin{aligned}    
    \left.FF^C_{s}\right|_{k_{\rm{off}}=0}&=1-\left(1-\frac{(2R-1)!!}{(2R)!!}\right)\frac{2e_G(k_1,...k_G)e_{G-2}(k_1,...,k_G)}{e_{G-1}(k_1,...,k_G)^2}\\
    &\geq 1-\left(1-\frac{(2R-1)!!}{(2R)!!}\right)\frac{\frac{G-1}{G}e_{G-1}(k_1,...,k_G)^2}{e_{G-1}(k_1,...,k_G)^2}\\
    &=\frac{1}{G}+(1-\frac{1}{G})\frac{(2R-1)!!}{(2R)!!},
\end{aligned}
\end{equation}
where Newton's inequality~\cite{hardy1952inequalities} was used in the second row
\begin{equation}\label{FFcmrna_Newton}
    e_G(k_1,...k_G)e_{G-2}(k_1,...,k_G)\leq\frac{G-1}{2G}e_{G-1}(k_1,...,k_G)^2,
\end{equation}
and the equality holds if and only if $k_1=k_2=...=k_G=k$. Hence, in cases 2 and 3 the minimum is achieved provided $k_1=k_2=...=k_G=k$.

Since we do not have explicit proof that the minimum in case 1 is smaller than the global minimum $\frac{1}{G}+(1-\frac{1}{G}\frac{(2R-1)!!}{(2R)!!})$, we checked it for many parameter values. We computed the Fano factor for two groups of parameters using the theoretical result, each group containing $10000$ points. In group 1, we set $G=3$ and $R=3$, and in group 2, we set $G=4$ and $R=6$. Other parameter values were selected from the interval $(0.1,100)$ with the constraint $k_1>k_1^\star$ and $k_{\rm{on}}>k_{\rm{on},2}^\star$ which is necessary to enforce case 1 (see Corollary~\ref{cor_cmrna_cases}). For all parameter sets, we found that the Fano factor was larger than Eq.~\eqref{cmrna_limit}, which suggests that this equation gives the global minimum Fano factor for all three cases. ~$\square$



\begin{proposition}
\label{prop_pro_FF}

Let $FF^P_{s}$ denote the Fano factor of protein number fluctuations with $G$ gene states. Under the assumption that the timescale of protein species is significantly larger than the timescales of all other species, the ssLNA predicts that in the steady state,
\begin{equation}
\label{pro_FF}
\begin{aligned}
    FF^P_{s}=1&+\frac{\lambda_1}{2\lambda}+\frac{\lambda_1}{2\lambda}\frac{e_{G-1}(k_2^2,...,k_G^2)(k_{\rm{off}}^2+2k_{\rm{off}}(k_{\rm{on}}+k_1))+k_{\rm{on}}^2e_{G-1}(k_1^2,...,k_G^2)}{(k_{\rm{off}}e_{G-1}(k_2,...,k_G)+k_{\rm{on}}e_{G-1}(k_1,...,k_G))^2},
\end{aligned}
\end{equation}
where $\lambda$ is the cytoplasmic mRNA degradation rate, and $\lambda_1$ is the protein production rate. The Fano factor of protein number fluctuations is always larger than $1$ and independent of the number of mRNA processing steps in the nucleus ($S$) and the cytoplasm ($R$). Note that for the special case $G=1$, this reduces to the much simpler result
\begin{equation}
   FF^P_{s}|_{G=1}= 1+\frac{\lambda_1}{\lambda}+\frac{\lambda_1}{\lambda}\frac{ k_{\rm{off}}k_1}{(k_{\rm{off}}+k_{\rm{on}})^2},
\end{equation}
which is precisely the same as that obtained using the standard three-stage model of gene expression ($U_{0} \xrightleftharpoons[k_{\rm{off}}]{k_{\rm{on}}} U_{1}\xrightarrow{k_1} U_1 + M, M \xrightarrow{\lambda} \varnothing,  M \xrightarrow{\lambda_1} M + P,  P\xrightarrow{\lambda_2}\varnothing$), when the timescale of protein fluctuations is much larger than of mRNA and gene fluctuations~\cite{shahrezaei2008analytical}. Generally, it can be shown that $FF^P_{s}|_{G=1} \ge FF^P_{s}$, implying that the three-state model overestimates the Fano factor of protein noise.
\end{proposition}

\noindent\textit{\textbf{Sketch of the derivation of Proposition~\ref{prop_pro_FF}.}} Similar to the proof of  Proposition~\ref{prop_cmrna_cov}, using the ssLNA one can write a reduced Lyapunov equation of the form
\begin{equation}\label{pro_lyp}
    \mathbf{J}_s^P \mathbf{C}_s^P+\mathbf{C}_s^P[\mathbf{J}_s^P]^T+\mathbf{D}_s^P=0,    
\end{equation}
where $\mathbf{C}_s^P$ is the covariance of protein number fluctuations (that we need to solve for), $\mathbf{J}_s^P$ is the reduced Jacobian matrix and $\mathbf{D}_s^P$ is the reduced diffusion matrix. Note that since in this case we only have one slow variable, all matrices in the Lyapunov equation reduce to scalars. The reduced Jacobian and diffusion matrices are given by
\begin{equation}
\label{pro_Jacob}
\begin{aligned}
    \mathbf{J}_s^P=-\lambda_2,
\end{aligned}
\end{equation}
and
\begin{equation}\label{pro_diff}
\begin{aligned}
    \mathbf{D}_s^P=&\frac{\lambda_1}{\lambda}\times\left(\frac{e_{G-1}(k_2^2,...,k_G^2)(k_{\rm{off}}^2+2k_{\rm{off}}(k_{\rm{on}}+k_1))+k_{\rm{on}}^2e_{G-1}(k_1^2,...,k_G^2)}{(k_{\rm{off}}e_{G-1}(k_2,...,k_G)+k_{\rm{on}}e_{G-1}(k_1,...,k_G))^3}\right)\\
    &\times e_{G+1}(k_{\rm{on}},k_1,...,k_G)+\biggl(\frac{\lambda_1^2}{\lambda}+\lambda_1\biggr)[M_1^C]+\lambda_2[P].
\end{aligned}
\end{equation}
The substitution of Eq.~(\ref{pro_Jacob}) and Eq.~(\ref{pro_diff}) into the Lyapunov equation Eq.~(\ref{pro_lyp}) immediately leads to the solution
\begin{equation}
    \mathbf{C}_s^P=\frac{\mathbf{D}_s^P}{2\lambda_2}.
\end{equation}
The Fano factor can then be calculated using the formula $\mathbf{C}_s^P/[P]$ where $[P]$ is given by Eq.~\eqref{mean}. For details of the calculations, see Appendix~\ref{app_pro_FF}.

\begin{corollary}
\label{cor_pro_thre}
The Fano factor of protein $FF^P_{s}$ varies with respect to $k_{\rm{off}}$ according to the following three cases. 

\begin{itemize}
   \item Case 1: When $k_1>k_1^\star$ and $k_{\rm{on}}>k_{\rm{on},2}^\star$, the Fano factor $FF^P_{s}$ decreases until it reaches the critical point $k_{\rm{off},1}$ and then increases, eventually approaching $1+\frac{\lambda_1}{\lambda}$ from below.
   
   \item Case 2: When $k_1\leq k_1^\star$ and $k_{\rm{on}}\geq k_{\rm{on},1}^\star$ or $k_1>k_1^\star$ and $k_{\rm{on},1}^\star\leq k_{\rm{on}}\leq k_{\rm{on},2}^\star$, the Fano factor $FF^P_{s}$ monotonically increases, eventually approaching $1+\frac{\lambda_1}{\lambda}$ from below. 
   
   \item Case 3: When $k_{\rm{on}}< k_{\rm{on},1}^\star$, the Fano factor $FF^P_{s}$ first increases until it reaches the critical point $k_{\rm{off},1}$ and then decreases, eventually approaching $1+\frac{\lambda_1}{\lambda}$ from above.
\end{itemize}
Definitions for $k_1^\star$, $k_{\rm{on},1}^\star$, $k_{\rm{on},2}^\star$ and $k_{\rm{off},1}$ have been introduced in Corollary~\ref{cor_cmrna_cases}. The proof can be found in Appendix~\ref{app_pro_cases}.
\end{corollary}

\begin{corollary}
\label{cor_pro_min}

The minimum of $FF^P_s$ is obtained by taking the limit of $k_1=k_2=...=k_G=k$ and $k_{\rm{off}}=0$, and is given by
\begin{equation}
    \label{FFPmin}
    \lim_{\substack{k_1=...=k_G=k \\ k_{\rm{off}}= 0}}FF^P_{s}=1+\frac{\lambda_1}{2\lambda}+\frac{\lambda_1}{2\lambda}\frac{k_{\rm{on}}^2Gk^{2(G-1)}}{k_{\rm{on}}^2G^2k^{2(G-1)}}=1+\frac{\lambda_1}{2\lambda}+\frac{\lambda_1}{2\lambda G}.
\end{equation}
\end{corollary}

\noindent\textit{\textbf{Proof of Corollary~\ref{cor_pro_min}}.} It follows from Corollary~\ref{cor_pro_thre} that when varying $k_{\rm{off}}$, the minimum is achieved at $k_{\rm{off}}=0$ in cases 2 and 3. We then fix $k_{\rm{off}}=0$, and show that the Fano factor achieves the minimum at $k_1=k_2=...=k_G=k$. In that case, the Fano factor is given by
\begin{equation}
\begin{aligned}
    FF^P_{s}\left|_{k_{\rm{off}=0}}\right.&=1+\frac{\lambda_1}{\lambda}-\frac{\lambda_1}{\lambda}\frac{e_{G}(k_1,...,k_G)e_{G-2}(k_1,...,k_G)}{e_{G-1}(k_1,...,k_G)^2}\\
    &\geq 1+\frac{\lambda_1}{\lambda}-\frac{\lambda_1}{\lambda}\frac{G-1}{2G}\\
    &=1+\frac{\lambda_1}{2\lambda}+\frac{\lambda_1}{2\lambda G}.
\end{aligned}
\end{equation}
where Newton's inequality Eq.~(\ref{FFcmrna_Newton}) was used in the second row. From here, it follows that the minimum is when $k_1=k_2=...=k_G=k$.

In case 1, we do not have explicit proof that the minimum is smaller than the global minimum $1+\frac{\lambda_1}{2\lambda}+\frac{\lambda_1}{2\lambda G}$, therefore we checked this for many parameter values. We computed the Fano factor for two groups of parameters using the theoretical result. $G=3$ was fixed in group 1, $G=4$ was fixed in group 2, and each group contained $10000$ points. Other parameter values were selected from the interval $(0.1,100)$ with the constraint $k_1>k_1^\star$ and $k_{\rm{on}}>k_{\rm{on},2}^\star$ which forces case 1 (see Corollary~\ref{cor_pro_thre}). For every pair of $\lambda$ and $\lambda_1$ values, the Fano factor was found to be larger than Eq.~\eqref{FFPmin} thus suggesting that this expression provides the global minimum for all three cases.~$\square$

\section{Confirmation of analytic results using stochastic simulations}
\label{sims}

In this section, we confirm our main theoretical results using stochastic simulations with the SSA~\cite{gillespie1977exact}. In Fig.~\ref{fig2} and Fig.~\ref{fig3}, we confirm the results of queueing theory and in Fig.~\ref{fig:modred} we confirm the results obtained using model reduction. We next discuss these figures in detail. 

\subsection{Queueing theory} 

\begin{figure}[hp]
    \centering
    \includegraphics[width=1.0\textwidth]{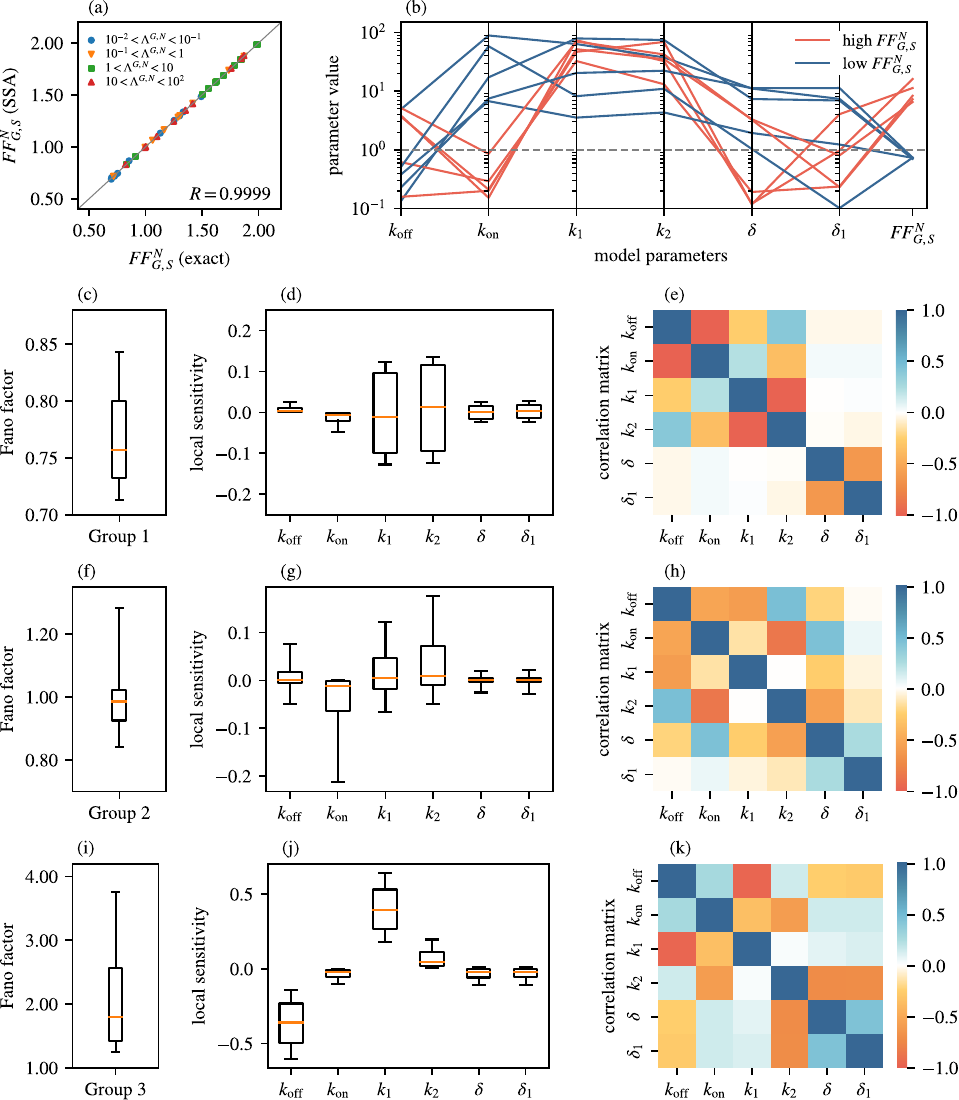}
    \caption{\label{fig2} \textbf{Testing the accuracy of the queueing theory’s predictions for the  
    Fano factor of the total nuclear mRNA and its sensitivity with respect to rate parameters}. (a) Comparison of the Fano factor obtained using results in Section~\ref{qtheory} and the stochastic simulation algorithm (SSA). Each point corresponds to one combination of parameters values. $\Lambda^{G,N}$ is the time separation parameter defined in Definition (\ref{def_timescales}). (b) Parallel coordinates plot showing how different values of model parameters lead to different values of the Fano factor $FF_{G,S}^{N}$. (c)-(k) Local sensitivity analysis for three groups of randomly chosen parameter sets, as explained in the main text: (c)-(e) is for Group $1$ ($FF_{G,S}^{N}<1$), (f)-(h) is for Group $2$ ($FF_{G,S}^{N}\approx 1$), and (i)-(k) is for Group $3$ ($FF_{G,S}^{N}>1$). (e), (h) and (k) are correlation matrices of the (local) logarithmic sensitivity defined in Eq.~(\ref{correlation-matrix-delta}). In all cases in this figure, we have fixed $G=2$ and $S=2$. See main text for parameter values.}
\end{figure}

In Fig.~\ref{fig2}(a), we compare the Fano factor of the total nuclear mRNA $FF_{G,S}^{N}$ computed from Eq.~(\ref{FF-N-G-S}) using Lemma~\ref{lemma_H-function} and Proposition~\ref{prop_B-1-time} to the one computed from stochastic simulations. Each point in Fig.~\ref{fig2}(a) corresponds to one set of model parameters. The number of gene states $G$ and the number of nuclear RNA states $S$ were fixed to $2$, whereas other parameters were selected randomly to achieve variation of the timescale separation parameter $\Lambda^{G,N}$ (see Definition~\ref{def_timescales}) over four orders of magnitude between $0.01$ and $100$. The agreement between theory and simulations shows that our result for the Fano factor is valid irrespective of the timescale separation parameter. 

In Fig.~\ref{fig2}(b), we use a parallel coordinates plot to explore how the Fano factor of the total nuclear mRNA for $G=2$ and $S=2$ depends on the model parameters $k_{\text{off}}$, $k_{\text{on}}$, $k_1$, $k_2$, $\delta$ and $\delta_1$. Each vertical line in Fig.~\ref{fig2}(b) represents one model parameter ($k_{\text{off}}, k_{\text{on}}, k_1, k_2, \delta, \delta_1$), except the last vertical line, which represents the Fano factor $FF_{G,S}^{N}$. Hence, one set of connected line segments across all vertical lines represents one parameter set, ending at the value of the Fano factor for that set. The line segments in the parallel coordinates plot suggest that high values of $FF_{G,S}^{N}$ are achieved when $k_{\text{off}}\gg k_{\text{on}}$, whereas low values of $FF_{G,S}^{N}$ are achieved when $k_{\text{off}}\ll k_{\text{on}}$. 

To explore this more systematically, we randomly selected $3000$ parameters that were grouped into three groups. Group 1 contained 1000 parameters sets in which the values of $k_{\text{off}}$ were selected from the interval $(0.1,10)$, whereas the values of $k_{\text{on}}, k_1$ and $k_2$ were selected from the interval $(10,100)$. Hence, in this group $k_{\text{off}}\ll k_{\text{on}},k_1,k_2$. The values of the Fano factor in this group were consistently below $1$, as shown in  Fig.~\ref{fig2}(c). Group 2 contained 1000 parameter sets with all parameters selected from the interval $(0.1,100)$, hence without \textit{a priori} restriction on the values of $k_{\text{on}}$ and $k_{\text{on}}$. Interestingly, the Fano factor in this group was sharply peaked around $1$, see Fig.~\ref{fig2}(f). Finally, Group 3 contained 1000 parameter sets in which the values of $k_{\text{off}}, k_1$ and $k_2$ were selected from the interval $(10,100)$, whereas the values of $k_{\text{on}}$ were selected from the interval $(0.1,1)$. Hence, in this group $k_{\text{off}}\gg k_{\text{on}}$, which yielded values of the Fano factor that were typically larger than $1$ (Fig.~\ref{fig2}(i)). In all three groups $\delta$ and $\delta_1$ were selected from the interval $(0.1,100)$, i.e. without further restrictions compared to other variables. These results corroborate our understanding from the parallel coordinates plot in Fig.~\ref{fig2}(b) that high values of $FF_{G,S}^{N}$ are achieved when $k_{\text{off}}\gg k_{\text{on}}$, whereas low values of $FF_{G,S}^{N}$ are achieved when $k_{\text{off}}\ll k_{\text{on}}$. 

Next, we wanted to understand how sensitive is the Fano factor $FF_{G,S}^{N}$ with respect to each of the model parameters. For each of the three aforementioned groups of parameters sets, we performed local sensitivity analysis by computing the (local) logarithmic sensitivity $\Delta(x)$~\cite{ingalls2008sensitivity} defined as,
\begin{equation}
    \label{Delta}
    \Delta(x)=\frac{x}{FF_{G,S}^N}\frac{\partial FF_{G,S}^N}{\partial x},
\end{equation}
where $x$ is any of the model parameters $k_{\text{off}}$, $k_{\text{on}}$, $k_1$, $k_2$, $\delta$ and $\delta_1$. The value of $\Delta(x)$ means that a change of $1$\% in $x$ causes a change of $\Delta(x)$\% in $FF_{G,S}^N$. Figs.~\ref{fig2} (d), (g) and (j) show box plots of $\Delta(x)$ for the three groups of parameters sets ($1$-$3$, respectively). On average, the values of $\Delta(x)$ are relatively small ($<0.1$) in all three groups, except for $\Delta(k_{\text{off}})$ and $\Delta(k_1)$ in Group 3, which are around $-0.36$ and $0.4$, respectively. 

Next, we explored how the values of $\Delta(x)$ correlate between each other for different choice of $x$. In Figs.~\ref{fig2} (e), (h) and (k), we plotted the correlation matrix
\begin{equation}
    \label{correlation-matrix-delta}
    \rho_{x,y}=\frac{\langle \Delta(x)\Delta(y)\rangle-\langle\Delta(x)\rangle\langle\Delta(y)}{\sqrt{\langle\Delta^2(x)\rangle-\langle\Delta(x)\rangle^2}\sqrt{\langle\Delta^2(y)\rangle-\langle\Delta(y)\rangle^2}}
\end{equation}
where $x$ and $y$ are any of the model parameters $k_{\text{off}}$, $k_{\text{on}}$, $k_1$, $k_2$, $\delta$ and $\delta_1$. In Group 1 ($FF_{G,S}^N<1$), there is a strong negative correlation between the pairs $\Delta(k_{\text{off}})$ and $\Delta(k_{\text{on}})$, $\Delta(k_1)$ and $\Delta(k_2)$, and $\Delta(\delta)$ and $\Delta(\delta_1)$ (Fig.~\ref{fig2}(e)). In contrast, there is almost no correlation between either $\Delta(\delta)$ or $\Delta(\delta_1)$ and the rest of $\Delta(x)$ for $x=k_{\text{off}}, k_{\text{on}}, k_1, k_2$. This means that the change in $FF_{G,S}^N$ due to a change in either $\delta$ or $\delta_1$ is practically independent of the change in $FF_{G,S}^N$ due to a change in any of the remaining parameters $k_{\text{off}}, k_{\text{on}}, k_1$ and $k_2$. Hence, the sets of parameters $\{\delta,\delta_1\}$ and $\{k_{\text{off}}, k_{\text{on}}, k_1, k_2\}$ can be considered as ``control knobs" for changing the value of the Fano factor, which may be of interest in synthetic biology. In the other two groups, the correlations in general increase and in some cases are even reversed. For example, $\Delta(k_{\text{off}})$ and $\Delta(k_{\text{on}})$ are negatively correlated in Group 1, but are positively correlated in Group 3. Similarly, $\Delta(\delta)$ and $\Delta(\delta_1)$ are negatively correlated in Group 1, but are positively correlated in Group 3. The least correlated pair of parameters in these two groups is $(k_1,k_2)$. 

\begin{figure}[hp]
    \centering
    \includegraphics[width=\textwidth]{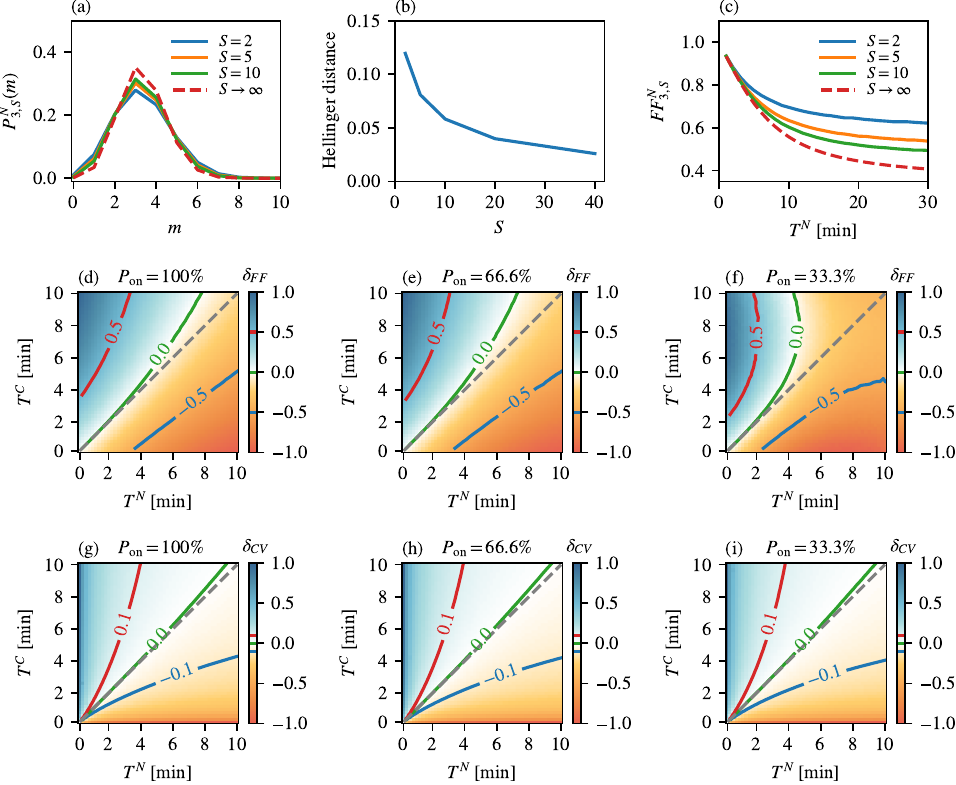}
    \caption{\label{fig3} \textbf{Testing the accuracy of the queueing theory's results for the distributions of nuclear mRNA and the difference between the Fano factors and coefficients of variation of nuclear and cytoplasmic mRNA, in the limit of large number of processing steps in the sub-cellular compartments.}
    (a) Probability distribution $P_{G,S}^{N}(m)$ obtained by SSA for various values of $S=2,5$ and $10$, compared to the probability distribution $P_{G,\infty}^{N}(m)$ computed in Eq.~(\ref{prob-nuclear-total-delay}) (shown by a red dashed line) for the delay model defined in Proposition~\ref{prop_nuclear-total-delay-prob}. (b) The Hellinger distance between the two distributions with respect to the number of states $S$. (c) Fano factor $FF_{G,S}^{N}$ as a function of the mean nuclear retention time $T^{N}$ computed by SSA for various values of $S=2,5$ and $10$, compared to the prediction of the delay model given by Eq. (\ref{FF-delay}) (shown by a red dashed line). Note that in (a)-(c) the gene is always on, i.e. $k_{\text{off}}=0$. (d)-(f) Heat plots of the relative difference $\delta_{FF}$ computed by SSA for $S=10$ and 2500 pairs of $T^N$ and $T^C$ (cytoplasmic retention time), where $\delta_{FF}$ is defined as $\Delta FF=FF_{G,S}^{N}-FF_{G,R}^{C}$ normalized by the maximum absolute difference $max\vert\Delta FF\vert$ for the whole dataset. In (d), $P_{\text{on}}=100\%$, in (e) $P_{\text{on}}=66\%$, whereas in (f) $P_{\text{on}}=33\%$, where $P_{\text{on}}=k_{\text{on}}/(k_{\text{on}}+k_{\text{off}})$ is the percentage of the time the gene is not in the off state. (g)-(h) The same as in (d)-(f) but for the relative difference $\delta_{CV}=\Delta CV/max\vert\Delta CV\vert$, where $\Delta CV=CV_{G,S}^{N}-CV_{G,R}^C$, and $max\vert \Delta CV\vert$ is the maximum absolute value of $\Delta CV$ for the whole dataset. In (d)-(i), the solid green line is a contour of constant difference value $0$, whereas the dashed gray line denotes $T^C=T^N$ for which the delay model predicts zero difference. See main text for parameter values.}
\end{figure}

In Fig.~\ref{fig3}, we explore how the delay model defined in Proposition~\ref{prop_nuclear-total-delay-prob} compares to the original model defined in reaction scheme (\ref{model_reactions}). We recall that the delay model is obtained by setting $k_{\text{off}}=0$, $k_1=\dots=k_G=k$ and $\delta=\delta_1$, and then taking the limit of $S,R,\delta,\delta_1,\lambda\rightarrow\infty$ such that $T^N=S/\delta$ and $T^C=R/\lambda$ are finite. This model describes an idealized scenario of constitutive gene expression in which nuclear retention and cytoplasmic degradation consist of many rate-limiting steps. In Fig.~\ref{fig3}(a), we compare the distribution of the total number of nuclear mRNA $P_{G,S}^{N}(m)$ obtained using stochastic simulations with the distribution $P_{G,\infty}^{N}(m)$ predicted by the delay model (Proposition~\ref{prop_nuclear-total-delay-prob}) for increasing values of $S$. In this example, we chose $G=3$ and $k=10$ min$^{-1}$ for which the mean time between successive mRNA production events was $G/k=0.3$ min$^{-1}$, whereas the mean nuclear retention times was set to $T^{N}=1$ minute. The Hellinger distance between the distributions (which varies between 0 and 1) is shown in Fig.~\ref{fig3}(b). As expected, the Hellinger distance is fairly large for $S=2$, and decays monotonically as $S$ is increased. 

In Fig.~\ref{fig3}(c), we inspect how the two models compare when the nuclear retention time $T^{N}$ is varied. We computed the Fano factor $FF_{G,S}^{N}$ from Eq.~(\ref{FF-N-G-S}) using Lemma~\ref{lemma_H-function} and Proposition~\ref{prop_B-1-time} for various values of $T^N$ and $S$, and compared it to the Fano factor $FF_{G,\infty}^{N}$ given by Eq.~(\ref{FF-delay}). We chose $G=3$ and $k=0.2$ min$^{-1}$, which yielded the mean time between successive mRNA production events of $15$ min. The results in Fig.~\ref{fig3}(c) confirm that $FF_{G,\infty}^{N}$ is monotonically decreasing with $T^{N}$, as stated in Proposition~\ref{prop_delay-nuclear-cytoplasmic-FF}, and we find that this monotonicity is preserved even for finite values of $S$. We further find that the agreement between the Fano factors is excellent for small values of $T^N$, even when $S$ is small. As $T^N$ is increased, the agreement gets progressively worse, but eventually saturates to a constant value for large values of $T_N$.

One of the predictions of the delay model, as stated in Corollary~\ref{cor_FF-N-C}, is that if $T^N < T^C$ then $FF_{G,\infty}^N>FF_{G,\infty}^C$ and \textit{vice versa}. In other words, which ever process has longer processing time, is prone to fewer fluctuations (relative to the mean). In Fig.~\ref{fig3}(d), we checked how well this statement holds for a finite value of $S=10$. We chose $k_{\text{off}}=0$, $G=3$ and $k=0.2$, and computed the Fano factors $FF_{G,S}^{N}$ and $FF_{G,R}^{C}$ using stochastic simulations for $2500$ pairs of $T^{N}$ and $T^C$ whose values were chosen equidistantly between $0.2$ and $10$ minutes. The figure shows the heatmap of $\delta_{FF}=\Delta FF/max\vert\Delta FF\vert$, where $\Delta FF$ is the difference $FF_{G,S}^{N}-FF_{G,R}^{C}$ and $max\vert\Delta FF\vert$ is the maximum absolute value of $\Delta FF$ for the whole dataset (2500 values). Hence, $\delta_{FF}$ is expected to be in the range between $-1$ and $1$, and the value of $0$ indicates that the values of Fano factors are equal. 
We find that the contour where $\delta_{FF}=0$ (green line) is reasonably close to the $y=x$ line predicted by the delay model (dashed gray line). Next, we asked how much this prediction of the delay model deviates when $k_{\text{off}}\neq 0$? We checked this for $k_{\text{off}}=0.02$ minutes and two values of $k_{\text{on}}$, $k_{\text{on}}=0.04$ for which the gene spends on average $P_{\text{on}}=66\%$ of the time in the on state (the on state is defined as any of the gene states $U_1,\dots,U_G$), and $k_{\text{on}}=0.01$ for which the gene spends on average $P_{\text{on}}=33\%$ of the time in the on state. Surprisingly, the relative difference $\delta_{FF}$ behaved very similarly for $P_{\text{on}}=66\%$ as it did for $P_{\text{on}}=100\%$, with only slight deviation from the prediction of the delay model (Fig.~\ref{fig3}(e)). On the other hand, for $P_{\text{on}}=33\%$ the difference was negative ($FF_{G,S}^{N}<FF_{G,R}^{C}$) in a much larger region of $T^{N}>T^{C}$ compared to the $P_{\text{on}}=100\%$ end $P_{\text{on}}=66\%$ cases, suggesting that the delay model is not an adequate approximation of the full model in the bursty regime (Fig.~\ref{fig3}(f)). In Figs.~{\ref{fig3}(g)-(i)}, we repeated this analysis using the same parameter sets, but for the relative difference in the coefficients of variation, $\delta_{CV}=\Delta CV/max\vert\Delta CV\vert$, where $\Delta CV=CV_{G,S}^{N}-CV_{G,R}^{C}$, and $max\vert\Delta CV\vert$ is the maximum absolute value of $\Delta CV$ for the whole dataset. In all three cases of $P_{\text{on}}=100\%$ (Fig.~\ref{fig3}(g)), $P_{\text{on}}=66.6\%$ (Fig.~\ref{fig3}(h)) and $P_{\text{on}}=33.3\%$ (Fig.~\ref{fig3}(i)), we find that the contour where $\delta_{CV}=0$ (green line) is very close to the $y=x$ line predicted by the delay model (dashed gray line). 

\subsection{Model reduction} 

\begin{figure}[hp]
    \centering
    \includegraphics[width=\textwidth]{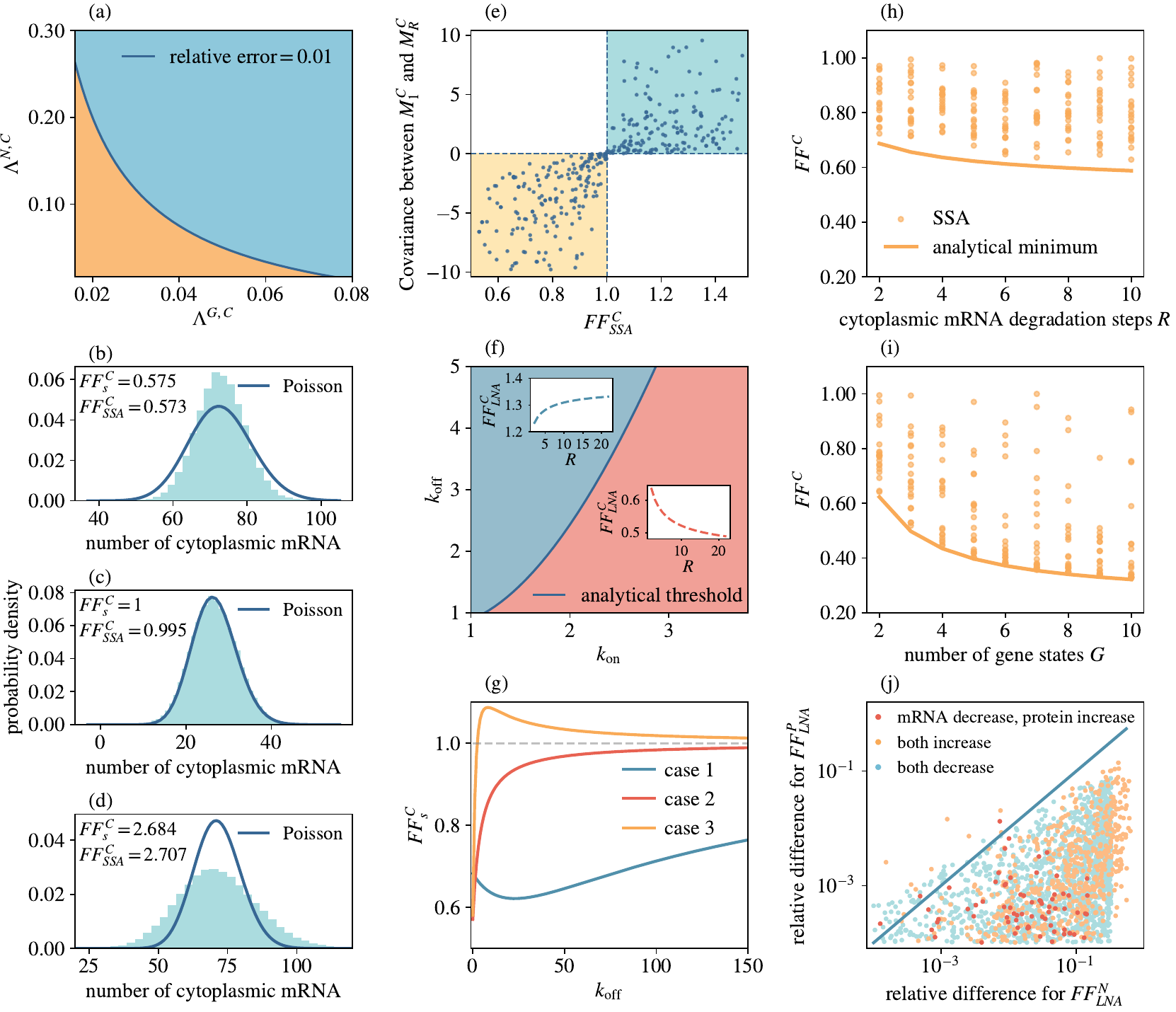}
    \caption{\label{fig:modred} \textbf{Simulations confirm the analytical results derived by model reduction.} (a) Relative error between the Fano factor of total cytoplasmic mRNA obtained using the ssLNA (Proposition~\ref{prop_cmrna_cov}) and the numerical results from LNA (exact). In the orange region, the relative error is smaller than $1\%$, while in the blue region it is greater than $1\%$. (b)-(d) Distributions of the total cytoplasmic mRNA (and the associated Fano factor) computed using the SSA for three different parameter sets. Corollary~\ref{cor_cmrna_thre} predicts the fluctuations to be sub-Poissonian (Fano factor $< 1$) for (b), Poissonian (Fano factor $=1$) for (c) and super-Poissonian (Fano factor $>1$) for (d). (e) Covariance between cytoplasmic mRNA species as a function of the Fano factor. Simulations using the SSA (each point is a parameter set) confirm the predictions of Corollary~\ref{cor_cmrna_cov_positive} which states that only the yellow and cyan regions should be inhabited. (f) Demarcation of parameter space according to whether the Fano factor of total cytoplasmic mRNA increases (blue region) or decreases (red region) with the number of processing steps $R$. These regions were constructed using the exact LNA. The solid blue line shows the analytical prediction for the case where the Fano factor does not change with $R$ (Corollary~\ref{cor_cmrna_FFmono}). Two insets show examples of how the Fano factor of total cytoplasmic mRNA varies with the number of processing steps $R$ in the blue and red regions, respectively. (g) Three different types of behaviour of the Fano factor of total cytoplasmic mRNA with respect to $k_{\rm{off}}$, as predicted by Corollary~\ref{cor_cmrna_cases}. (h)-(i) The minimum of the Fano factor of total cytoplasmic mRNA obtained in Corollary~\ref{cor_cmrna_limit} (solid orange line) bounds from below the Fano factors computed for several parameter sets using the LNA (points). (j) Absolute value of the relative difference between the Fano factors of total nuclear mRNA and protein at $S=2$ and $S=12$ steps of nuclear mRNA processing (computed using the exact LNA). The blue line $y=x$ is a visual aid, showing that the dependence of the Fano factor of protein noise on $S$ is comparable or weaker than that of total nuclear mRNA (Proposition~\ref{prop_pro_FF}). See main text for parameter values.}
\end{figure}

In Proposition~\ref{prop_cmrna_cov}, we obtained the Fano factor of total cytoplasmic mRNA using the ssLNA, which assumes that the timescale of this species is slower than the gene and nuclear mRNA timescales. To check the accuracy of this approximation, we compared the ssLNA results with a direct numerical solution of the equations describing the time-evolution of the first and second moments of the molecule numbers of the species in reaction system~\eqref{model_reactions}---we note that because the system is composed purely of first-order reactions, these equations when derived from the CME are exactly the same as the rate equations (for the mean) and the matrix Lyapunov equation (for the variance and covariances) given by the conventional LNA~\cite{elf2003fast}. In Fig.~\ref{fig:modred}(a), we show the relative error between the analytical (ssLNA) and the exact (numerical LNA) predictions for the Fano factor of total cytoplasmic mRNA. Here we scanned parameter space by fixing some of the parameters ($G=5, S=3, R=4, k_{\rm{off}}=0.8, k_1=...=k_5=300, \lambda=0.08$) and varying the rest ($1\leq k_{\rm{on}}\leq 4.5, 0.08\leq\delta=\delta_1\leq 5$). This confirms that the analytical Fano factor is highly accurate when the timescales are well separated, specifically when $\Lambda^{G,C}$ and $\Lambda^{N,C}$ (see Definition~\ref{def_timescales}) are sufficiently small.

In Corollary~\ref{cor_cmrna_thre}, we used the ssLNA to predict the value of the deactivation rate at which the Fano factor crosses the threshold value $1$, i.e. when fluctuations change from sub-Poissonian to super-Poissonian. To validate this prediction, we used the SSA to simulate the reaction system~\eqref{model_reactions} and to calculate the probability distributions of total cytoplasmic mRNA, below, at and above the theoretical threshold given by Eq.~\eqref{thre1}. The results are shown in Fig.~\ref{fig:modred}(b)-(d). Note that here we fixed the parameters $G=4, S=1, R=2, \delta_1=15, \lambda=0.3$, and varied the values of $k_{\rm{off}}$, $k_{\rm{on}}$, and $k_1=k_2=k_3=k_4=k$ (specifically $k_{\rm{off}}=3, k_{\rm{on}}=15, k=46$ for Fig.~\ref{fig:modred}(b); $k_{\rm{off}}=60, k_{\rm{on}}=10, k=40$ for Fig.~\ref{fig:modred}(c); 
$k_{\rm{off}}=50, k_{\rm{on}}=5, k=150$ for Fig.~\ref{fig:modred}(d)). In each case we compare the distribution obtained from simulations with a Poisson distribution having the same mean, thus clearly showing that the distributions are narrower than Poisson, same as Poisson and wider than Poisson, respectively, in accordance with the theoretical threshold. 

In Corollary~\ref{cor_cmrna_cov_positive}, we showed that the covariance between any two cytoplasmic mRNA species is positive when the Fano factor of total cytoplasmic mRNA is greater than $1$ and negative otherwise. This is confirmed using stochastic simulations in Fig.~\ref{fig:modred}(e). Parameters were generated randomly and uniformly on the log scale 
within two regions (1) $G=3, S=1, R=2, 1\leq k_{\rm{off}}\leq 60,1\leq k_{\rm{on}}\leq 20,10\leq k_1=k_2=k_3\leq 100,10\leq\delta_1\leq 20, 0.13\leq\lambda
\leq 0.27$ and (2) $G=4, S=1, R=3, 1\leq k_{\rm{off}}\leq 60, 1\leq k_{\rm{on}}\leq 20,10\leq k_1=k_2=k_3=k_4\leq 100,10\leq \delta_1\leq 20, 0.08\leq\lambda\leq 0.1$. 

In Corollary~\ref{cor_cmrna_FFmono}, we showed that the Fano factor of total cytoplasmic mRNA increases with the number of processing steps $R$ in the cytoplasm when the Fano factor is greater than $1$, and decreases when the Fano factor is below $1$. We test this analytical result in Fig~\ref{fig:modred}(f). The dark blue
line represents the analytical threshold when the Fano factor equals $1$. The exact Fano factor is then computed for all points in parameter space using the LNA such that the regions are coloured red when the Fano factor decreases with $R$ and blue when the opposite occurs. Note that the analytical threshold separates the two regions, thus verifying our result. Parameters were selected uniformly in the parameter region: $G=4, S=3, 1\leq k_{\rm{off}}\leq 5,1\leq k_{\rm{on}}\leq 3.8,k_1=k_2=k_3=k_4=3k_{\rm{on}}+10, \delta=\delta_1=15, \lambda_0=0.1$. In the insets of Fig~\ref{fig:modred}(f), we show specific examples of how the Fano factors vary with $R$ in the two regions.

In Corollary~\ref{cor_cmrna_cases}, we showed that the analytical Fano factor of total cytoplasmic mRNA from ssLNA has three different behaviours with respect to the deactivation rate $k_{\rm{off}}$. These three behaviours are shown in Fig.~\ref{fig:modred}(g). Parameters in these examples are $G=3, R=3, k_{\rm{on}}=8, k_1=5, k_2=4.5$ and $k_3=1$ for case 1, $G=3, R=3, k_{\rm{on}}=2, k_1=3, k_2=5$, and $k_3=2$ for case 2, and $G=3, R=3, k_{\rm{on}}=0.9, k_1=4, k_2=2$, and $k_3=5$ for case 3. 

In Corollary~\ref{cor_cmrna_limit}, we derived a lower bound of the Fano factor of total cytoplasmic mRNA, in particular showing that it depends only on the number of gene states $G$ and the number of cytoplasmic processing reactions $R$. To assess the accuracy of this theoretical result, we computed the Fano factor using the LNA for a number of parameter sets. In Fig.~\ref{fig:modred}(h)-(i) we show that the Fano factors from simulations are always larger than the theoretical minimum (shown by a solid orange line). The parameters for this study were sampled uniformly on the log scale across the parameter region: $1\leq k_{\rm{off}}\leq 31.6, 2\leq k_{\rm{on}}\leq 31.6, 10\leq k_1,k_2=...=k_G\leq 100, 10\leq\delta,\delta_1 \leq 31.6, 0.04\leq\lambda\leq 0.1$, and $S$ was fixed to $3$. Furthermore, $G$ was fixed to $2$ in Fig.~\ref{fig:modred}(h) and $R$ was fixed to 5 in Fig.~\ref{fig:modred}(i). Note that here we use $FF^C$ as the notation for Fano factor of total cytoplasmic mRNA on the y-axis, as both $FF^C_s$ (the solid line) and $FF^C_{LNA}$ (points) are shown in the figure.

Finally, we use the LNA to study how the Fano factor of proteins and total nuclear mRNA varies with the number of nuclear mRNA processing steps $S$. The results are shown in Fig.~\ref{fig:modred}(j) where we plot the relative difference in the Fano factors for $S=2$ and $S=12$ steps for $2000$ parameter sets. We find three types of monotonic behaviours: (1) the Fano factors of both total nuclear mRNA and protein increase with $S$ (orange points), (2) the Fano factor of total nuclear mRNA decreases while that of protein increases with $S$ (red points) and (3) the Fano factors of both total nuclear mRNA and protein decrease with $S$ (blue points). Note that most points fall below the line $y=x$, indicating that the relative differences in Fano factors for proteins are typically smaller than those for total nuclear mRNA, hence proteins are not significantly impacted by the number of nuclear mRNA processing steps. This is in agreement with the prediction of Proposition~\ref{prop_pro_FF}. Note that $\delta$ was scaled to $\delta=(S-1)\delta_0$ in the simulations, where $S$ is the number of nuclear processing states and $\delta_0$ is the scaling factor, to maintain the same mean molecule numbers as the number of nuclear mRNA processing states increases.
The parameters in this case were carefully chosen to reflect two important natural constraints: (1) the mean protein molecule numbers are larger than the mean total mRNA molecule numbers; (2) proteins exhibit longer half-lives than mRNA~\cite{schwanhausser2011global}. To fulfil these criteria, parameter values were sampled uniformly on the log scale across the region $(0.1,100)$, while $\lambda_2$ was constrained on the interval $(0.001,1.02)$.
 
\section{Summary and Conclusion}
\label{Conc}

In this paper, we have studied a complex multi-stage, two-compartment model of stochastic gene expression using two distinct mathematical tools, queueing theory and model reduction. This allowed us to analytically probe the statistics of nuclear mRNA, cytoplasmic mRNA and protein counts in steady-state conditions, which we then verified using stochastic simulations.  

While multi-stage models of the mRNA lifecycle are not very common, they have been previously constructed and studied~\cite{pedraza2008effects,filatova2022modulation,gorin2022modeling,szavits2022mean,shi2023stochastic,hansen2018cytoplasmic}. A speciality of these models is that since they describe the birth or death of mRNA or proteins via several reaction steps, they explicitly account for molecular memory between individual events, i.e. the time between successive birth/death reactions is random but not sampled from the exponential (memory-less) distribution. However, these models describe exclusively super-Poissonian fluctuations which are characteristic of bursty transcription~\cite{dar2012transcriptional} and therefore cannot describe sub-Poissonian fluctuations that have been measured for some genes~\cite{weidemann2023minimal,sun2020size,muthukrishnan2012dynamics,lionnet2010nuclear}. A multi-stage model was constructed in Ref.~\cite{weidemann2023minimal} to explain sub-Poisson mRNA fluctuations in some genes, but it cannot explain super-Poisson fluctuations in other genes. The distinction of our model from these other multi-stage models in the literature is that it is the first one which can describe both sub-Poissonian and super-Poissonian mRNA fluctuations and therefore can be seen as a generalization of existing models that can explain the gamut of available gene expression data. 

We note that while to date most studies have found super-Poissonian noise, this is in part because often these do not correct for extrinsic noise due to the coupling of the transcription rate to cell volume which artificially increases the Fano factor~\cite{padovan2015single,weidemann2023minimal,jia2023coupling,foreman2020mammalian}; for single-cell sequencing data, this is further exacerbated by the large amount of technical noise, particularly that due to the cell-to-cell variation in capture efficiency (the probability of any individual mRNA molecule being sampled)~\cite{klein2015droplet}. We expect that as methods to correct for these factors become more widely used, a significant fraction of gene expression data with apparent Fano factors a bit larger than 1 will be reinterpreted as being due to sub-Poissonian noise, hence the development of models that can be fitted to this type of data will increasingly become crucial to obtain a more refined understanding of gene expression. 

Our model reduction theory clearly shows that the transition from sub- to super-Poisson mRNA behaviour occurs as the deactivation rate increases beyond a certain threshold. Interestingly, this implies that the vast majority of previous models (which can only predict super-Poisson fluctuations) are in reality only correct for large enough deactivation rates. This threshold varies with the number of rate-limiting steps in transcriptional initiation and the speed of this process, as well as with the magnitude of the activation rate. Curiously, while the Fano factor of mRNA in a compartment increases with the number of processing steps in that compartment when mRNA fluctuations are super-Poissonian, the reverse occurs when fluctuations are sub-Poissonian; this explains the seemingly contradictory observations in Refs.~\cite{weidemann2023minimal} and~\cite{hansen2018cytoplasmic}. We also showed that the lower bound on the Fano factor of mRNA fluctuations is achieved when the gene is always on and the rate of moving from one transcriptional initiation stage $U_i$ to the next $U_{i+1}$ is independent of $i$. This case, of course, is only rarely met because the rates of RNAP binding, opening the DNA double helix and of RNAP leaving the proximal-promoter paused state are not generally similar. While the lower bound was previously computed numerically~\cite{weidemann2023minimal} here we go further by providing simple expressions that clarify the explicit dependence of the minimum on the number of rate-limiting steps in initiation ($G$) and the number of processing steps in a compartment ($S$ or $R$, depending on if it is the nucleus or the cytoplasm, respectively). In contrast to what we found for mRNA fluctuations, the lower bound for the Fano factor of protein fluctuations is greater than one, implying super-Poissonian fluctuations, even when the fluctuations of the mRNA from which it is translated, are sub-Poissonian. In addition, we found that the Fano factor of proteins is not strongly modulated by the number of mRNA processing steps and that it is smaller than that predicted by the standard three-stage model of gene expression~\cite{shahrezaei2008analytical}.

The aforementioned results were all derived using reduction of the stochastic model and therefore are strictly only valid in the limit that timescales of protein number fluctuations are longer than those of mRNA number fluctuations, and the latter longer than those of gene state fluctuations. Since in mammalian cells, gene timescales are typically quite short, of the order of seconds to few minutes~\cite{lammers2020matter}, nuclear and cytoplasmic retention times for mRNA vary from minutes to many hours~\cite{bartman2019transcriptional,schwanhausser2011global}, while protein degradation times are often longer than the cell-cycle duration which is many hours long~\cite{schwanhausser2011global}, it follows that the timescale separation ansatz that we assumed is valid in many cases of practical interest. Nevertheless, to develop a more general theory, we employed queueing theory, which enabled the derivation of a number of exact and approximate results for mRNA statistics. In particular, we obtained an exact (though complex) expression for the Fano factor of total nuclear mRNA fluctuations whose numerical computation is efficient compared to its estimation using stochastic simulations, since the ensemble averaging step is bypassed. By the use of this formula, we performed an extensive parameter scan that calculated the (local) logarithmic sensitivity of the nuclear mRNA Fano factor to variation in the rate parameter values. The theory also allowed us to compute in closed-form approximate formulae for the sub-Poissonian distributions of total nuclear and total cytoplasmic mRNA that are accurate in the limit of small deactivation rates and quasi-deterministic nuclear and cytoplasmic retention times (which naturally follow when the processing of transcripts in the nucleus or cytoplasm occurs in many steps). These formulae maybe useful for maximum likelihood or Bayesian estimation of rate parameters from experimental data. 

We also showed that under the same conditions that we assumed to derive the mRNA count distributions, the Fano factor of nuclear mRNA is larger (smaller) than that of cytoplasmic mRNA, if the nuclear retention is smaller (larger) than the cytoplasmic retention time (the time for a transcript to degrade in the cytoplasm)---the same result holds for the coefficient of variation of mRNA fluctuations. 
Using stochastic simulations, we showed that this prediction was approximately true even if the number of processing steps is not very large and if the deactivation and activation rates are comparable. Unsurprisingly, the theory partially breaks down when the gene spends most of its time in the off state, i.e. the deactivation rate is much larger than the activation rate. In this case, simulations show that the cytoplasmic Fano factor is greater than the nuclear one, only when the nuclear retention time  is larger than the cytoplasmic retention time, in agreement with the theory. But they also show that the opposite case of larger Fano factor in the nucleus can be obtained both when the nuclear retention time is larger than the cytoplasmic one and vice versa, which disagrees with the theory. Experiments measure cases where the Fano factor is larger or smaller in the nucleus compared to the cytoplasm~\cite{hansen2018cytoplasmic,weidemann2023minimal}, likely indicating  that the ratio of the retention times in the nuclear and cytoplasmic compartments varies considerably in living cells. 

Concluding, here we have constructed and analysed a stochastic model of gene expression that encompasses and extends existing models to provide a nuanced quantitative description of gene expression that aligns with various experimental results. Our dualistic approach, using two distinctly different analytical tools, shows that analytical insight into complex biochemical models with large numbers of molecular species is not impossible, and while the calculations are laborious, the resulting final expressions offer invaluable insight that is difficult to obtain otherwise.        
 
\section*{Acknowledgments}
M. M. acknowledges support from a scholarship provided by the Darwin Trust. J. S-N and R. G. acknowledge support from the Leverhulme Trust (RPG-2020-327). 

\section*{Appendix}

\begin{appendices}

\setcounter{equation}{0}
\renewcommand\theequation{A.\arabic{equation}} 

\addcontentsline{toc}{section}{Appendices}
\renewcommand{\thesubsection}{\Alph{subsection}} 

\subsection{The slow-scale linear noise approximation (ssLNA)}\label{app_ssLNA}

Here we provide a brief, self-contained summary of the ssLNA; for details the reader is referred to the original publication~\cite{thomas2012slow}. We shall specifically focus on a general chemical system composed purely of reactions with first-order kinetics, since the system with reaction scheme~\eqref{model_reactions} is of this type. 

Suppose we have a reaction system composed of a set of chemical species $X_i, i=1,2,...,M$ interacting via a set of chemical reactions $j=1,2,...,N$, and that the $j$th reaction has the form
\begin{equation}\label{app_ssLNA_reaction}
s_{1j}X_1+...+s_{Mj}X_M\xrightarrow[]{k_j} r_{1j}X_1+...+r_{Mj}X_M,
\end{equation}
where $s_{ij}$ and $r_{ij}$ are stoichiometric  coefficients and $k_j$ is the rate of reaction $j$ with units of inverse time. Note that since each reaction is first-order, it follows that for any $j$, the value of $s_{ij}$ can only be 1 for one particular value of $i$ and is zero for all other values of $i$. 

One can then construct the stoichiometric matrix $\bm{S}$ with elements $S_{ij}=r_{ij}-s_{ij}$ and the rate vector $\vec{f}$ with elements $f_{j}=k_j\prod_{i=1}^M [X_i]^{s_{ij}}$, where $[X_i]$ denotes the mean number of molecules of chemical species $X_i$. Note that the latter follows directly from the law of mass action. The deterministic rate equations of the reaction system Eq.~(\ref{app_ssLNA_reaction}) are then given by
\begin{equation}
    \frac{d[\vec{X}]}{dt}=\mathbf{S}\vec{f},
\end{equation}
and the elements in the associated Jacobian matrix can be written as $\mathbf{J}_{ij}=\partial_j(\mathbf{S}\vec{f})_i,$ where $\partial_j$ denotes the partial derivative with respect to $[X_j]$. 

It then follows from the linear noise approximation (LNA)~\cite{van2007stochastic,elf2003fast} that the covariance matrix $\mathbf{C}$ of molecule numbers of each molecular species can be obtained by solving the Lyapunov equation
\begin{equation}
    \mathbf{J} \mathbf{C}+\mathbf{C}\mathbf{J^T}+\mathbf{D}=\mathbf{0},
\end{equation}
where $\mathbf{D}$ is the diffusion matrix given by
\begin{equation}
    \mathbf{D}=\mathbf{S}\mathbf{F}\mathbf{S^T},
\end{equation}
and $\mathbf{F}$ is a diagonal matrix whose non-zero diagonal elements are given by the rate vector $\vec{f}$. Note that the volume of the system does not appear in any of these equations because we are specifically considering a system of first-order reactions (in the traditional formulation where other types of reactions are allowed, the volume explicitly appears). 

Under timescale separation conditions~\cite{thomas2012slow,thomas2012rigorous}, it can be shown that the Lyapunov equation above simplifies to a different Lyapunov equation which is only in terms of the covariance of the molecule numbers of the slow species, $\mathbf{C}_s$. This is given by:
\begin{equation}
\mathbf{J}_s\mathbf{C}_s+\mathbf{C}_s\mathbf{J}_s^T+\mathbf{D}_s=\mathbf{0}.    
\end{equation}
This is the ssLNA. In what follows, we explain how to obtain the matrices of constants in this equation.

The reduced Jacobian matrix is defined as $\mathbf{J}_s=\mathbf{J}_{ss}-\mathbf{J}_{sf}\mathbf{J}_{ff}^{-1}\mathbf{S}_{ff}\sqrt{\mathbf{F}}$. The reduced diffusion matrix $\mathbf{D}_s$ is defined as
\begin{equation}\label{ssLNA_diff}
    \mathbf{D}_s=(\mathbf{W}-\mathbf{V})(\mathbf{W}-\mathbf{V})^T,
\end{equation}
where $\mathbf{W}=\mathbf{S}_s\sqrt{\mathbf{F}}$ and $\mathbf{V}=\mathbf{J}_{sf}\mathbf{J}_{ff}^{-1}\mathbf{S}_f\sqrt{\mathbf{F}}$. Note that the Jacobian and stoichiometric matrix have the partitions
\begin{equation}
\mathbf{J}=
\left[
\begin{array}{c|c}
 \mathbf{J}_{ss}&\mathbf{J}_{sf} \\\hline
 \mathbf{J}_{fs}&\mathbf{J}_{ff} 
\end{array} 
\right],\;
\mathbf{S}=
\left[\frac{\mathbf{S}_s}{\mathbf{S}_f}\right].
\end{equation}
These partitions of the matrices follow the partitioning of the species into slow and fast. $\mathbf{S}_s$ is the stoichiometric matrix of the slow species only; $\mathbf{S}_f$ is the stoichiometric matrix of the fast species only; $\mathbf{J}_{ss}$ is the Jacobian of the slow species with respect to the slow species, i.e. the derivative of the RHS of the rate equations of the means of the slow species with respect to the means of the slow species; $\mathbf{J}_{sf}$ is the Jacobian of the slow species with respect to the fast species, i.e. the derivative of the RHS of the rate equations of the means of the slow species with respect to the means of the fast species; $\mathbf{J}_{ff}$ is the Jacobian of the fast species with respect to the fast species, i.e. the derivative of the RHS of the rate equations of the means of the fast species with respect to the means of the fast species; $\mathbf{J}_{fs}$ is the Jacobian of the fast species with respect to the slow species, i.e. the derivative of the RHS of the rate equations of the means of the fast species with respect to the means of the slow species.

\subsection{Solution of Lyapunov equations for cytoplasmic mRNA fluctuations}\label{app_cmrna_FF}

Here, we will utilize the ssLNA notation developed in Appendix~\ref{app_ssLNA} to derive the Lyapunov equation~\eqref{cmrna_Lyapunov}. 

To construct the rate function matrix $\mathbf{F}$, we need to specify the order in which we consider the reactions in~\eqref{model_reactions}. We shall choose the reactions in the following order (left to right and top to bottom): 
\begin{equation}
\label{model_reactions_ord}
\begin{aligned}
    &U_{0} \xrightarrow{k_{\rm{on}}} U_{1},\\
    &U_{1}\xrightarrow{k_1} U_2 \xrightarrow{k_2} ...\xrightarrow{k_{G-1}} U_G \xrightarrow{k_{G}} U_{1}+M^{N}_1,\\
    &U_{1} \xrightarrow{k_{\rm{off}}} U_{0},\\
    &M^N_1\xrightarrow{\delta}M^N_2\xrightarrow{\delta}...\xrightarrow{\delta}M^{N}_{S-1}\xrightarrow{\delta}M^N_{S},\\
    &M^N_S\xrightarrow{\delta_1}M^C_1,\\
    &M^C_1\xrightarrow{\lambda}M^C_2\xrightarrow{\lambda}...\xrightarrow{\lambda}M^C_{R-1}\xrightarrow{\lambda}M^C_R\xrightarrow{\lambda}\varnothing,\\
\end{aligned}
\end{equation}
Note that here we have excluded the species $M_0^N$ and $P$ since they do not influence the steady-state fluctuations of the mRNA and gene species. Given this reaction ordering, it then follows from the law of mass action that the rate function matrix $\mathbf{F}$ is given by
\begin{equation}\label{cmrna_rateF}
\begin{aligned}
\mathbf{F}=&\text{diag}(\vec{f})\\
=&(k_{\rm{on}}[U_0],k_1[U_1],...,k_G[U_G],k_{\rm{off}}[U_1],\delta [M^N_1],...,\delta_1 [M^N_S],\lambda [M^C_1],...,\lambda [M^C_R]).     
\end{aligned}
\end{equation}

We next formulate the stoichiometric matrix $\mathbf{S}$. We order the species as follows: the slow species $M_1^C, M_2^C, \ldots, M_R^C$, followed by the fast species $U_0, U_1, \ldots, U_G, M_1^N, \ldots, M_S^N$. Hence, the upper $R$ rows of the stoichiometric matrix represent the slow species and the rest the fast species. Note that each column corresponds to a reaction defined in~(\ref{model_reactions_ord}), maintaining the same order as we previously used for Eq.~(\ref{cmrna_rateF}). Given the chosen order of reactions and species, the stoichiometric matrix is given by

\begin{equation}
\begin{aligned}
&\mathbf{S}_{(G+S+R)\times(G+S+R+2)}=
\left(\frac{\mathbf{S}_s}{\mathbf{S}_f}\right)=\\&
\left(
\begin{array}{ccccccccccccccccccccccc}
    \multicolumn{8}{c}{\multirow{5}[2]{*}{\huge $0$}}\\
     & & & & & & & &0 &0 &0 &... &1 &-1 &0 &... &0 &0 &0\\
     & & & & & & & &0 &0 &0 &... &0 &1 &-1 &... &0 &0 &0\\
     & & & & & & & &\vdots & & & & &\vdots & & & & & \\
     & & & & & & & &0 &0 &0 &... &0 &0 &0 &... &0 &1 &-1\\
     \hline
    -1 &0 &0 &...&0 &0 &0 &1 &0 &0 &0 &... &0 &%
    \multicolumn{6}{c}{\multirow{9}[2]{*}{\huge $0$}}\\
    1 &-1 &0 &...&0 &0 &1 &-1 &0 &0 &0 &... &0 & & & & & &\\
    0 & 1 & -1 &.. &0 &0 &0 &0 &0 &0 &0 &... &0 & & & & & &\\
    \vdots & & & & & & & & & & & & & & & & & &\\
    0 &0 &0 &...&1 &-1 &0 &0 &0 &0 &0 &... &0 & & & & & &\\
    0 &0 &0 &...&0 &0 &1 &0 &-1 &0 &0 &... &0 & & & & & &\\
    0 &0 &0 &...&0 &0 &0 &0 &1 &-1 &0 &... &0 & & & & & &\\
    \vdots & & & & & & & & & & & & & & & &\\
    0 &0 &0 &...&0 &0 &0 &0 &0 &0 &... &1 &-1 & & & & & &\\ 
\end{array}
\right).  
\end{aligned}
\end{equation}

We can then obtain the partitioned Jacobian matrix
\begin{equation}
\begin{aligned}
&\begin{aligned}[c]
\mathbf{J}_{ss}=\left(
\begin{array}{cccccccc}
-\lambda &0 &0 &0 &0 &0 &...\\
\lambda &-\lambda &0 &0 &0 &0 &...\\
0 &\lambda &-\lambda &0 &0 &0 &...\\
... & & & & & &\\
0 &0 &0 &0 &... &\lambda &-\lambda \\
\end{array}\right),  
\end{aligned}
\quad
\begin{aligned}[c]
\mathbf{J}_{sf}=\left(
\begin{array}{ccccc}
0 &0 &0 &... &\delta_1\\ 
0 &0 &0 &... &0\\ 
0 &0 &0 &... &0\\ 
... & & & &\\ 
0 &0 &0 &... &0\\ 
\end{array}\right),
\end{aligned}\quad
\begin{aligned}
\mathbf{J}_{fs}=\underline{\boldsymbol{0}},
\end{aligned}\\
&\begin{aligned}[c]
\mathbf{J}_{ff}=\left(
\begin{array}{cccccccccccc}
-k_{\rm{on}} &k_{\rm{off}} &0 &... &0 &0 &0 &0 &0 &0 &...\\
k_{\rm{on}}-k_G &-k_{\rm{off}}-k_1-k_G &-k_G &... &-k_G &-k_G &0 &0 &0 &0 &...\\
0 &k_1 &-k_2 &... &0 &0 &0 &0 &0 &0 &...\\
...& & & & & & & & & &\\
0 &0 &0 &... &k_{G-2} &-k_{G-1} &0 &0 &0 &0 &...\\
-k_G &-k_G &-k_G &... &-k_G &-k_G &-\delta &0 &0 &0 &...\\
0 &0 &0 &... &0 &0 &\delta &-\delta &0 &0 &...\\
...& & & & & & & & & &\\
0 &0 &0 &0 &0 &0 &0 &0 &... &\delta &-\delta_1\\
\end{array}
\right).
\end{aligned}
\end{aligned}
\end{equation}
Given the partitioned Jacobian matrix above, we can obtain the reduced Jacobian matrix $\mathbf{J}_s^C$ in Eq.~(\ref{cmrna_Lyapunov}) as
\begin{equation}\label{app_cmrna_Jacob}
    \mathbf{J}^C_{s}=\mathbf{J}_{ss}-\mathbf{J}_{sf}\mathbf{J}_{ff}^{-1}\mathbf{J}_{fs}=\mathbf{J}_{ss},
\end{equation}
which trivially follows from $\mathbf{J}_{fs}=0$.

Now we can calculate the $\mathbf{W}$ and $\mathbf{V}$ matrices in Eq.~(\ref{ssLNA_diff}):
\begin{equation}
\begin{aligned}
&\mathbf{W}_{R\times(G+S+R+2)}=\mathbf{S}_s \sqrt{\mathbf{F}}\\
&=\left(
    \begin{array}{ccccccccccccccc}
    0 &0 &... &\sqrt{\delta_1 [M^N_S]} &-\sqrt{\lambda [M^C_1]} &0 &... &0 &0\\
    0 &0 &... &0 &\sqrt{\lambda [M^C_1]} &-\sqrt{\lambda [M^C_2]} &... &0 &0\\
    ... & & & & & & & &\\
    0 &0 &... &0 &0 &0 &... &\sqrt{\lambda [M^C_{R-1}]} &-\sqrt{\lambda [M^C_R]} \\
    \end{array}
    \right),    
\end{aligned}
\end{equation}

\begin{equation}
\mathbf{V}_{R\times(G+S+R+2)}=\mathbf{J}_{sf}\mathbf{J}_{ff}^{-1}\mathbf{S}_f\sqrt{\mathbf{F}}=\left(
\begin{array}{ccc}
\vec{b}_1 & \vec{b}_2\\
\multicolumn{2}{c}{\multirow{2}[2]{*}{\Large $0$}}\\
 & & \\
 & & \\
\end{array}
\right),
\end{equation}
where
\begin{equation}
\begin{aligned}
&\vec{b}_1=\left(
    \begin{array}{cccccccccccc}
    (a_1-a_0)\sqrt{k_{\rm{on}}[U_0]} &... &(a_{G-1}-a_{G-2})\sqrt{k_{G-2}[U_{G-2}]} &-a_{G-1}\sqrt{k_{G-1}[U_{G-1}]}\\
    \end{array}
    \right),    
\end{aligned}
\end{equation}

\begin{equation}
\begin{aligned}
&\vec{b}_{2}=\left(
    \begin{array}{cccccccc}
    &(a_1-1)\sqrt{k_G[U_G]}&(a_0-a_1)\sqrt{k_{\rm{off}}[U_1]}&0 &... &\sqrt{\delta_1 [M^N_S]} &0 &...
    \end{array}
    \right).  
\end{aligned}
\end{equation}
The variables $a_g$ above ($g=1,2,...,G-1$) are defined as
\begin{equation}\label{cmrna_recursive_a}
\begin{aligned}
     &a_0=\frac{k_Gk_{\rm{on}} e_{G-2}(k_{1},...,k_{G-1})+e_{G}(k_1,...,k_{G})+k_{\rm{off}}e_{G-1}(k_2,...,k_G)}{k_{\rm{on}} e_{G-1}(k_1,...,k_{G})+k_{\rm{off}}e_{G-1}(k_2,...,k_G)},\\ 
     &a_1=\frac{k_G k_{\rm{on}} e_{G-2}(k_{1},...,k_{G-1})+k_{\rm{off}}e_{G-1}(k_2,...,k_G)}{k_{\rm{on}} e_{G-1}(k_1,...,k_{G})+k_{\rm{off}}e_{G-1}(k_2,...,k_G)},\\
     &a_2=a_1-\frac{k_G k_{\rm{on}} e_1^{\text{max}}+k_{\rm{off}}e_{G-1}(k_2,...,k_G)}{k_{\rm{on}} e_{G-1}(k_1,...,k_{G})+k_{\rm{off}}e_{G-1}(k_2,...,k_G)},\\  &a_g=a_{g-1}-\frac{k_G k_{\rm{on}}e_{g-1}^{\text{max}}}{k_{\rm{on}} e_{G-1}(k_1,...,k_{G})+k_{\rm{off}}e_{G-1}(k_2,...,k_G)},\ g=3,4,...,G-1.
\end{aligned}
\end{equation}
Note that in the above equations we have introduced a new variable $e_g^{\text{max}}$. Based on Definition~\ref{sym_poly}, we have
\begin{equation}
\begin{aligned}
e_{G-2}(k_{1},...,k_{G-1})&=\sum_{1\leq j_1<j_2...<j_{G-2}\leq G-1}k_{j_1}\cdot\cdot\cdot k_{j_{G-2}},
\end{aligned}
\end{equation}
and $e_g^{\text{max}}$ in Eq.~(\ref{cmrna_recursive_a}) denotes the element $k_{j_1}\cdot\cdot\cdot k_{j_{G-2}}$ with the $g^{\text{th}}$ maximal sum of indices in $e_{G-2}(k_{1},...,k_{G-1})$. An example of this notation is as follows. If we consider the case $G=4$, we have
\begin{equation}
e_{2}(k_{1},...,k_3) = k_1k_2+k_1k_3+k_2k_3,
\end{equation}
and the element with the $g^{\text{th}}$ ($g=1,2,3$) maximal sum of indices are
\begin{equation}
\begin{aligned}
e_1^{\text{max}}=k_2k_3,\quad e_2^{\text{max}}=k_1k_3,\quad e_3^{\text{max}}=k_1k_2.
\end{aligned}
\end{equation}

Given matrices $\mathbf{W}$ and $\mathbf{V}$, we can now calculate the effective diffusion matrix $\mathbf{D}^C_{s}$ according to Eq.~(\ref{ssLNA_diff}). The non-zero elements in $\mathbf{D}^C_{s}$ are given by
\begin{equation}\label{app_cmrna_diff}
\begin{aligned}
&\mathbf{D}^C_{s(1,1)}=\sum_{g=1}^{G-1}(a_{g-1}-a_g)^2k_{g-1}[U_{g-1}]+a_{G-1}^2k_{G-1}[U_{G-1}]+(1-a_1)^2k_G[U_G]\\
&\qquad\qquad+(a_1-a_0)^2k_{\rm{off}}[U_1]+\lambda [M^C_1],\\ 
&\mathbf{D}^C_{s(1,2)}=-\lambda[M_1^C],\\ 
&\mathbf{D}^C_{s(i-1,i)}=-\lambda[M_{i-1}^C],\; \mathbf{D}^C_{s(i,i)}=\lambda([M_{i-1}^C]+[M_{i}^C]),\; \mathbf{D}^C_{s(i,i+1)}=-\lambda[M_i^C],\\
&\text{for}\; i=2,3,...,R-1,\\
&\mathbf{D}^C_{s(R,R-1)}=-\lambda[M_{R-1}^C],\; \mathbf{D}^C_{s(R,R)}=\lambda([M_{R-1}^C]+[M_{R}^C]).
\end{aligned}
\end{equation}
Furthermore from Eq.~(\ref{cmrna_recursive_a}) we have
\begin{equation}
\begin{aligned}
&a_0-a_1=\frac{e_{G}(k_1,...,k_G)}{k_{\rm{on}} e_{G-1}(k_1,...,k_{G})+k_{\rm{off}}e_{G-1}(k_2,...,k_G)},\\
&a_1-a_2=\frac{k_{\rm{off}}e_{G-1}(k_2,...,k_G)+k_{\rm{on}}k_G e_1^{\text{max}}}{k_{\rm{on}} e_{G-1}(k_1,...,k_{G})+k_{\rm{off}}e_{G-1}(k_2,...,k_G)},\\
&a_{g-1}-a_g=\frac{k_G k_{\rm{on}}e_{g-1}^{\text{max}}}{k_{\rm{on}} e_{G-1}(k_1,...,k_{G})+k_{\rm{off}}e_{G-1}(k_2,...,k_G)}, g=3,4,...,G-2,\\
&a_{G-1}=\frac{k_G k_{\rm{on}}e_{G-2}(k_1,...,k_{G-2})}{k_{\rm{on}} e_{G-1}(k_1,...,k_{G})+k_{\rm{off}}e_{G-1}(k_2,...,k_G)},\\
&a_1-1=\frac{k_{\rm{on}}(k_Ge_{G-2}(k_1,...,k_{G-1})-e_{G-1}(k_1,...,k_G))}{k_{\rm{on}} e_{G-1}(k_1,...,k_{G})+k_{\rm{off}}e_{G-1}(k_2,...,k_G)}\\
&\quad\quad\;\;\,= -\frac{k_{\rm{on}}e_{G-1}(k_1,...,k_{G-1})}{k_{\rm{on}} e_{G-1}(k_1,...,k_{G})+k_{\rm{off}}e_{G-1}(k_2,...,k_G)}.
\end{aligned}
\end{equation}
Substitution into Eq.~(\ref{app_cmrna_diff}) we find that $\mathbf{D}^C_{s(1,1)}$ simplifies to 
\begin{equation}
\begin{aligned}
\mathbf{D}^C_{s(1,1)}=&\left(\frac{e_G(k_1^2,...,k_G^2)\frac{2k_{\rm{off}}}{k_1}+\sum_{g=1}^{G-1}k_G^2k_{\rm{on}}^2(e_{g}^{\text{max}})^2+k_{\rm{off}}^2e_{G-1}(k_2^2,...,k_G^2)}{(k_{\rm{on}} e_{G-1}(k_1,...,k_{G})+k_{\rm{off}}e_{G-1}(k_2,...,k_G))^3}\right.\\
&\left.+\frac{2k_{\rm{off}}k_{\rm{on}}e_{G-1}(k_2^2,...,k_G^2)+k_{\rm{on}}^2e_{G-1}(k_1^2,...,k_{G-1}^2)}{(k_{\rm{on}}e_{G-1}(k_1,...,k_{G})+k_{\rm{off}}e_{G-1}(k_2,...,k_G))^3}\right)\\
&\times e_{G+1}(k_{\rm{on}},k_1,...,k_G)+\lambda[M_1^C]\\
=&\lambda[M_1^C]+e_{G+1}(k_{\rm{on}},k_1,...,k_G)\times\\
&\frac{\left(k_{\rm{on}}^2 e_{G-1}(k_1^2,...,k_G^2)+e_{G-1}(k_2^2,...,k_G^2)(k_{\rm{off}}^2+2k_{\rm{off}}k_{\rm{on}}+2k_{\rm{off}}k_1)\right)}{(k_{\rm{off}}e_{G-1}(k_2,...,k_G)+k_{\rm{on}}e_{G-1}(k_1,...,k_G))^3}.
\end{aligned}
\end{equation}

\subsection{Dependence of the Fano factor of cytoplasmic mRNA on the deactivation rate \texorpdfstring{$k_{\rm{off}}$}{koff}} \label{app_cmrna_cases}
\noindent\textit{\textbf{Proof of Corollary~\ref{cor_cmrna_cases}}.} Firstly, a Taylor series expansion of $FF^C_{s}$ with respect to $1/k_{\rm{off}}$ gives  
\begin{equation}
\label{cmrna_series}
\begin{aligned}
   FF^C_{s}=&1+2\left(1-\frac{(2R-1)!!}{(2R)!!}\right)\frac{k_1 (e_{G-1}(k_2,...,k_G)-k_{\rm{on}}e_{G-2}(k_2,...,k_G))}{e_{G-1}(k_2,...,k_G)}\frac{1}{k_{\rm{off}}}\\
   &+O\left(\frac{1}{k_{\rm{off}}^2}\right). 
\end{aligned}
\end{equation}
The second term in Eq.~(\ref{cmrna_series}) is negative when $e_{G-1}(k_2,...,k_G)-k_{\rm{on}}e_{G-2}(k_2,...,k_G)<0$, and positive when $e_{G-1}(k_2,...,k_G)-k_{\rm{on}}e_{G-2}(k_2,...,k_G)>0$. On the other hand, when $e_{G-1}(k_2,...,k_G)-k_{\rm{on}}e_{G-2}(k_2,...,k_G)=0$, Eq.~(\ref{cmrna_series}) becomes
\begin{equation}
\label{cmrna_series_equal}
\begin{aligned}
    FF^C_{s}=&1-2\left(1-\frac{(2R-1)!!}{(2R)!!}\right)\frac{e_{G}(k_1,...,k_G)e_{G-2}(k_1,...,k_G)}{e_{G-2}(k_2,...,k_G)^2}\frac{1}{k_{\rm{off}}^2}\\
    &+O\left(\frac{1}{k_{\rm{off}}^3}\right). 
\end{aligned}
\end{equation}
Therefore, the Fano factor goes to $1$ when $k_{\rm{off}}\rightarrow\infty$. Specifically, if $k_{\rm{on}}$ is large enough such that $e_{G-1}(k_2,...,k_G)-k_{\rm{on}}e_{G-2}(k_2,...,k_G)\leq 0$, i.e. $k_{\rm{on}}\geq k_{\rm{on},1}^\star$ holds, then the Fano factor goes to $1$ from below; if $k_{\rm{on}}$ is small enough that $e_{G-1}(k_2,...,k_G)-k_{\rm{on}}e_{G-2}(k_2,...,k_G)>0$, i.e. $k_{\rm{on}}< k_{\rm{on},1}^\star$ holds, then it goes to $1$ from above.
 
Secondly, when $k_{\rm{off}}=0$, we have
\begin{equation}
\label{koff=0}
\begin{aligned}
   \left.FF^C_{s}\right|_{k_{\rm{off}}=0}
    &=1-\left(1-\frac{(2R-1)!!}{(2R)!!}\right)\frac{2e_G(k_1,...k_G)e_{G-2}(k_1,...,k_G)}{e_{G-1}(k_1,...,k_G)^2}<1,
\end{aligned}
\end{equation}
where we have used the property 
\begin{equation}
    e_{G-1}(k_1^2,...,k_G^2)=e_{G-1}(k_1,...,k_G)^2-2e_G(k_1,...k_G)e_{G-2}(k_1,...,k_G).
\end{equation}

Thirdly, the critical point $k_{\rm{off},1}$ is obtained by solving $\frac{\partial FF^C}{\partial k_{\rm{off}}}=0$, which gives
\begin{equation}
\label{koff1}
\begin{aligned}
   k_{\rm{off},1}=&\frac{k_{\rm{on}}e_{G-1}(k_2,...,k_G)e_{G-1}(k_1,...,k_G)-k_{\rm{on}}^2k_1e_{G-2}(k_2^2,...,k_G^2)}{e_{G-1}(k_2,...,k_G)(e_{G-1}(k_2,...,k_G)-k_{\rm{on}}e_{G-2}(k_2,...,k_G))}\\
    &+\frac{k_{\rm{on}}^2e_{G-1}(k_2,...,k_G)e_{G-2}(k_2,...,k_G)}{e_{G-1}(k_2,...,k_G)(e_{G-1}(k_2,...,k_G)-k_{\rm{on}}e_{G-2}(k_2,...,k_G))}.
\end{aligned}
\end{equation}
For the sake of simplicity, we denote the numerator and denominator of Eq.~(\ref{koff1}) by $N(k_{\rm{off},1})$ and $D(k_{\rm{off},1})$ respectively. When $k_{\rm{on}}\geq k_{\rm{on},1}^\star$, it follows from Eq.~(\ref{def_conditions}) that $e_{G-1}(k_2,...,k_G)-k_{\rm{on}}e_{G-2}(k_2,...,k_G)<0$, hence $D(k_{\rm{off},1})<0$. We consider two cases, $k_{\rm{off},1}>0$ and $k_{\rm{off},1}<0$, separately.

\textbf{Case 1}: Suppose $k_{\rm{on}}\geq k_{\rm{on},1}^\star$ and $k_{\rm{off},1}>0$. Since $D(k_{\rm{off},1})<0$, it follows that $N(k_{\rm{off},1})<0$. Solving $N(k_{\rm{off},1})<0$ gives
\begin{equation}
\label{Nkoff1<0}
\begin{aligned}
    &k_{\rm{on}}(e_{G-1}(k_2,...,k_G)e_{G-2}(k_2,...,k_G)-k_1e_{G-2}(k_2^2,...,k_G^2))\\
     <&-e_{G-1}(k_2,...,k_G) e_{G-1}(k_1,...,k_G).
\end{aligned}      
\end{equation}
From here it follows that if $k_1<k_1^\star$ (i.e. $e_{G-1}(k_2,...,k_G)e_{G-2}(k_2,...,k_G)-k_1e_{G-2}(k_2^2,...,k_G^2)>0$), then $k_{\rm{on}}<0$, which contradicts $k_{\rm{on}}\geq k_{\rm{on},1}^\star$, and therefore $k_1$ can only be larger than $k_1^\star$. Consequently, Eq.~(\ref{Nkoff1<0}) becomes
\begin{equation}
    k_{\rm{on}}>\frac{e_{G-1}(k_2,...,k_G)e_{G-1}(k_1,...,k_G)}{k_1e_{G-2}(k_2^2,...,k_G^2)-e_{G-1}(k_2,...,k_G)e_{G-2}(k_2,...,k_G)}=k_{\rm{on},2}^\star.
\end{equation}
In addition, $k_{\rm{on},2}^\star-k_{\rm{on},1}^\star$ gives
\begin{equation}
\label{kon2-kon1}
\begin{aligned}
    \frac{2e_{G-1}(k_2^2,...,k_G^2)e_{G-2}(k_1,...,k_G)}{e_{G-2}(k_2,...,k_G)(k_1e_{G-2}(k_2^2,...,k_G^2)-e_{G-1}(k_2,...,k_G)e_{G-2}(k_2,...,k_G))},    
\end{aligned}   
\end{equation}
which is positive since $k_1> k_1^\star$. Hence, we have $k_{\rm{on},2}^\star>k_{\rm{on},1}^\star$. Therefore, the conditions for case 1 can be re-expressed as
\begin{equation}
    \label{cond_case1}
    k_1> k_1^\star\quad \text{and} \quad k_{\rm{on}}>k_{\rm{on},2}^\star.
\end{equation}
    
Next, we show that $k_{\rm{off},1}$ gives the local minimum of $FF_{s}^{C}$. We take the second derivative of $FF^C_{s}$ and substitute $k_{\rm{off}}=k_{\rm{off},1}$, which gives
\begin{equation}
\begin{aligned}
    \left.\frac{\partial^2FF^C_{s}}{\partial k_{\rm{off}}^2}\right|_{k_{\rm{off}}=k_{\rm{off},1}}=&\frac{k_1e_{G-1}(k_2^2,...,k_G^2)((2R)!!-(2R-1)!!)}{4k_{\rm{on}}^3 (2R)!!}\\
    &\frac{(e_{G-1}(k_2,...,k_G)-k_{\rm{on}}e_{G-2}(k_2,...,k_G))^4}{\beta^3},
\end{aligned}
\end{equation}
where
\begin{equation}
\label{beta}
\begin{aligned}
    \beta=&k_{\rm{on}}e_{G-3}(k_2,...,k_G)e_{G}(k_1,...,k_G)+k_{\rm{on}}k_1e_{G-2}(k_2^2,...,k_G^2)\\
    &-e_{G-1}(k_2,...,k_G)e_{G-1}(k_1,...,k_G).
\end{aligned}
\end{equation}
The only thing left to show is that $\beta>0$. If $\beta>0$, then
\begin{equation}
\label{cmrna_inequalityf1}
    k_{\rm{on}}>\frac{e_{G-1}(k_2,...,k_G)e_{G-1}(k_1,...,k_G)}{e_G(k_1,...,k_G)e_{G-3}(k_2,...,k_G)+k_1e_{G-2}(k_2^2,...,k_G^2)}:=f_1. 
\end{equation}
As $k_{\rm{on}}>k_{\rm{on},2}^\star$ according to Eq.~(\ref{cond_case1}), $\beta$ is automatically larger than $0$ if we can show that $k_{\rm{on},2}^\star>f_1$. Indeed,
\begin{equation}
\begin{aligned}
    k_{\rm{on},2}^\star-f_1=&\frac{e_{G-1}(k_2^2,...k_G^2)e_{G-1}(k_1,...,k_G)e_{G-2}(k_1,...,k_G)}{(e_G(k_1,...,k_G)e_{G-3}(k_2,...,k_G)+k_1e_{G-2}(k_2^2,...,k_G^2))}\\
    &\times\frac{1}{(k_1e_{G-2}(k_2^2,...,k_G^2)-e_{G-1}(k_2,...,k_G)e_{G-2}(k_2,...,k_G))}>0,
\end{aligned}
\end{equation}
since $k_1>k_1^\star$. Therefore, $\beta>0$ is true and $k_{\rm{off},1}$ gives the local minimum of $FF_{s}^{C}$. In case 1, the Fano factor decreases until it reaches the critical point $k_{\rm{off},1}$ and then increases, eventually approaching $1$ from below. The Fano factor is sub-Poissonian for all values of $k_{\rm{off}}$.

\textbf{Case 2}: Suppose $k_{\rm{on}}\geq k_{\rm{on},1}^\star$ and $k_{\rm{off},1}<0$. Since $D(k_{\rm{off},1})<0$, $N(k_{\rm{off},1})$ satisfies $N(k_{\rm{off},1})\geq 0$, which yields
\begin{equation}
\label{Nkoff1>0}
\begin{aligned}
    &k_{\rm{on}}(e_{G-1}(k_2,...,k_G)e_{G-2}(k_2,...,k_G)-k_1e_{G-2}(k_2^2,...,k_G^2))\\
    \geq &-e_{G-1}(k_2,...,k_G) e_{G-1}(k_1,...,k_G).
\end{aligned}      
\end{equation}
If $k_1<k_1^\star$, then
\begin{equation}
    k_{\rm{on}}\geq \frac{-e_{G-1}(k_2,...,k_G) e_{G-1}(k_1,...,k_G)}{e_{G-1}(k_2,...,k_G)e_{G-2}(k_2,...,k_G)-k_1e_{G-2}(k_2^2,...,k_G^2)}=k_{\rm{on},2}^\star,
\end{equation}
where $k_{\rm{on},2}^\star$ is negative. So if $k_1<k_1^\star$, then the inequalities $k_{\rm{on}}\geq k_{\rm{on},1}^\star$ and $k_{\rm{on}}\geq k_{\rm{on},2}^\star$, are reduced to $k_{\rm{on}}\geq k_{\rm{on},1}^\star$.

If $k_1>k_1^\star$, then
\begin{equation}
    k_{\rm{on}}\leq \frac{-e_{G-1}(k_2,...,k_G) e_{G-1}(k_1,...,k_G)}{e_{G-1}(k_2,...,k_G)e_{G-2}(k_2,...,k_G)-k_1e_{G-2}(k_2^2,...,k_G^2)}=k_{\rm{on},2}^\star.
\end{equation}
In addition, it follows from Eq.~(\ref{kon2-kon1}) that $k_{\rm{on},2}^\star>k_{\rm{on},1}^\star$ when $k_1> k_1^\star$. Hence, if $k_1> k_1^\star$ then $k_{\rm{on},1}^\star \leq k_{\rm{on}}\leq k_{\rm{on},2}^\star$. Note that if $k_1=k_1^\star$, then $k_{\rm{off},1}<0$. Under this condition, $k_{\rm{on}}\geq k_{\rm{on},1}^\star$. Therefore, the conditions for case 2 can be re-expressed as
\begin{equation}
    k_1\leq k_1^\star\; \text{and}\; k_{\rm{on}}\geq k_{\rm{on},1}^\star \quad  \text{or} \quad k_1>k_1^\star\; \text{and} \; k_{\rm{on},1}^\star \leq k_{\rm{on}}\leq k_{\rm{on},2}^\star.
\end{equation}
Since $\left.FF^C_{s}\right|_{k_{\rm{off}}=0}<1$, there is no critical point if $k_{\rm{off}}>0$, and $FF^C\rightarrow 1$ from below when $k_{\rm{off}}\rightarrow \infty$. From this, we conclude that in case 2, the Fano factor monotonically increases and is always sub-Poissonian.

\textbf{Case 3:} Suppose $k_{\rm{on}}<k_{\rm{on},1}^\star$. If $k_{\rm{off},1}>0$, then $N(k_{\rm{off},1})>0$ since $D(k_{\rm{off},1})>0$. Similar to case 2, from Eq.~(\ref{Nkoff1>0}) it follows that if $k_1<k_1^\star$, then
\begin{equation}
    k_{\rm{on}}>\frac{-e_{G-1}(k_2,...,k_G) e_{G-1}(k_1,...,k_G)}{e_{G-1}(k_2,...,k_G)e_{G-2}(k_2,...,k_G)-k_1e_{G-2}(k_2^2,...,k_G^2)}=k_{\rm{on},2}^\star, 
\end{equation}
where $k_{\rm{on},2}^\star<0$. Hence, $k_{\rm{on}}< k_{\rm{on},1}^\star$ if $k_1<k_1^\star$. On the other hand, if $k_1>k_1^\star$, then
\begin{equation}
    k_{\rm{on}}<\frac{-e_{G-1}(k_2,...,k_G) e_{G-1}(k_1,...,k_G)}{e_{G-1}(k_2,...,k_G)e_{G-2}(k_2,...,k_G)-k_1e_{G-2}(k_2^2,...,k_G^2)}=k_{\rm{on},2}^\star.   
\end{equation}
It follows from Eq.~(\ref{kon2-kon1}) that $k_{\rm{on},2}^\star>k_{\rm{on},1}^\star$ if $k_1>k_1^\star$. Therefore, if $k_1>k_1^\star$, then $k_{\rm{on}}< k_{\rm{on},1}^\star$. In addition, if $k_1= k_1^\star$, then $k_{\rm{off},1}>0$, so under this condition, $k_{\rm{on}}< k_{\rm{on},1}^\star$. Hence, the conditions for case 3 can be re-expressed as
\begin{equation}
    k_1\leq k_1^\star,\; k_{\rm{on}}< k_{\rm{on},1}^\star \quad \text{and} \quad k_1>k_1^\star,\; k_{\rm{on}}< k_{\rm{on},1}^\star,
\end{equation}
i.e. if $k_{\rm{off},1}>0$, $k_1$ can take any positive real values given $k_{\rm{on}}< k_{\rm{on},1}^\star$.
Hence, $k_{\rm{off},1}>0$ as long as $k_{\rm{on}}< k_{\rm{on},1}^\star$.
    
Next, we show that $k_{\rm{off},1}$ gives the maximum of the Fano factor $FF_{s}^{C}$. To achieve that, we need to show that $\beta<0$ ($\beta$ was introduced in Eq.~(\ref{beta})). If $\beta<0$, then
\begin{equation}
   k_{\rm{on}}<f_1. 
\end{equation}
Since $k_{\rm{on}}< k_{\rm{on},1}^\star$ in case 3, we conclude that $\beta<0$ if $f_1>k_{\rm{on},1}^\star$. Indeed, $f_1-k_{\rm{on},1}^\star$ gives 
\begin{equation}
    \frac{e_{G-1}(k_2^2,...,k_G^2)e_{G-2}(k_1,...,k_G)}{k_1e_{G-2}(k_2,...,k_G)(e_{G-2}(k_2^2,...,k_G^2)+e_{G-1}(k_2,...,k_G)e_{G-3}(k_2,...,k_G))}>0.
\end{equation}
Therefore $\beta<0$ holds for the case $k_{\rm{on}}<k_{\rm{on},1}^\star$, and $k_{\rm{off},1}$ gives the local maximum of $FF_{s}^{C}$.

When $k_{\rm{on}}<k_{\rm{on},1}^\star$, the Fano factor first increases until it reaches the critical point $k_{\rm{off},1}$, after which it decreases, eventually approaching $1$ from above. The Fano factor changes from sub-Poissonian to super-Poissonian as $k_{\rm{off}}$ crosses the threshold $k_{\rm{off}}^\star$, which is given in Corollary~\ref{cor_cmrna_thre}.~$\square$

\subsection{Solution of Lyapunov equations for proteins fluctuations}\label{app_pro_FF}

\textit{\textbf{Details of calculation in Proposition~\ref{prop_pro_FF}.}}
Similar to Appendix~\ref{app_cmrna_FF}, we utilize the ssLNA in Appendix~\ref{app_ssLNA} to derive the Lyapunov equation Eq.~\eqref{pro_lyp}.

We follow the order of reactions used in Appendix~\ref{app_cmrna_FF}, but additionally we include the protein production and protein degradation reactions to construct the rate function matrix:
\begin{equation}\label{pro_rateF}
\begin{aligned}
\mathbf{F}=&\text{diag}(\vec{f})\\
=&\left(k_{\rm{on}}[u_0],k_1[u_1],...,k_G[u_G],k_{\rm{off}}[u_1],\delta [M^N_1],...,\delta_1 [M^N_s],\lambda [M^C_1],...,\right.\\
&\left.\lambda [M^C_R],\lambda_1 [M^C_1],\lambda_2 [P]\right).    
\end{aligned}
\end{equation}
We next formulate the stoichiometric matrix $\mathbf{S}$. We order the species as follows: the slow species $P$, followed by the fast species $U_0, U_1, \ldots, U_G, M_1^N, \ldots, M_S^N, M_1^C, M_2^C, \ldots, M_R^C$. Hence, the first row of the stoichiometric matrix represents the slow species and the rest the fast species. Given the chosen order of reactions and species, the stoichiometric matrix is given by

\begin{equation}
\begin{aligned}
&\mathbf{S}_{(G+S+2)\times(G+S+5)}=
\left(\frac{\mathbf{S}_s}{\mathbf{S}_f}\right)=\\&
\left(
\begin{array}{ccccccccccccccccccccccc}
    0&0 &0 &... &0 &0 &0 &0 &0 &0 &0 &... &0 &0 &0 &... &0 &0 &0 &1 &-1
    \\\hline
    -1 &0 &0 &...&0 &0 &0 &1 &0 &0 &0 &... &0 &%
    \multicolumn{6}{c}{\multirow{9}[2]{*}{\huge $0$}}\\
    1 &-1 &0 &...&0 &0 &1 &-1 &0 &0 &0 &... &0 & & & & & & &&\\
    0 & 1 & -1 &... &0 &0 &0 &0 &0 &0 &0 &... &0 & & & & & & & &\\
    \vdots & & & & & & & & & & & & & & & & & & & &\\
    0 &0 &0 &...&1 &-1 &0 &0 &0 &0 &0 &... &0 & & & & & & & &\\
    0 &0 &0 &...&0 &0 &1 &0 &-1 &0 &0 &... &0 & & & & & & & &\\
    0 &0 &0 &...&0 &0 &0 &0 &1 &-1 &0 &... &0 & & & & & & & &\\
    \vdots & & & & & & & & & & & & & & & & &&\\
    0 &0 &0 &...&0 &0 &0 &0 &0 &0 &... &1 &-1 & & & & & & & &\\
    \multicolumn{8}{c}{\multirow{5}[2]{*}{\huge $0$}}\\
     & & & & & & & &0 &0 &0 &... &1 &-1 &0 &... &0 &0 &0 &0 &0\\
     & & & & & & & &0 &0 &0 &... &0 &1 &-1 &... &0 &0 &0 &0 &0\\
     & & & & & & & &\vdots & & & & &\vdots & & & & & &0 &0\\
     & & & & & & & &0 &0 &0 &... &0 &0 &0 &... &0 &1 &-1 &0 &0\\
\end{array}
\right).  
\end{aligned}
\end{equation}

From the Jacobian $\mathbf{J}_{ij}=\partial_j(\mathbf{S}\vec{f})_i$, we can readily obtain its four partitioned submatrices
\begin{equation}\label{pro_Jacob_part}
\begin{aligned}
&\begin{aligned}[c]
\mathbf{J}_{ss}=-\lambda_2,
\end{aligned}
\quad
\begin{aligned}[c]
\mathbf{J}_{sf}=\left(
\begin{array}{ccccccccc}
0&0&...&0 &\lambda_1\\ 
\end{array}\right),
\end{aligned}\quad
\begin{aligned}
\mathbf{J}_{fs}=\boldsymbol{0},
\end{aligned}\\
&\begin{aligned}[c]
\mathbf{J}_{ff}=\left(
\mathbf{J}_{ff,1} \mathbf{J}_{ff,2}\right),
\end{aligned}
\end{aligned}
\end{equation}
where
\begin{equation}
\begin{aligned}
&\mathbf{J}_{ff,1}=\left(\begin{array}{ccccccccccccccc}
-k_{\rm{on}} &k_{\rm{off}} &0 &... &0 &0 &0 &0\\
k_{\rm{on}}-k_G &-k_{\rm{off}}-k_1-k_G &-k_G &... &-k_G &-k_G &0 &0\\
0 &k_1 &-k_2 &... &0 &0 &0 &0 \\
...& & & & & & & \\
0 &0 &0 &... &k_{G-2} &-k_{G-1} &0 &0 \\
-k_G &-k_G &-k_G &... &-k_G &-k_G &-\delta &0\\
0 &0 &0 &... &0 &0 &\delta &-\delta\\
...& & & & & & & \\
0 &0 &0 &0 &0 &0 &0 &0\\
0 &0 &0 &0 &0 &0 &0 &0\\
0 &0 &0 &0 &0 &0 &0 &0 \\
...& & & & & & & \\
0 &0 &0 &0 &0 &0 &0 &0 \\
\end{array}\right), \\
&\mathbf{J}_{ff,2}=\left(\begin{array}{ccccccccccccccc}
&... &0 &...&0 &0 &0 &...\\
&... &0 &...&0 &0 &0 &...\\
&... &0 &...&0 &0 &0 &...\\
&... & & & & & &\\
&... &0 &... &0 &0 &0 &...\\
&... &0 &... &0 &0 &0 &...\\
&... &0 &... &0 &0 &0 &...\\
&... & & & & & &\\
&... &\delta &-\delta_1 &0 &0 &0 &... \\
&... &0 &\delta_1 &-\lambda &0 &0 &...\\
&... &0 &0 &\lambda &-\lambda &0&...\\
&... & & & & & &\\
&... &0 &0 &0 &0 &\lambda&-\lambda\\
\end{array}\right). 
\end{aligned}    
\end{equation}
Given the partitioned Jacobian matrix above, we can obtain the Jacobian matrix $\mathbf{J}_s^P$ in Eq.~(\ref{pro_lyp}) as
\begin{equation}\label{Jacobianproteins}
    \mathbf{J}_s^P=\mathbf{J}_{ss}-\mathbf{J}_{sf}\mathbf{J}_{ff}^{-1}\mathbf{J}_{fs}=\mathbf{J}_{ss}=-\lambda_2,
\end{equation}
which is trivial as $\mathbf{J}_{fs}=\underline{\boldsymbol{0}}$.

Now we can calculate the diffusion matrix $D_s^P$ in Eq.~(\ref{pro_lyp}):
\begin{equation}
    \mathbf{D}_s^P=(\mathbf{W}-\mathbf{V})(\mathbf{W}-\mathbf{V})^T,
\end{equation}
where
\begin{equation}
\begin{aligned}
&\mathbf{W}_{1\times(G+S+5)}=\left(
    \begin{array}{ccccccccccccccc}
    0 &0 &0 &... &0 &\sqrt{\lambda_1[M_1^C]} &-\sqrt{\lambda_2[P]}\\
    \end{array}
    \right),   
\end{aligned}
\end{equation}
and
\begin{equation}\label{app_pro_diff_B}
\begin{aligned}
&\mathbf{V}_{1\times(G+S+5)}=\\
&\left((a_{g+1}-a_g)\sqrt{k_g[U_g]}\right.\quad
-a_{G-1}\sqrt{k_{G-1}[U_{G-1}]}\quad (a_1-\frac{\lambda_1}{\lambda})\sqrt{k_{G}[U_{G}]}\quad\\
&(a_0-a_1)\sqrt{k_{\rm{off}}[U_{1}]}\quad
0\quad...\quad 0 \quad
\frac{\lambda_1}{\lambda}\sqrt{\lambda[M_1^C]} \left.\quad 0 \quad 0 \right).
\end{aligned}
\end{equation}
Note that $g=0,1,...,G-2$; $s=1,2,...,S-1$; $r=1,2,...,R$.

The variables $a_g$ ($g=0,1,2,...,G-1$) are defined as:
\begin{equation}\label{pro_recursive_a}
\begin{aligned}
     &a_0=\frac{\lambda_1(k_{\rm{off}}e_{G-1}(k_2,...,k_{G})+e_{G}(k_{1},...,k_{G})+k_{\rm{on}}k_Ge_{G-2}(k_1,...,k_{G-1}))}{\lambda(k_{\rm{on}} e_{G-1}(k_1,...,k_{G})+k_{\rm{off}}e_{G-1}(k_2,...,k_G))},\\ &a_1=\frac{\lambda_1(k_{\rm{off}}e_{G-1}(k_2,...,k_{G})+k_{\rm{on}}k_Ge_{G-2}(k_1,...,k_{G-1}))}{\lambda(k_{\rm{on}} e_{G-1}(k_1,...,k_{G})+k_{\rm{off}}e_{G-1}(k_2,...,k_G))},\\
     &a_2=a_1-\frac{\lambda_1(k_{\rm{off}}e_{G-1}(k_2,...,k_G)+k_{\rm{on}}k_G e_1^{\text{max}})}{\lambda(k_{\rm{on}} e_{G-1}(k_1,...,k_{G})+k_{\rm{off}}e_{G-1}(k_2,...,k_G))},\\
     &a_g=a_{g-1}-\frac{\lambda_1(k_G k_{\rm{on}}e_{g-1}^{\text{max}})}{\lambda(k_{\rm{on}} e_{G-1}(k_1,...,k_{G})+k_{\rm{off}}e_{G-1}(k_2,...,k_G))}, g=3,4,...,G-1.
\end{aligned}
\end{equation}
Note that the notation $e_g^{\text{max}}$ was previously introduced in Appendix~\ref{app_cmrna_FF}.

Given matrices $\mathbf{W}$ and $\mathbf{V}$, we can now calculate the effective diffusion matrix $\mathbf{D}^P_{s}$ according to Eq.~(\ref{ssLNA_diff}): 
\begin{equation}\label{pro_diff1}
\begin{aligned}
\mathbf{D}_s^P=&\sum_{g=0}^{G-2}(a_g-a_{g+1})^2k_g[U_g]+a_{G-1}^2k_{G-1}[U_{G-1}]+(a_1-\frac{\lambda_1}{\lambda})^2k_{G}[U_{G}]\\&+(a_1-a_0)^2k_{\rm{off}}[U_1]+(\frac{\lambda_1^2}{\lambda}+\lambda_1)[M_1^C]+\lambda_2[P].
\end{aligned}
\end{equation}
Furthermore, from Eq.~(\ref{pro_recursive_a}) we have
\begin{equation}
\begin{aligned}
&a_0-a_1=\frac{\lambda_1e_{G}(k_1,...,k_G)}{\lambda(k_{\rm{on}} e_{G-1}(k_1,...,k_{G})+k_{\rm{off}}e_{G-1}(k_2,...,k_G))},\\
&a_1-a_2=\frac{\lambda_1(k_{\rm{off}}e_{G-1}(k_2,...,k_G)+k_{\rm{on}}k_G e_1^{\text{max}})}{\lambda(k_{\rm{on}} e_{G-1}(k_1,...,k_{G})+k_{\rm{off}}e_{G-1}(k_2,...,k_G))},\\
&a_{g-1}-a_g=\frac{\lambda_1(k_G k_{\rm{on}}e_{g-1}^{\text{max}})}{\lambda(k_{\rm{on}} e_{G-1}(k_1,...,k_{G})+k_{\rm{off}}e_{G-1}(k_2,...,k_G))}, g=3,4,...,G-2,\\
&a_{G-1}=\frac{\lambda_1k_G k_{\rm{on}}e_{G-2}(k_1,...,k_{G-2})}{\lambda(k_{\rm{on}} e_{G-1}(k_1,...,k_{G})+k_{\rm{off}}e_{G-1}(k_2,...,k_G))},\\
&a_1-\frac{\lambda_1}{\lambda}=\frac{\lambda_1(k_{\rm{on}}k_Ge_{G-2}(k_1,...,k_{G-1})-k_{\rm{on}} e_{G-1}(k_1,...,k_{G}))}{\lambda(k_{\rm{on}} e_{G-1}(k_1,...,k_{G})+k_{\rm{off}}e_{G-1}(k_2,...,k_G))}\\
&\quad\quad\quad\;\, = -\frac{\lambda_1k_{\rm{on}}e_{G-1}(k_1,...,k_{G-1})}{\lambda(k_{\rm{on}} e_{G-1}(k_1,...,k_{G})+k_{\rm{off}}e_{G-1}(k_2,...,k_G))}.
\end{aligned}
\end{equation}
Substitution of these into Eq.~(\ref{pro_diff1}) gives
\begin{equation}
\begin{aligned}
\mathbf{D}_s^P=&\left(\frac{\lambda_1^2(e_G(k_1^2,...,k_G^2)\frac{2k_{\rm{off}}}{k_1}+\sum_{g=1}^{G-1}k_G^2k_{\rm{on}}^2(e_{g}^{\text{max}})^2+k_{\rm{off}}^2e_{G-1}(k_2^2,...,k_G^2))}{\lambda^2(k_{\rm{on}} e_{G-1}(k_1,...,k_{G})+k_{\rm{off}}e_{G-1}(k_2,...,k_G))^3}\right.\\
&\left.+\frac{\lambda_1^2(2k_{\rm{off}}k_{\rm{on}}e_{G-1}(k_2^2,...,k_G^2)+k_{\rm{on}}^2e_{G-1}(k_1^2,...,k_{G-1}^2))}{\lambda^2(k_{\rm{on}}e_{G-1}(k_1,...,k_{G})+k_{\rm{off}}e_{G-1}(k_2,...,k_G))^2})\right)\\
&\times e_{G+1}(k_{\rm{on}},k_1,...,k_G)+(\frac{\lambda_1^2}{\lambda}+\lambda_1)[M_1^C]+\lambda_2[P] \\
=&\frac{\lambda_1}{\lambda}\frac{e_{G-1}(k_2^2,...,k_G^2)(k_{\rm{off}}^2+2k_{\rm{off}}(k_{\rm{on}}+k_1))+k_{\rm{on}}^2e_{G-1}(k_1^2,...,k_G^2)}{(k_{\rm{off}}e_{G-1}(k_2,...,k_G)+k_{\rm{on}}e_{G-1}(k_1,...,k_G))^3}\\
&\times e_{G+1}(k_{\rm{on}},k_1,...,k_G)+(\frac{\lambda_1^2}{\lambda}+\lambda_1)[M_1^C]+\lambda_2[P].       
\end{aligned}
\end{equation}

\subsection{Dependence of the Fano factor of proteins on the deactivation rate \texorpdfstring{$k_{\rm{off}}$}{koff}}\label{app_pro_cases}
\noindent\textit{\textbf{Proof of Corollary~\ref{cor_pro_thre}}.} 
Our method for analysing the dependence of $FF_s^P$ on $k_{\rm{off}}$ is similar to that used in Appendix~\ref{app_cmrna_cases}, where we calculate its values at $k_{\rm{off}}=0$ and $k_{\rm{off}}\rightarrow \infty$, and then determine the behaviour between these two extreme points.

Firstly, a series expansion of $FF^P_{s}$ with respect to $1/k_{\rm{off}}$ gives,    
\begin{equation}\label{pro_series}
\begin{aligned}
   FF^P_{s}=&1+\frac{\lambda_1}{\lambda}+\frac{\lambda_1k_1(e_{G-1}(k_2,...,k_G)-k_{\rm{on}}e_{G-2}(k_2,...,k_G))}{\lambda e_{G-1}(k_2,...,k_G)}\frac{1}{k_{\rm{off}}}\\
   &+O\left(\frac{1}{k_{\rm{off}}^2}\right). 
\end{aligned}
\end{equation}
When $k_{\rm{on}}=k_{\rm{on},1}^\star$, Eq.~(\ref{pro_series}) becomes
\begin{equation}\label{pro_series_equal}
\begin{aligned}
   FF^P_{s}=&1+\frac{\lambda_1}{\lambda}-\frac{\lambda_1 e_{G}(k_1,...,k_G)e_{G-2}(k_1,...,k_G)}{\lambda e_{G-2}(k_2,...,k_G)^2}\frac{1}{k_{\rm{off}}^2}\\
   &+O\left(\frac{1}{k_{\rm{off}}^3}\right). 
\end{aligned}
\end{equation}
Similar to Appendix~\ref{app_cmrna_cases}, we can conclude that the Fano factor goes to $1+\frac{\lambda_1}{\lambda}$ when $k_{\rm{off}}\rightarrow\infty$. Specifically, if $k_{\rm{on}}\geq k_{\rm{on},1}^\star$, it goes to $1+\frac{\lambda_1}{\lambda}$ from below; if $k_{\rm{on}}< k_{\rm{on},1}^\star$, it goes to $1+\frac{\lambda_1}{\lambda}$ from above.

Secondly, when $k_{\rm{off}}=0$, the Fano factor $FF^P_{s}$ becomes
\begin{equation}\label{koff=0p}
\begin{aligned}
    FF^P_{s}\left|_{k_{\rm{off}=0}}\right.&=1+\frac{\lambda_1}{\lambda}\frac{e_{G-1}(k_1^2,...,k_G^2)}{e_{G-1}(k_1,...,k_G)^2}\\
    &=1+\frac{\lambda_1}{2\lambda}+\frac{\lambda_1}{2\lambda}\frac{e_{G-1}(k_1,...,k_G)^2-2e_{G}(k_1,...,k_G)e_{G-2}(k_1,...,k_G)}{e_{G-1}(k_1,...,k_G)^2}\\
    &=1+\frac{\lambda_1}{\lambda}-\frac{\lambda_1}{\lambda}\frac{e_{G}(k_1,...,k_G)e_{G-2}(k_1,...,k_G)}{e_{G-1}(k_1,...,k_G)^2}< 1+\frac{\lambda_1}{\lambda},
\end{aligned}
\end{equation}
where we have used the property 
\begin{equation}
    e_{G-1}(k_1^2,...,k_G^2)=e_{G-1}(k_1,...,k_G)^2-2e_G(k_1,...k_G)e_{G-2}(k_1,...,k_G).
\end{equation}
Thirdly, to find the monotonic behaviour between $k_{\rm{off}}=0$ and $k_{\rm{off}}\rightarrow \infty$, we need to find the critical point. As in  Appendix~\ref{app_cmrna_cases}, the critical point is $k_{\rm{off},1}$. Taking the second derivative of $FF^P_{s}$ and substituting $k_{\rm{off}}=k_{\rm{off},1}$ gives
\begin{equation}
\begin{aligned}
    \left.\frac{\partial^2FF^P_{s}}{\partial k_{\rm{off}}^2}\right|_{k_{\rm{off}}=k_{\rm{off},1}}=&\frac{\lambda_1 k_1e_{G-1}(k_2^2,...,k_G^2)}{8k_{\rm{on}}^3 \lambda}\\
    &\frac{(e_{G-1}(k_2,...,k_G)-k_{\rm{on}}e_{G-2}(k_2,...,k_G))^4}{\beta^3},
\end{aligned}
\end{equation}
where $\beta$ has been defined in Appendix~\ref{app_cmrna_cases}. From the value of $\beta$, the sign of the second derivative of $FF^P_{s}$ and the type of critical point (local maximum or minimum) can be deduced.

In Appendix~\ref{app_cmrna_cases}, the conditions for the three cases were obtained from the expression of $k_{\rm{off},1}$ and $\beta$. Since $FF^C_s$ and $FF^P_s$ have the same dependence on $k_{\rm{off},1}$ and $\beta$, it follows that $FF^P_s$ should also have the same three cases, with the exception that in the limit $k_{\rm{off}}$ approaches infinity, the Fano factor of proteins approaches the value $1+\frac{\lambda_1}{\lambda}$ (while that of cytoplasmic mRNA approaches the value 1).

\end{appendices}

\bibliographystyle{unsrturl} 
\bibliography{sn-bibliography_doi2.bib}

\begin{thebibliography}{10}

\bibitem{van2007stochastic}
Nicolaas~Godfried Van~Kampen.
\newblock {\em Stochastic processes in physics and chemistry}.
\newblock Elsevier, Amsterdam, 2007.
\newblock \href {https://doi.org/10.1063/1.2915501} {\path{doi:10.1063/1.2915501}}.

\bibitem{gillespie1977exact}
Daniel~T Gillespie.
\newblock Exact stochastic simulation of coupled chemical reactions.
\newblock {\em The journal of physical chemistry}, 81(25):2340--2361, 1977.
\newblock \href {https://doi.org/10.1021/j100540a008} {\path{doi:10.1021/j100540a008}}.

\bibitem{gillespie2007stochastic}
Daniel~T Gillespie.
\newblock Stochastic simulation of chemical kinetics.
\newblock {\em Annu. Rev. Phys. Chem.}, 58:35--55, 2007.
\newblock \href {https://doi.org/10.1146/annurev.physchem.58.032806.104637} {\path{doi:10.1146/annurev.physchem.58.032806.104637}}.

\bibitem{milo2015cell}
Ron Milo and Rob Phillips.
\newblock {\em Cell biology by the numbers}.
\newblock Garland Science, 2015.
\newblock \href {https://doi.org/10.1201/9780429258770} {\path{doi:10.1201/9780429258770}}.

\bibitem{taniguchi2010quantifying}
Yuichi Taniguchi, Paul~J Choi, Gene-Wei Li, Huiyi Chen, Mohan Babu, Jeremy Hearn, Andrew Emili, and X~Sunney Xie.
\newblock Quantifying e. coli proteome and transcriptome with single-molecule sensitivity in single cells.
\newblock {\em Science}, 329(5991):533--538, 2010.
\newblock \href {https://doi.org/10.1126/science.1188308} {\path{doi:10.1126/science.1188308}}.

\bibitem{schwanhausser2011global}
Bj{\"o}rn Schwanh{\"a}usser, Dorothea Busse, Na~Li, Gunnar Dittmar, Johannes Schuchhardt, Jana Wolf, Wei Chen, and Matthias Selbach.
\newblock Global quantification of mammalian gene expression control.
\newblock {\em Nature}, 473(7347):337--342, 2011.
\newblock \href {https://doi.org/10.1038/nature10098} {\path{doi:10.1038/nature10098}}.

\bibitem{swain2002intrinsic}
Peter~S Swain, Michael~B Elowitz, and Eric~D Siggia.
\newblock Intrinsic and extrinsic contributions to stochasticity in gene expression.
\newblock {\em Proceedings of the National Academy of Sciences}, 99(20):12795--12800, 2002.
\newblock \href {https://doi.org/10.1073/pnas.162041399} {\path{doi:10.1073/pnas.162041399}}.

\bibitem{golding2005real}
Ido Golding, Johan Paulsson, Scott~M Zawilski, and Edward~C Cox.
\newblock Real-time kinetics of gene activity in individual bacteria.
\newblock {\em Cell}, 123(6):1025--1036, 2005.
\newblock \href {https://doi.org/10.1016/j.cell.2005.09.031} {\path{doi:10.1016/j.cell.2005.09.031}}.

\bibitem{peccoud1995markovian}
Jean Peccoud and Bernard Ycart.
\newblock Markovian modeling of gene-product synthesis.
\newblock {\em Theoretical population biology}, 48(2):222--234, 1995.
\newblock \href {https://doi.org/10.1006/tpbi.1995.1027} {\path{doi:10.1006/tpbi.1995.1027}}.

\bibitem{raj2006stochastic}
Arjun Raj, Charles~S Peskin, Daniel Tranchina, Diana~Y Vargas, and Sanjay Tyagi.
\newblock Stochastic {mRNA} synthesis in mammalian cells.
\newblock {\em PLoS biology}, 4(10):e309, 2006.
\newblock \href {https://doi.org/10.1371/journal.pbio.0040309} {\path{doi:10.1371/journal.pbio.0040309}}.

\bibitem{zenklusen2008single}
Daniel Zenklusen, Daniel~R Larson, and Robert~H Singer.
\newblock Single-{RNA} counting reveals alternative modes of gene expression in yeast.
\newblock {\em Nature structural \& molecular biology}, 15(12):1263--1271, 2008.
\newblock \href {https://doi.org/10.1038/nsmb.1514} {\path{doi:10.1038/nsmb.1514}}.

\bibitem{halpern2015bursty}
Keren~Bahar Halpern, Sivan Tanami, Shanie Landen, Michal Chapal, Liran Szlak, Anat Hutzler, Anna Nizhberg, and Shalev Itzkovitz.
\newblock Bursty gene expression in the intact mammalian liver.
\newblock {\em Molecular cell}, 58(1):147--156, 2015.
\newblock \href {https://doi.org/10.1016/j.molcel.2015.01.027} {\path{doi:10.1016/j.molcel.2015.01.027}}.

\bibitem{jiao2015distribution}
Feng Jiao, Qiwen Sun, Moxun Tang, Jianshe Yu, and Bo~Zheng.
\newblock Distribution modes and their corresponding parameter regions in stochastic gene transcription.
\newblock {\em SIAM Journal on Applied Mathematics}, 75(6):2396--2420, 2015.
\newblock \href {https://doi.org/10.1137/151005567} {\path{doi:10.1137/151005567}}.

\bibitem{singer2014dynamic}
Zakary~S Singer, John Yong, Julia Tischler, Jamie~A Hackett, Alphan Altinok, M~Azim Surani, Long Cai, and Michael~B Elowitz.
\newblock Dynamic heterogeneity and dna methylation in embryonic stem cells.
\newblock {\em Molecular cell}, 55(2):319--331, 2014.
\newblock \href {https://doi.org/10.1016/j.molcel.2014.06.029} {\path{doi:10.1016/j.molcel.2014.06.029}}.

\bibitem{cao2020analytical}
Zhixing Cao and Ramon Grima.
\newblock Analytical distributions for detailed models of stochastic gene expression in eukaryotic cells.
\newblock {\em Proceedings of the National Academy of Sciences}, 117(9):4682--4692, 2020.
\newblock \href {https://doi.org/10.1073/pnas.1910888117} {\path{doi:10.1073/pnas.1910888117}}.

\bibitem{foreman2020mammalian}
Robert Foreman and Roy Wollman.
\newblock Mammalian gene expression variability is explained by underlying cell state.
\newblock {\em Molecular systems biology}, 16(2):e9146, 2020.
\newblock \href {https://doi.org/10.15252/msb.20199146} {\path{doi:10.15252/msb.20199146}}.

\bibitem{jia2023coupling}
Chen Jia and Ramon Grima.
\newblock Coupling gene expression dynamics to cell size dynamics and cell cycle events: Exact and approximate solutions of the extended telegraph model.
\newblock {\em Iscience}, 26(1), 2023.
\newblock \href {https://doi.org/10.1016/j.isci.2022.105746} {\path{doi:10.1016/j.isci.2022.105746}}.

\bibitem{ham2020extrinsic}
Lucy Ham, Rowan~D Brackston, and Michael~PH Stumpf.
\newblock Extrinsic noise and heavy-tailed laws in gene expression.
\newblock {\em Physical review letters}, 124(10):108101, 2020.
\newblock \href {https://doi.org/10.1103/PhysRevLett.124.108101} {\path{doi:10.1103/PhysRevLett.124.108101}}.

\bibitem{peterson2015effects}
Joseph~R Peterson, John~A Cole, Jingyi Fei, Taekjip Ha, and Zaida~A Luthey-Schulten.
\newblock Effects of dna replication on {mRNA} noise.
\newblock {\em Proceedings of the National Academy of Sciences}, 112(52):15886--15891, 2015.
\newblock \href {https://doi.org/10.1073/pnas.1516246112} {\path{doi:10.1073/pnas.1516246112}}.

\bibitem{hilfinger2011separating}
Andreas Hilfinger and Johan Paulsson.
\newblock Separating intrinsic from extrinsic fluctuations in dynamic biological systems.
\newblock {\em Proceedings of the National Academy of Sciences}, 108(29):12167--12172, 2011.
\newblock \href {https://doi.org/10.1073/pnas.1018832108} {\path{doi:10.1073/pnas.1018832108}}.

\bibitem{tunnacliffe2020transcriptional}
Edward Tunnacliffe and Jonathan~R Chubb.
\newblock What is a transcriptional burst?
\newblock {\em Trends in Genetics}, 36(4):288--297, 2020.
\newblock \href {https://doi.org/10.1016/j.tig.2020.01.003} {\path{doi:10.1016/j.tig.2020.01.003}}.

\bibitem{gandhi2011transcription}
Saumil~J Gandhi, Daniel Zenklusen, Timoth{\'e}e Lionnet, and Robert~H Singer.
\newblock Transcription of functionally related constitutive genes is not coordinated.
\newblock {\em Nature structural \& molecular biology}, 18(1):27--34, 2011.
\newblock \href {https://doi.org/10.1038/nsmb.1934} {\path{doi:10.1038/nsmb.1934}}.

\bibitem{sun2020size}
Xi-Ming Sun, Anthony Bowman, Miles Priestman, Francois Bertaux, Amalia Martinez-Segura, Wenhao Tang, Chad Whilding, Dirk Dormann, Vahid Shahrezaei, and Samuel Marguerat.
\newblock Size-dependent increase in {RNA} polymerase {II} initiation rates mediates gene expression scaling with cell size.
\newblock {\em Current Biology}, 30(7):1217--1230, 2020.
\newblock \href {https://doi.org/10.1016/j.cub.2020.01.053} {\path{doi:10.1016/j.cub.2020.01.053}}.

\bibitem{muthukrishnan2012dynamics}
Anantha-Barathi Muthukrishnan, Meenakshisundaram Kandhavelu, Jason Lloyd-Price, Fedor Kudasov, Sharif Chowdhury, Olli Yli-Harja, and Andre~S Ribeiro.
\newblock Dynamics of transcription driven by the teta promoter, one event at a time, in live escherichia coli cells.
\newblock {\em Nucleic acids research}, 40(17):8472--8483, 2012.
\newblock \href {https://doi.org/10.1093/nar/gks583} {\path{doi:10.1093/nar/gks583}}.

\bibitem{lionnet2010nuclear}
T~Lionnet, Bin Wu, D~Gr{\"u}nwald, RH~Singer, and DR~Larson.
\newblock Nuclear physics: quantitative single-cell approaches to nuclear organization and gene expression.
\newblock In {\em Cold Spring Harbor symposia on quantitative biology}, volume~75, page 113, 2010.
\newblock \href {https://doi.org/10.1101/sqb.2010.75.057} {\path{doi:10.1101/sqb.2010.75.057}}.

\bibitem{weidemann2023minimal}
Douglas~E Weidemann, James Holehouse, Abhyudai Singh, Ramon Grima, and Silke Hauf.
\newblock The minimal intrinsic stochasticity of constitutively expressed eukaryotic genes is sub-poissonian.
\newblock {\em Science Advances}, 9(32):eadh5138, 2023.
\newblock \href {https://doi.org/10.1126/sciadv.adh5138} {\path{doi:10.1126/sciadv.adh5138}}.

\bibitem{ramos2015gene}
Alexandre~F Ramos, Jos{\'e} Eduardo~M Hornos, and John Reinitz.
\newblock Gene regulation and noise reduction by coupling of stochastic processes.
\newblock {\em Physical Review E}, 91(2):020701, 2015.
\newblock \href {https://doi.org/10.1103/PhysRevE.91.020701} {\path{doi:10.1103/PhysRevE.91.020701}}.

\bibitem{szavitsnossangrima2023}
Juraj Szavits-Nossan and Ramon Grima.
\newblock Uncovering the effect of {RNA} polymerase steric interactions on gene expression noise: Analytical distributions of nascent and mature rna numbers.
\newblock {\em Phys. Rev. E}, 108:034405, Sep 2023.
\newblock \href {https://doi.org/10.1103/PhysRevE.108.034405} {\path{doi:10.1103/PhysRevE.108.034405}}.

\bibitem{voliotis2008fluctuations}
Margaritis Voliotis, Netta Cohen, Carmen Molina-Par{\'\i}s, and Tanniemola~B Liverpool.
\newblock Fluctuations, pauses, and backtracking in dna transcription.
\newblock {\em Biophysical journal}, 94(2):334--348, 2008.
\newblock \href {https://doi.org/10.1529/biophysj.107.105767} {\path{doi:10.1529/biophysj.107.105767}}.

\bibitem{bokes2012exact}
Pavol Bokes, John~R King, Andrew~TA Wood, and Matthew Loose.
\newblock Exact and approximate distributions of protein and {mRNA} levels in the low-copy regime of gene expression.
\newblock {\em Journal of mathematical biology}, 64:829--854, 2012.
\newblock \href {https://doi.org/10.1007/s00285-011-0433-5} {\path{doi:10.1007/s00285-011-0433-5}}.

\bibitem{paulsson2005models}
Johan Paulsson.
\newblock Models of stochastic gene expression.
\newblock {\em Physics of life reviews}, 2(2):157--175, 2005.
\newblock \href {https://doi.org/10.1016/j.plrev.2005.03.003} {\path{doi:10.1016/j.plrev.2005.03.003}}.

\bibitem{shahrezaei2008analytical}
Vahid Shahrezaei and Peter~S Swain.
\newblock Analytical distributions for stochastic gene expression.
\newblock {\em Proceedings of the National Academy of Sciences}, 105(45):17256--17261, 2008.
\newblock \href {https://doi.org/10.1073/pnas.0803850105} {\path{doi:10.1073/pnas.0803850105}}.

\bibitem{popovic2016geometric}
Nikola Popovi{\'c}, Carsten Marr, and Peter~S Swain.
\newblock A geometric analysis of fast-slow models for stochastic gene expression.
\newblock {\em Journal of mathematical biology}, 72:87--122, 2016.
\newblock \href {https://doi.org/10.1007/s00285-015-0876-1} {\path{doi:10.1007/s00285-015-0876-1}}.

\bibitem{jia2011intrinsic}
Tao Jia and Rahul~V Kulkarni.
\newblock Intrinsic noise in stochastic models of gene expression with molecular memory and bursting.
\newblock {\em Physical review letters}, 106(5):058102, 2011.
\newblock \href {https://doi.org/10.1103/PhysRevLett.106.058102} {\path{doi:10.1103/PhysRevLett.106.058102}}.

\bibitem{kumar2015transcriptional}
Niraj Kumar, Abhyudai Singh, and Rahul~V Kulkarni.
\newblock Transcriptional bursting in gene expression: analytical results for general stochastic models.
\newblock {\em PLoS computational biology}, 11(10):e1004292, 2015.
\newblock \href {https://doi.org/10.1371/journal.pcbi.1004292} {\path{doi:10.1371/journal.pcbi.1004292}}.

\bibitem{szavits2023charting}
Juraj Szavits-Nossan and Ramon Grima.
\newblock Charting the landscape of stochastic gene expression models using queueing theory.
\newblock {\em arXiv}, page 2307.03253, 2023.
\newblock \href {https://doi.org/10.48550/arXiv.2307.03253} {\path{doi:10.48550/arXiv.2307.03253}}.

\bibitem{thomas2012slow}
Philipp Thomas, Arthur~V Straube, and Ramon Grima.
\newblock The slow-scale linear noise approximation: an accurate, reduced stochastic description of biochemical networks under timescale separation conditions.
\newblock {\em BMC systems biology}, 6(1):1--23, 2012.
\newblock \href {https://doi.org/10.1186/1752-0509-6-39} {\path{doi:10.1186/1752-0509-6-39}}.

\bibitem{thomas2012rigorous}
Philipp Thomas, Ramon Grima, and Arthur~V Straube.
\newblock Rigorous elimination of fast stochastic variables from the linear noise approximation using projection operators.
\newblock {\em Physical Review E}, 86(4):041110, 2012.
\newblock \href {https://doi.org/10.1103/PhysRevE.86.041110} {\path{doi:10.1103/PhysRevE.86.041110}}.

\bibitem{eilertsen2022stochastic}
Justin Eilertsen, Kashvi Srivastava, and Santiago Schnell.
\newblock Stochastic enzyme kinetics and the quasi-steady-state reductions: Application of the slow scale linear noise approximation {\`a} la fenichel.
\newblock {\em Journal of Mathematical Biology}, 85(1):3, 2022.
\newblock \href {https://doi.org/10.1007/s00285-022-01768-6} {\path{doi:10.1007/s00285-022-01768-6}}.

\bibitem{fuda2009defining}
Nicholas~J Fuda, M~Behfar Ardehali, and John~T Lis.
\newblock Defining mechanisms that regulate {RNA} polymerase {II} transcription in vivo.
\newblock {\em Nature}, 461(7261):186--192, 2009.
\newblock \href {https://doi.org/10.1038/nature08449} {\path{doi:10.1038/nature08449}}.

\bibitem{voss2014dynamic}
Ty~C Voss and Gordon~L Hager.
\newblock Dynamic regulation of transcriptional states by chromatin and transcription factors.
\newblock {\em Nature Reviews Genetics}, 15(2):69--81, 2014.
\newblock \href {https://doi.org/10.1038/nrg3623} {\path{doi:10.1038/nrg3623}}.

\bibitem{friedman2012mechanism}
Larry~J Friedman and Jeff Gelles.
\newblock Mechanism of transcription initiation at an activator-dependent promoter defined by single-molecule observation.
\newblock {\em Cell}, 148(4):679--689, 2012.
\newblock \href {https://doi.org/10.1016/j.cell.2012.01.018} {\path{doi:10.1016/j.cell.2012.01.018}}.

\bibitem{lloyd2016dissecting}
Jason Lloyd-Price, Sofia Startceva, Vinodh Kandavalli, Jerome~G Chandraseelan, Nadia Goncalves, Samuel~MD Oliveira, Antti H{\"a}kkinen, and Andre~S Ribeiro.
\newblock Dissecting the stochastic transcription initiation process in live escherichia coli.
\newblock {\em DNA Research}, 23(3):203--214, 2016.
\newblock \href {https://doi.org/10.1093/dnares/dsw009} {\path{doi:10.1093/dnares/dsw009}}.

\bibitem{core2008transcription}
Leighton~J Core and John~T Lis.
\newblock Transcription regulation through promoter-proximal pausing of {RNA} polymerase {II}.
\newblock {\em Science}, 319(5871):1791--1792, 2008.
\newblock \href {https://doi.org/10.1126/science.1150843} {\path{doi:10.1126/science.1150843}}.

\bibitem{bartman2019transcriptional}
Caroline~R Bartman, Nicole Hamagami, Cheryl~A Keller, Belinda Giardine, Ross~C Hardison, Gerd~A Blobel, and Arjun Raj.
\newblock Transcriptional burst initiation and polymerase pause release are key control points of transcriptional regulation.
\newblock {\em Molecular cell}, 73(3):519--532, 2019.
\newblock \href {https://doi.org/10.1016/j.molcel.2018.11.004} {\path{doi:10.1016/j.molcel.2018.11.004}}.

\bibitem{shao2017paused}
Wanqing Shao and Julia Zeitlinger.
\newblock Paused {RNA} polymerase {II} inhibits new transcriptional initiation.
\newblock {\em Nature genetics}, 49(7):1045--1051, 2017.
\newblock \href {https://doi.org/10.1038/ng.3867} {\path{doi:10.1038/ng.3867}}.

\bibitem{roussel2006validation}
Marc~R Roussel and Rui Zhu.
\newblock Validation of an algorithm for delay stochastic simulation of transcription and translation in prokaryotic gene expression.
\newblock {\em Physical biology}, 3(4):274, 2006.
\newblock \href {https://doi.org/10.1088/1478-3975/3/4/005} {\path{doi:10.1088/1478-3975/3/4/005}}.

\bibitem{larson2011real}
Daniel~R Larson, Daniel Zenklusen, Bin Wu, Jeffrey~A Chao, and Robert~H Singer.
\newblock Real-time observation of transcription initiation and elongation on an endogenous yeast gene.
\newblock {\em science}, 332(6028):475--478, 2011.
\newblock \href {https://doi.org/10.1126/science.1202142} {\path{doi:10.1126/science.1202142}}.

\bibitem{alberts2022molecular}
Bruce Alberts, Rebecca Heald, Alexander Johnson, David Morgan, Martin Raff, Keith Roberts, and Peter Walter.
\newblock {\em Molecular biology of the cell 7th Edition}.
\newblock W. W. Norton, New York, NY, 2022.

\bibitem{khodor2011nascent}
Yevgenia~L Khodor, Joseph Rodriguez, Katharine~C Abruzzi, Chih-Hang~Anthony Tang, Michael~T Marr, and Michael Rosbash.
\newblock Nascent-seq indicates widespread cotranscriptional pre-{mRNA} splicing in drosophila.
\newblock {\em Genes \& development}, 25(23):2502--2512, 2011.
\newblock \href {https://doi.org/10.1101/gad.178962.111} {\path{doi:10.1101/gad.178962.111}}.

\bibitem{coulon2014kinetic}
Antoine Coulon, Matthew~L Ferguson, Valeria de~Turris, Murali Palangat, Carson~C Chow, and Daniel~R Larson.
\newblock Kinetic competition during the transcription cycle results in stochastic {RNA} processing.
\newblock {\em Elife}, 3:e03939, 2014.
\newblock \href {https://doi.org/10.7554/eLife.03939} {\path{doi:10.7554/eLife.03939}}.

\bibitem{cao2001computational}
Dan Cao and Roy Parker.
\newblock Computational modeling of eukaryotic {mRNA} turnover.
\newblock {\em {RNA}}, 7(9):1192--1212, 2001.
\newblock \href {https://doi.org/10.1017/S1355838201010330} {\path{doi:10.1017/S1355838201010330}}.

\bibitem{deneke2013complex}
Carlus Deneke, Reinhard Lipowsky, and Angelo Valleriani.
\newblock Complex degradation processes lead to non-exponential decay patterns and age-dependent decay rates of messenger {RNA}.
\newblock {\em PloS one}, 8(2):e55442, 2013.
\newblock \href {https://doi.org/10.1371/journal.pone.0055442} {\path{doi:10.1371/journal.pone.0055442}}.

\bibitem{coller2005general}
Jeff Coller and Roy Parker.
\newblock General translational repression by activators of {mRNA} decapping.
\newblock {\em Cell}, 122(6):875--886, 2005.
\newblock \href {https://doi.org/10.1016/j.cell.2005.07.012} {\path{doi:10.1016/j.cell.2005.07.012}}.

\bibitem{huch2014interrelations}
Susanne Huch and Tracy Nissan.
\newblock Interrelations between translation and general {mRNA} degradation in yeast.
\newblock {\em Wiley Interdisciplinary Reviews: {RNA}}, 5(6):747--763, 2014.
\newblock \href {https://doi.org/10.1002/wrna.1244} {\path{doi:10.1002/wrna.1244}}.

\bibitem{fralix2023markovian}
Brian Fralix, Mark Holmes, and Andreas L{\"o}pker.
\newblock A markovian arrival stream approach to stochastic gene expression in cells.
\newblock {\em Journal of Mathematical Biology}, 86(5):79, 2023.
\newblock \href {https://doi.org/10.1007/s00285-023-01913-9} {\path{doi:10.1007/s00285-023-01913-9}}.

\bibitem{dean2022noise}
Justin Dean and Ayalvadi Ganesh.
\newblock Noise dissipation in gene regulatory networks via second order statistics of networks of infinite server queues.
\newblock {\em Journal of Mathematical Biology}, 85(2):14, 2022.
\newblock \href {https://doi.org/10.1007/s00285-022-01781-9} {\path{doi:10.1007/s00285-022-01781-9}}.

\bibitem{elgart2010applications}
Vlad Elgart, Tao Jia, and Rahul~V. Kulkarni.
\newblock Applications of little's law to stochastic models of gene expression.
\newblock {\em Phys. Rev. E}, 82:021901, Aug 2010.
\newblock \href {https://doi.org/10.1103/PhysRevE.82.021901} {\path{doi:10.1103/PhysRevE.82.021901}}.

\bibitem{choubey2018nascent}
Sandeep Choubey.
\newblock Nascent {RNA} kinetics: Transient and steady state behavior of models of transcription.
\newblock {\em Physical Review E}, 97(2):022402, 2018.
\newblock \href {https://doi.org/10.1103/PhysRevE.97.022402} {\path{doi:10.1103/PhysRevE.97.022402}}.

\bibitem{thattai2016universal}
Mukund Thattai.
\newblock Universal poisson statistics of mrnas with complex decay pathways.
\newblock {\em Biophysical journal}, 110(2):301--305, 2016.
\newblock \href {https://doi.org/10.1016/j.bpj.2015.12.001} {\path{doi:10.1016/j.bpj.2015.12.001}}.

\bibitem{mather2013translational}
William~H Mather, Jeff Hasty, Lev~S Tsimring, and Ruth~J Williams.
\newblock Translational cross talk in gene networks.
\newblock {\em Biophysical journal}, 104(11):2564--2572, 2013.
\newblock \href {https://doi.org/10.1016/j.bpj.2013.04.049} {\path{doi:10.1016/j.bpj.2013.04.049}}.

\bibitem{Takacs_1958}
L.~Tak{\'a}cs.
\newblock On a coincidence problem concerning telephone traffic.
\newblock {\em Acta Mathematica Academiae Scientiarum Hungarica}, 9(1):45--81, 3 1958.
\newblock \href {https://doi.org/10.1007/BF02023865} {\path{doi:10.1007/BF02023865}}.

\bibitem{Liu_1990}
L.~Liu, B.~R.~K. Kashyap, and J.~G.~C. Templeton.
\newblock On the gix/g/infinity system.
\newblock {\em Journal of Applied Probability}, 27(3):671--683, 1990.
\newblock \href {https://doi.org/10.2307/3214550} {\path{doi:10.2307/3214550}}.

\bibitem{Cox_1967}
D.R. Cox.
\newblock {\em Renewal theory}.
\newblock Methuen, London, 1967.

\bibitem{Szavits_2023}
Juraj Szavits-Nossan and Ramon Grima.
\newblock Steady-state distributions of nascent {RNA} for general initiation mechanisms.
\newblock {\em Phys. Rev. Res.}, 5:013064, 1 2023.
\newblock \href {https://doi.org/10.1103/PhysRevResearch.5.013064} {\path{doi:10.1103/PhysRevResearch.5.013064}}.

\bibitem{kim2014validity}
Jae~Kyoung Kim, Kre{\v{s}}imir Josi{\'c}, and Matthew~R Bennett.
\newblock The validity of quasi-steady-state approximations in discrete stochastic simulations.
\newblock {\em Biophysical journal}, 107(3):783--793, 2014.
\newblock \href {https://doi.org/10.1016/j.bpj.2014.06.012} {\path{doi:10.1016/j.bpj.2014.06.012}}.

\bibitem{kim2017reduction}
Jae~Kyoung Kim and Eduardo~D Sontag.
\newblock Reduction of multiscale stochastic biochemical reaction networks using exact moment derivation.
\newblock {\em PLoS computational biology}, 13(6):e1005571, 2017.
\newblock \href {https://doi.org/10.1371/journal.pcbi.1005571} {\path{doi:10.1371/journal.pcbi.1005571}}.

\bibitem{mastny2007two}
Ethan~A Mastny, Eric~L Haseltine, and James~B Rawlings.
\newblock Two classes of quasi-steady-state model reductions for stochastic kinetics.
\newblock {\em The Journal of chemical physics}, 127(9), 2007.
\newblock \href {https://doi.org/10.1063/1.2764480} {\path{doi:10.1063/1.2764480}}.

\bibitem{herath2018reduced}
Narmada Herath and Domitilla Del~Vecchio.
\newblock Reduced linear noise approximation for biochemical reaction networks with time-scale separation: The stochastic tqssa+.
\newblock {\em The Journal of Chemical Physics}, 148(9), 2018.
\newblock \href {https://doi.org/10.1063/1.5012752} {\path{doi:10.1063/1.5012752}}.

\bibitem{kang2019quasi}
Hye-Won Kang, Wasiur~R KhudaBukhsh, Heinz Koeppl, and Grzegorz~A Rempa{\l}a.
\newblock Quasi-steady-state approximations derived from the stochastic model of enzyme kinetics.
\newblock {\em Bulletin of mathematical biology}, 81:1303--1336, 2019.
\newblock \href {https://doi.org/10.1007/s11538-019-00574-4} {\path{doi:10.1007/s11538-019-00574-4}}.

\bibitem{macdonald1998symmetric}
Ian~Grant Macdonald.
\newblock {\em Symmetric functions and Hall polynomials}.
\newblock Oxford university press, Oxford, 1998.
\newblock \href {https://doi.org/10.1093/oso/9780198534891.001.0001} {\path{doi:10.1093/oso/9780198534891.001.0001}}.

\bibitem{battich2015control}
Nico Battich, Thomas Stoeger, and Lucas Pelkmans.
\newblock Control of transcript variability in single mammalian cells.
\newblock {\em Cell}, 163(7):1596--1610, 2015.
\newblock \href {https://doi.org/10.1016/j.cell.2015.11.018} {\path{doi:10.1016/j.cell.2015.11.018}}.

\bibitem{lammers2020matter}
Nicholas~C Lammers, Yang~Joon Kim, Jiaxi Zhao, and Hernan~G Garcia.
\newblock A matter of time: Using dynamics and theory to uncover mechanisms of transcriptional bursting.
\newblock {\em Current opinion in cell biology}, 67:147--157, 2020.
\newblock \href {https://doi.org/10.1016/j.ceb.2020.08.001} {\path{doi:10.1016/j.ceb.2020.08.001}}.

\bibitem{hardy1952inequalities}
Godfrey~Harold Hardy, John~Edensor Littlewood, and George P{\'o}lya.
\newblock {\em Inequalities}.
\newblock Cambridge university press, Cambridge, 1952.
\newblock \href {https://doi.org/10.1007/978-3-642-61983-0_6} {\path{doi:10.1007/978-3-642-61983-0_6}}.

\bibitem{ingalls2008sensitivity}
Olaf Wolkenhauer, Peter Wellstead, Kwang-Hyun Cho, and Brian Ingalls.
\newblock {Sensitivity analysis: from model parameters to system behaviour}.
\newblock {\em Essays in Biochemistry}, 45:177--194, 09 2008.
\newblock \href {https://doi.org/10.1042/bse0450177} {\path{doi:10.1042/bse0450177}}.

\bibitem{elf2003fast}
Johan Elf and M{\aa}ns Ehrenberg.
\newblock Fast evaluation of fluctuations in biochemical networks with the linear noise approximation.
\newblock {\em Genome research}, 13(11):2475--2484, 2003.
\newblock \href {https://doi.org/10.1101/gr.1196503} {\path{doi:10.1101/gr.1196503}}.

\bibitem{pedraza2008effects}
Juan~M Pedraza and Johan Paulsson.
\newblock Effects of molecular memory and bursting on fluctuations in gene expression.
\newblock {\em Science}, 319(5861):339--343, 2008.
\newblock \href {https://doi.org/10.1126/science.1144331} {\path{doi:10.1126/science.1144331}}.

\bibitem{filatova2022modulation}
Tatiana Filatova, Nikola Popovi{\'c}, and Ramon Grima.
\newblock Modulation of nuclear and cytoplasmic {mRNA} fluctuations by time-dependent stimuli: Analytical distributions.
\newblock {\em Mathematical biosciences}, 347:108828, 2022.
\newblock \href {https://doi.org/10.1016/j.mbs.2022.108828} {\path{doi:10.1016/j.mbs.2022.108828}}.

\bibitem{gorin2022modeling}
Gennady Gorin and Lior Pachter.
\newblock Modeling bursty transcription and splicing with the chemical master equation.
\newblock {\em Biophysical Journal}, 121(6):1056--1069, 2022.
\newblock \href {https://doi.org/10.1016/j.bpj.2022.02.004} {\path{doi:10.1016/j.bpj.2022.02.004}}.

\bibitem{szavits2022mean}
Juraj Szavits-Nossan and Ramon Grima.
\newblock Mean-field theory accurately captures the variation of copy number distributions across the {mRNA} life cycle.
\newblock {\em Physical Review E}, 105(1):014410, 2022.
\newblock \href {https://doi.org/10.1103/PhysRevE.105.014410} {\path{doi:10.1103/PhysRevE.105.014410}}.

\bibitem{shi2023stochastic}
Changhong Shi, Xiyan Yang, Jiajun Zhang, and Tianshou Zhou.
\newblock Stochastic modeling of the {mRNA} life process: A generalized master equation.
\newblock {\em Biophysical Journal}, 122(20):4023--4041, 2023.
\newblock \href {https://doi.org/10.1016/j.bpj.2023.08.024} {\path{doi:10.1016/j.bpj.2023.08.024}}.

\bibitem{hansen2018cytoplasmic}
Maike~MK Hansen, Ravi~V Desai, Michael~L Simpson, and Leor~S Weinberger.
\newblock Cytoplasmic amplification of transcriptional noise generates substantial cell-to-cell variability.
\newblock {\em Cell systems}, 7(4):384--397, 2018.
\newblock \href {https://doi.org/10.1016/j.cels.2018.08.002} {\path{doi:10.1016/j.cels.2018.08.002}}.

\bibitem{dar2012transcriptional}
Roy~D Dar, Brandon~S Razooky, Abhyudai Singh, Thomas~V Trimeloni, James~M McCollum, Chris~D Cox, Michael~L Simpson, and Leor~S Weinberger.
\newblock Transcriptional burst frequency and burst size are equally modulated across the human genome.
\newblock {\em Proceedings of the National Academy of Sciences}, 109(43):17454--17459, 2012.
\newblock \href {https://doi.org/10.1073/pnas.1213530109} {\path{doi:10.1073/pnas.1213530109}}.

\bibitem{padovan2015single}
Olivia Padovan-Merhar, Gautham~P Nair, Andrew~G Biaesch, Andreas Mayer, Steven Scarfone, Shawn~W Foley, Angela~R Wu, L~Stirling Churchman, Abhyudai Singh, and Arjun Raj.
\newblock Single mammalian cells compensate for differences in cellular volume and dna copy number through independent global transcriptional mechanisms.
\newblock {\em Molecular cell}, 58(2):339--352, 2015.
\newblock \href {https://doi.org/10.1016/j.molcel.2015.03.005} {\path{doi:10.1016/j.molcel.2015.03.005}}.

\bibitem{klein2015droplet}
Allon~M Klein, Linas Mazutis, Ilke Akartuna, Naren Tallapragada, Adrian Veres, Victor Li, Leonid Peshkin, David~A Weitz, and Marc~W Kirschner.
\newblock Droplet barcoding for single-cell transcriptomics applied to embryonic stem cells.
\newblock {\em Cell}, 161(5):1187--1201, 2015.
\newblock \href {https://doi.org/10.1016/j.cell.2015.04.044} {\path{doi:10.1016/j.cell.2015.04.044}}.

\end{thebibliography}

\end{document}